\documentclass[structabstract]{aa}  
\usepackage{graphicx}
\usepackage{amsmath,amssymb}
\usepackage{txfonts}
\begin{document}

\def\Arrow{\mathop{\longrightarrow}\limits}
\def\Harpoons{\mathop{\rightleftharpoons}\limits}

   \title{A (sub)millimetre study of dense cores in Orion B9\thanks{This publication is based on data acquired with the Atacama Pathfinder EXperiment (APEX) under programmes 079.F-9313A, 084.F-9304A, and 084.F-9312A. APEX is a collaboration between the Max-Planck-Institut f\"{u}r Radioastronomie, the European Southern Observatory, and the Onsala Space Observatory.}\fnmsep \thanks{Our SABOCA map 
shown in Fig.~2 is available electronically in FITS format at the CDS.}}

   \author{O. Miettinen\inst{1}, J. Harju\inst{2,1}, L.~K. Haikala\inst{2,1}, \and M. Juvela\inst{1}}

 \offprints{O. Miettinen}

   \institute{Department of Physics, P.O. Box 64, FI-00014 University of 
Helsinki, Finland\\ \email{oskari.miettinen@helsinki.fi} \and Finnish Centre 
for Astronomy with ESO (FINCA), University of Turku, V\"ais\"al\"antie 20, 
FI-21500 Piikki\"o, Finland}

   \date{Received ; accepted}

% \abstract{}{}{}{}{} 
% 5 {} token are mandatory
\authorrunning{Miettinen et al.}
\titlerunning{Dense cores in Orion B9}

  \abstract
   {Studies of dense molecular-cloud 
cores at (sub)millimetre wavelengths are 
needed to understand the early stages of star formation.}
   {We aim to further constrain the properties and evolutionary stages of 
dense cores in Orion B9.  
The prime objective of this study is to examine the dust emission of the cores 
near the peak of their spectral energy distributions, and to determine the 
degrees of CO depletion, deuterium fractionation, and ionisation.}
   {The central part of Orion B9 was mapped at 350 $\mu$m with APEX/SABOCA. 
A sample of nine cores in the region were observed in C$^{17}$O$(2-1)$, 
H$^{13}$CO$^+(4-3)$ (towards 3 sources), DCO$^+(4-3)$, N$_2$H$^+(3-2)$, 
and N$_2$D$^+(3-2)$ with APEX/SHFI. These data are used in conjunction with 
our previous APEX/LABOCA 870-$\mu$m dust continuum data.} 
   {All the LABOCA cores in the region covered by our SABOCA map were detected 
at 350 $\mu$m. The strongest 350 $\mu$m emission is seen towards 
the Class 0 candidate SMM 3. Many of the LABOCA cores show 
evidence of substructure in the higher-resolution SABOCA image. 
In particular, we report on the discovery of multiple very low-mass
condensations in the prestellar core SMM 6. Based on the 350-to-870 $\mu$m 
flux density ratios, we determine dust temperatures of 
$T_{\rm dust}\simeq7.9-10.8$ K, and dust emissivity indices of  
$\beta\sim0.5-1.8$. The CO depletion factors are in the range 
$f_{\rm D}\sim 1.6-10.8$. The degree of deuteration in N$_2$H$^+$ is 
$\simeq 0.04-0.99$, where the highest value (seen towards the 
prestellar core SMM 1) is, to our knowledge, the most 
extreme level of N$_2$H$^+$ deuteration reported so far. The level of HCO$^+$ 
deuteration is about 1--2\%. The fractional ionisation and cosmic-ray 
ionisation rate of H$_2$ could be determined only towards two sources with 
the lower limits of $\sim2-6\times10^{-8}$ and 
$\sim2.6\times10^{-17}-4.8\times10^{-16}$ s$^{-1}$, respectively.  
We also detected D$_2$CO towards two sources.}
   {The detected protostellar cores are classified as Class 0 objects, in 
agreement with our previous SED results. The detection of subcondensations 
within SMM 6 shows that core fragmentation can already take place during the 
prestellar phase. The origin of this substructure is likely caused by thermal 
Jeans fragmentation of the elongated parent core. Varying levels of $f_{\rm D}$ 
and deuteration among the cores suggest that they are evolving chemically at 
different rates. A low $f_{\rm D}$ value and the presence of gas-phase D$_2$CO 
in SMM 1 suggest that the core chemistry is affected by the nearby outflow. 
The very high N$_2$H$^+$ deuteration in SMM 1 is likely to be remnant of the 
earlier CO-depleted phase.}


   \maketitle
%
%________________________________________________________________

\section{Introduction}

New stars in our Galaxy form predominantly in the so-called 
giant molecular clouds (GMCs). The nearest GMC to the Sun is the Orion 
molecular cloud complex\footnote{In this paper we adopt a distance of 450 pc 
to the Orion GMC (\cite{genzel1989}). The actual distance may be somewhat 
smaller as, for example, Menten et al. (2007) determined the distance to the 
Orion Nebula to be $414\pm7$ pc.}. This cloud complex contains a number of 
star-forming regions, one example being the Orion B9 region in the central 
part of Orion B mole\-cular cloud. Orion B9 represents a very early stage
of star formation which is manifest in the fact that most ``dense
cores'' in the cloud are either prestellar or contain Class 0 protostars. 
These objects are cold, and their study requires observations at 
(sub)millimetre wavelengths.

\begin{figure*}
\begin{center}
\includegraphics[width=0.85\textwidth]{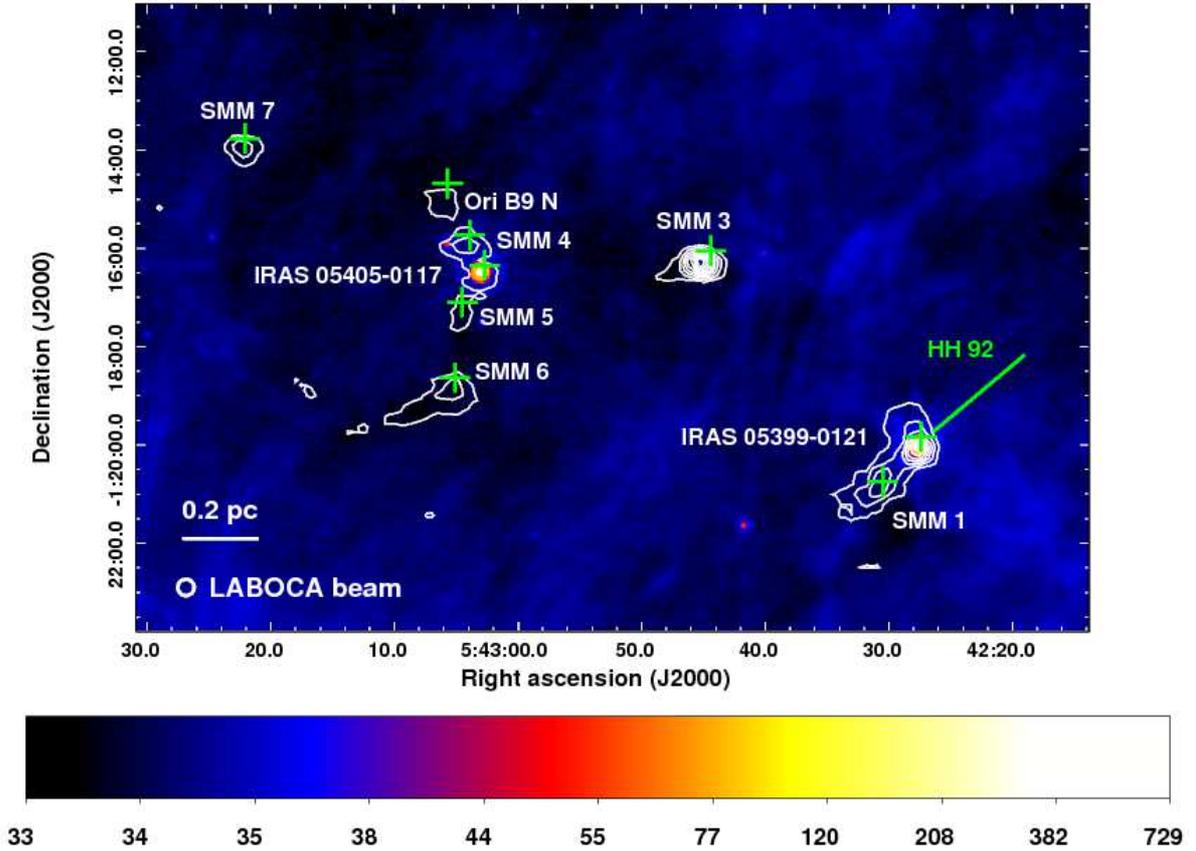}
\caption{\textit{Spitzer}/MIPS 24-$\mu$m image of the central part 
of Orion B9 overlaid with contours showing the LABOCA 870-$\mu$m dust continuum
emission. The contours go from 0.1 ($\sim3.3\sigma$) to 1.0 
Jy~beam$^{-1}$ in steps of 0.1 Jy~beam$^{-1}$. The 24-$\mu$m image is shown 
with a logarithmic scaling to improve the contrast between bright and faint 
features. The colour bar indicates the 24-$\mu$m intensity scale in units of 
MJy~sr$^{-1}$. The green plus signs show the target positions of our 
molecular-line observations (i.e., the submm peak positions of the LABOCA map 
before adjusting the pointing; see text). The green line shows the base and 
northwest tip of the HH 92 jet driven by IRAS 05399-0121 (\cite{bally2002}). 
The 0.2-pc scale bar and the effective LABOCA beam HPBW ($\sim20\arcsec$) are 
shown in the bottom left (Paper I).}
\label{figure:laboca}
\end{center}
\end{figure*}

Using the LABOCA bolometer array on APEX (Atacama Pathfinder EXperiment), 
we mapped that region at 870 $\mu$m (\cite{miettinen2009}; hereafter Paper I). 
The dust continuum mapping resulted in the discovery of twelve dense cores 
of which four were found to be associated with IRAS point sources, and eight 
of them appeared previously unknown sources. Of the newly disco\-vered cores, 
two (we called SMM 3 and 4) were found to be associated with 
\textit{Spitzer} 24- and 70-$\mu$m sources, and were tentatively classified 
as Class-0 protostellar candidates. Six cores showed no signs of embedded 
infrared emission, and were thus classified as starless. 
Figure~\ref{figure:laboca} shows the 870 $\mu$m emission as contours overlaid 
on the \textit{Spitzer}/MIPS 24-$\mu$m image of the central part of Orion 
B9\footnote{Comparison of our LABOCA map to the new SABOCA map 
together with \textit{Spitzer} images showed that the LABOCA observations were 
not well-pointed. The SABOCA peaks appeared to systematically
lie southeast from the LABOCA peak positions. The difference in angular 
resolution between SABOCA and LABOCA data did not appear to be the cause 
of the offset in peak positions; by smoothing the SABOCA map to correspond 
the resolution of our LABOCA data, and regridding the maps onto the same 
grid, the offsets still remained. We used the brightest SABOCA 350-$\mu$m 
source in the region, SMM3, as a reference to adjust the pointing of our 
LABOCA map. The pointing of the LABOCA map was shifted by 
$(\Delta \alpha,\, \Delta \delta)=(+1\farcs72, \, -9\farcs97)$. 
Unfortunately, the target positions of our molecular-line observations, which 
were chosen to be the LABOCA peak positions from our previous map, are now 
slightly offset from the dust maxima (but still within the beam size for most 
line observations).}. In Paper I, we also derived the degree of deuteration, 
or the $N({\rm N_2D^+})/N({\rm N_2H^+})$ co\-lumn density ratio, towards 
selected positions and found the values in the range 0.03--0.04, comparable 
to those seen in other low-mass star-forming regions 
(e.g., \cite{crapsi2005}; \cite{emprechtinger2009}; \cite{friesen2010b}). 
Taking advantage of the H$_2$D$^+$ data from Harju et al. (2006), the 
ionisation degree and the cosmic-ray ionisation rate of H$_2$ towards the 
same target positions were estimated to be $x({\rm e})\sim10^{-7}$ and 
$\zeta_{\rm H_2}\sim1-2\times10^{-16}$ s$^{-1}$, respectively. 

The physical properties of dense cores in Orion B9 
were studied further by Miettinen et al. (2010; hereafter Paper II). Using 
the observations of the NH$_3$ $(1,\,1)$ and $(2,\,2)$ inversion transitions 
performed with the Effelsberg 100-m telescope, we determined the gas kinetic 
temperature, kinematical properties, and dynamical state of the cores. 
The gas kinetic temperature of the cores were found to be 
$T_{\rm kin}\sim9.4-13.9$ K, and the internal non-thermal motions in the 
cores appeared to be subsonic, or at most transonic. The virial-parameter 
analysis showed that the starless cores in the region are likely to be 
gravitationally bound, and thus prestellar objects. Interestingly, some of 
the cores were found to have significantly (by $\sim5-7.5$ km~s$^{-1}$) lower 
radial velocity than the systemic velocity of the region 
(${\rm v_{\rm LSR}}\sim9$ km~s$^{-1}$). This suggests that they belong to 
the ``low-velocity part'' of Orion B which is likely to originate from the 
feedback from the massive stars of the nearby Ori OB 1b association.

In this paper, we present the results of our new APEX observations of the 
Orion B9 region. These include the dust continuum mapping with the SABOCA 
(Submillimetre APEX Bolometer Camera) bolometer array at 350 $\mu$m, and 
observations of the molecular-line transitions C$^{17}$O$(2-1)$, 
H$^{13}$CO$^+(4-3)$, DCO$^+(4-3)$, N$_2$H$^+(3-2)$, and N$_2$D$^+(3-2)$. 
The SABOCA 350-$\mu$m observations allow us to probe the peak of the core's 
spectral energy distributions (SEDs). We also aim to study the degrees 
of CO depletion and deuterium fractionation, and the fractional ionisation in 
the cores. These are useful parameters to further constrain the physical 
and chemical properties of the sources, and their evolutionary stages.

This paper is structured as follows. The observations and data-reduction 
procedures are described in Sect.~2. The direct observational results are 
presented in Sect.~3. In Sect.~4, we des\-cribe the analysis and present the 
results of the physical and chemical properties of the cores. Discussion of 
our results is presented in Sect.~5, and in Sect.~6, we summarise 
the main conclusions of this study.

\section{Observations and data reduction}

\subsection{Submillimetre dust continuum observations}

The central $18\farcm0 \times 14\farcm5$ part of Orion B9 was 
mapped with SABOCA (\cite{siringo2010}) on the APEX 12-m telescope 
(\cite{gusten2006}) at Llano de Chajnantor (Chilean Andes). 
SABOCA is a 37-channel on-sky bolometer array operating at 350 $\mu$m, with 
a nominal re\-solution of $\sim7\farcs5$ (HPBW). The effective field of 
view of the array is 1\farcm5. The SABOCA passband has an equivalent width 
of about 120 GHz centred on an effective frequency of 852 GHz.

The observations were carried out on 5 October and 10 November 2009. The 
atmospheric zenith opacity at 350 $\mu$m was measured using the sky-dip 
method, and was found to be in the range 
$\tau_{\rm z}^{350\,\mu{\rm m}}=0.8-1.1$. The amount of precipitable water 
vapour (PWV) was in the range 0.4--0.6 mm. 
The telescope pointing and focus checks were made at regular intervals using 
the pla\-nets Mars, Jupiter, and Uranus, and several secondary ca\-librators 
(e.g., CRL618, N2071IR, and CW Leo). The absolute calibration 
uncertainty for SABOCA is 25--30\%. Mapping was performed using a 
'fast-scanning' method without chopping the secondary mirror 
(\cite{reichertz2001}). The mosaic was constructed by combining 147 individual 
fast-scanning (typically 1\farcm2~s$^{-1}$) maps. The total integration time 
spent on the area was 9.7 hr.

Data reduction was done with the CRUSH-2 (Comprehensive Reduction Utility for 
SHARC-2) (version 2.03-2) software package (\cite{kovacs2008}), which includes 
calibration data covering the period of our SABOCA observations. We used the 
pipeline iterations with specifying the 'deep' option, which is appropriate for 
point-like sources. To improve the image appearance and 
increase the signal-to-noise (S/N) ratio, a beam-smoothing was applied, i.e., 
the map was smoothed with a Gaussian kernel of the size 
7\farcs5 (FWHM). Therefore, the angular resolution of the final image is 
10\farcs6 (0.02 pc at 450 pc). The gridding was done with a cell size of 
1\farcs5. The resulting $1\sigma$ rms noise level in the final co-added map 
is $\sim0.06$ Jy~beam$^{-1}$.

\subsection{Molecular-line observations}

The spectral-line observations of C$^{17}$O$(2-1)$, DCO$^+(4-3)$, 
N$_2$H$^+(3-2)$, and N$_2$D$^+(3-2)$ towards nine Orion B9 cores, and 
H$^{13}$CO$^+(4-3)$ observations towards three cores, were carried out on 
2--3, 8--9, and 11--12 September, and 13 and 18--19 
November 2009 with APEX. The target cores with their 
physical pro\-perties derived in Paper II are listed in 
Table~\ref{table:sources}. In Table~\ref{table:sources} we give the 
LABOCA submm peak positions of the cores after adjusting the pointing. The 
target positions of our molecular-line observations can be found in Table~1 
of Paper II. The observed transitions, their spectroscopic 
properties, and observational parameters are listed in 
Table~\ref{table:lines}. The N$_2$H$^+(3-2)$ observations were already 
presented in Paper II, but are re-analysed in the present paper. 

As frontend for the C$^{17}$O$(2-1)$ and N$_2$D$^+(3-2)$ observations, we used 
APEX-1 of the SHFI (Swedish Heterodyne Facility Instrument; 
\cite{belitsky2007}; \cite{vassilev2008a},b). 
APEX-1 operates in single-sideband (SSB) mode using sideband separation 
mixers, and it has a sideband rejection ratio $>10$ dB. The SHFI 
sidebands are separated by 12 GHz, i.e., $\pm6$ GHz around the local 
oscillator (LO) frequency. Therefore, the centre frequency for the image band 
is 12 GHz above or below the observing frequency depending on whether the 
receiver tuning is optimised for operation in the upper or lower sideband 
(USB or LSB), respectively. For the H$^{13}$CO$^+(4-3)$, DCO$^+(4-3)$, and 
N$_2$H$^+(3-2)$ observations the frontend used was APEX-2, which has similar 
characteristics as APEX-1. The backend for all observations was the Fast 
Fourier Transfrom Spectrometer (FFTS; \cite{klein2006}) with two 1 
GHz units. Both units were divided into 8\,192 channels. The two units were 
connected to one receiver each, thus providing 1 GHz bandwidth for two 
receivers simultaneously.

The observations were performed in the wobbler-switching mode with a 
$100\arcsec$ azimuthal throw (symmetric offsets) and a chopping rate of 0.5 Hz.
The telescope pointing and focus corrections were checked by 
continuum scans on the planets Mars and Uranus, and 
the pointing was found to be accurate to $\sim3\arcsec$. Calibration was 
done by the chopper-wheel technique, and the output intensity scale given by 
the system is $T_{\rm A}^*$, the antenna temperature corrected for atmospheric
attenuation. The observed intensities were converted to the main-beam 
brightness tem\-perature scale by $T_{\rm MB}=T_{\rm A}^*/\eta_{\rm MB}$, where 
$\eta_{\rm MB}=B_{\rm eff}/F_{\rm eff}$ is the main beam efficiency, and 
$B_{\rm eff}$ and $F_{\rm eff}$ are the beam and forward efficiencies, 
respectively. The absolute calibration uncertainty is estimated to be 10\%.

The spectra were reduced using the CLASS90 programme of the IRAM's GILDAS 
software package\footnote{{\tt http://www.iram.fr/IRAMFR/GILDAS}}. 
The individual spectra were averaged and the resulting spectra were 
Hanning-smoothed in order to improve the S/N ratio of the 
data. A first- or third-order polynomial was applied to correct the baseline 
in the final spectra. The resulting $1\sigma$ rms noise levels are 
$\sim20-80$ mK at the smoothed resolutions. As shown in the second last 
column of Table~\ref{table:lines}, the on-source integration times were 
diffe\-rent between different sources. This is particularly the case for 
C$^{17}$O$(2-1)$, N$_2$D$^+(3-2)$, and H$^{13}$CO$^+(4-3)$ where $t_{\rm int}$ 
varies by a factor of $\sim3-7$. This explains the variations in the rms 
noise values (up to a factor of 2.4 in the C$^{17}$O data).

The $J=2-1$ transition of C$^{17}$O contains nine hyperfine (hf) components. 
We fitted this hf structure using ``method hfs'' of CLASS90 to derive the LSR 
velocity (${\rm v}_{\rm LSR}$) of the emission, and FWHM linewidth 
($\Delta {\rm v}$). The hf line fitting can also be used to derive the line 
optical thickness, $\tau$. However, in all spectra the hf components are 
mostly blended together and thus the optical thickness could not be reliably 
determined. For the rest frequencies of the hf components, we used the values 
from Ladd et al. (1998; Table 6 therein). The adopted central frequency, 
224\,714.199 MHz, is that of the $J_F = 2_{9/2} \rightarrow 1_{7/2}$ hf 
component which has a relative intensity of $R_i=\frac{1}{3}$.

Also, rotational lines of H$^{13}$CO$^+$ and DCO$^+$ also have 
hf structure (e.g., \cite{schmid2004}; \cite{caselli2005}). 
The $J=4-3$ transition of DCO$^+$ is split up into six hf components. 
To fit this hf structure, we used the rest frequencies from the 
CDMS database (\cite{muller2005}). The adopted central frequency of 
DCO$^+(4-3)$, 288\,143.855 MHz, is that of the $J_F = 4_5 \rightarrow 3_4$ hf 
component which has a relative intensity of $R_i=\frac{11}{27}$. We note that 
this frequency is 2.7 kHz lower than the value determined by Caselli \& Dore 
(2005; their Table~5). The frequency interval of the hf components for 
DCO$^+(4-3)$ is very small. Therefore, the lines overlap significantly 
which causes the hf structure to be heavily blended. In general, for the 
$J_{\rm upper}\geq3$ lines of DCO$^+$ the hf components are so heavily 
blended that, even determined through a single Gaussian fit, the linewidth 
is not expected to be significantly overestimated (\cite{caselli2005}). 
The hf structure of H$^{13}$CO$^+$ lines is more complicated because both 
the $^{1}$H and $^{13}$C nuclei have a nuclear spin of $I=1/2$, and the 
nuclear magnetic spin couples to rotation. To our knowledge, the rest 
frequencies of the H$^{13}$CO$^+(4-3)$ hf components have not been published. 
We thus fitted these lines using a single Gaussian fit to derive the values 
of ${\rm v}_{\rm LSR}$ and $\Delta {\rm v}$. As in the case of 
DCO$^+(J_{\rm upper}\geq3)$, the $\Delta {\rm v}$ thus determined is not 
expected to be significantly overestimated because of the 
strong blending of the hf components. The central frequency used was 
346\,998.344 MHz (CDMS), which is 3.0 kHz lower than the value determined by 
Schmid-Burgk et al. (2004; Table 3 therein).

The $J=3-2$ transitions of both N$_2$H$^+$ and N$_2$D$^+$ contain 38 hf
components. The hf lines were fitted using the rest frequencies 
from Pagani et al. (2009b; Tables 4 and 10 therein). The adopted central 
frequencies of N$_2$H$^+(3-2)$ and N$_2$D$^+(3-2)$, 279\,511.832 and 
231\,321.912 MHz, are those of the $J_{F_1 F} = 3_{45} \rightarrow 2_{34}$
hf component which has a relative intensity of $R_i=\frac{11}{63}$.
Also in these cases, the hf components are blended and thus the value 
of $\tau$ could not be reliably determined through hf fitting.

\begin{table*}
\caption{Source list.}
\begin{minipage}{2\columnwidth}
\centering
\renewcommand{\footnoterule}{}
\label{table:sources}
\begin{tabular}{c c c c c c c c}
\hline\hline 
Source & $\alpha_{2000.0}$\tablefootmark{a} & $\delta_{2000.0}$\tablefootmark{a} & $T_{\rm kin}$ & $M$ & $N({\rm H_2})$\tablefootmark{b} & $\langle n({\rm H_2}) \rangle$ & Class\\
       & [h:m:s] & [$\degr$:$\arcmin$:$\arcsec$] & [K] & [M$_{\sun}$] & [$10^{22}$ cm$^{-2}$] & [$10^4$ cm$^{-3}$] & \\
\hline
IRAS 05399-0121 & 05 42 27.5 & -01 20 00 & $13.5\pm1.6$ & $7.8\pm0.5$ & $4.2\pm0.9$/$2.8\pm0.6$ & $5.5\pm1.3$ & 0/I\\
SMM 1 & 05 42 30.5 & -01 20 55 & $11.9\pm0.9$ & $11.1\pm0.3$ & $2.8\pm0.4$/$2.7\pm0.4$ & $5.3\pm0.9$ & prestellar\\
SMM 3 & 05 42 45.2 & -01 16 13 & $11.3\pm0.8$ & $7.8\pm0.6$ & $8.4\pm1.1$/$2.5\pm0.3$ & $10.5\pm2.1$ & 0\\
IRAS 05405-0117 & 05 43 02.7 & -01 16 31 & $11.3\pm0.6$ & $2.8\pm0.4$ & $1.4\pm0.1$/$1.2\pm0.1$ & $3.8\pm0.5$ & 0\\
SMM 4 & 05 43 04.0 & -01 15 54 & $13.9\pm0.8$ & $2.8\pm0.3$ & $1.4\pm0.1$/$1.1\pm0.1$ & $3.8\pm0.4$ & 0\\
SMM 5 & 05 43 04.6 & -01 17 17 & $11.3\pm0.7$ & $1.9\pm0.4$ & $1.2\pm0.1$/$0.9\pm0.1$ & $2.5\pm0.5$ & prestellar ?\\
SMM 6 & 05 43 05.2 & -01 18 48 & $11.0\pm0.4$ & $8.2\pm1.1$ & $2.0\pm0.1$/$1.7\pm0.1$ & $2.2\pm0.3$ & prestellar\\
Ori B9 N & 05 43 05.7 & -01 14 51 & $13.4\pm1.3$ & $2.3\pm0.4$ & $0.9\pm0.1$/$0.9\pm0.1$ & $2.1\pm0.4$ & prestellar ?\\
SMM 7 & 05 43 22.2 & -01 13 56 & $9.4\pm1.1$ & $3.6\pm1.0$ & $3.4\pm0.9$/$2.1\pm0.5$ & $4.8\pm1.3$ & prestellar\\
\hline 
\end{tabular} 
\tablefoot{Columns (2) and (3) give the equatorial coordinates 
[$(\alpha, \,\delta)_{2000.0}$]. Columns (4)--(7) list the gas kinetic 
temperature, core mass, beam-averaged peak H$_2$ column density, and the 
volume-averaged H$_2$ number density, respectively. In the last column we give 
the comments on the source classification. The virial parameter of 
the starless cores SMM 5 and Ori B9 N is $\alpha_{\rm vir}\gtrsim2$, and thus 
it is unclear if they are prestellar (Paper II). \tablefoottext{a}{These 
coordinates refer to the LABOCA peak positions after adjusting the pointing by 
$(\Delta \alpha,\, \Delta \delta)=(+1\farcs72, \, -9\farcs97)$. The 
coordinates of our molecular-line observation target positions can be 
found in Table~1 of Paper II.} \tablefoottext{b}{The first value is calculated 
towards the revised LABOCA peak position by using the temperature derived 
towards the line observation target position. The latter $N({\rm H_2})$ value 
refers to the position used in molecular-line observations.}}
\end{minipage}
\end{table*}

\begin{table*}
\caption{Observed spectral-line transitions and observational parameters.}
\begin{minipage}{2\columnwidth}
\centering
\renewcommand{\footnoterule}{}
\label{table:lines}
\begin{tabular}{c c c c c c c c c c c c}
\hline\hline 
Transition & $\nu$ & $E_{\rm u}/k_{\rm B}$ & $n_{\rm crit}$ & HPBW & $\eta_{\rm MB}$ & $T_{\rm sys}$ & PWV & \multicolumn{2}{c}{Channel spacing\tablefootmark{a}} & $t_{\rm int}$ & rms\\
      & [MHz] & [K] & [cm$^{-3}$] & [\arcsec] & & [K] & [mm] & [kHz] & [km~s$^{-1}$] & [min] & [mK]\\
\hline        
C$^{17}$O$(2-1)$ & 224\,714.199\tablefootmark{b} & 16.2 & $9.0\times10^3$ & 27.8 & 0.75 & 255--288 & 0.8--1.4 & 122.07 & 0.16 & 2.9--19.5 & 24--58\\
N$_2$D$^+(3-2)$ & 231\,321.912\tablefootmark{c} & 22.2 & $1.7\times10^6$& 27.0 & 0.75 & 235--335 & 0.2--3.8 & 122.07 & 0.16 & 11.5--37.5 & 22--35\\ 
N$_2$H$^+(3-2)$ & 279\,511.832\tablefootmark{c} & 26.8 & $2.9\times10^6$ & 22.3 & 0.74 & 211--359 & 0.04--1.4 & 122.07 & 0.13 & 2.7--5.5 & 32--84\\
DCO$^+(4-3)$ & 288\,143.855\tablefootmark{d} & 34.6 & $4.7\times10^6$ & 21.7 & 0.74 & 170--172 & 0.01--0.5 & 122.07 & 0.13 & 4.1--6.5 & 23--29\\
H$^{13}$CO$^+(4-3)$ & 346\,998.344\tablefootmark{e} & 41.6 & $8.2\times10^6$ & 18.0 & 0.73 & 362--521 & 0.9--1.3 & 122.07 & 0.11 & 8.5--22 & 25--55\\
\hline 
\end{tabular} 
\tablefoot{Columns (2)--(4) give the rest frequencies of the observed 
transitions ($\nu$), their upper-state energies ($E_{\rm u}/k_{\rm B}$, 
where $k_{\rm B}$ is the Boltzmann constant), and critical densities. Critical 
densities were calculated at $T=10$ K using the collisional-rate data 
available in the Leiden Atomic and Molecular Database (LAMDA; 
{\tt http://www.strw.leidenuniv.nl/$\sim$moldata/}) 
(\cite{schoier2005}). For N$_2$D$^+$, we used the Einstein $A-$coefficient 
from Pagani et al. (2009b) and the same collisional rate as for N$_2$H$^+$. 
Columns (5)--(12) give the APEX beamsize (HPBW) and the main beam efficiency 
($\eta_{\rm MB}$) at the observed frequencies, the SSB system temperatures 
during the observations ($T_{\rm sys}$ in $T_{\rm MB}$ scale, see text), 
the amount of PWV, channel widths (both in kHz and km~s$^{-1}$) of the 
original data, the on-source integration times per position ($t_{\rm int}$), 
and the $1\sigma$ rms noise at the smoothed resolution.\\
\tablefoottext{a}{The original channel spacings. The final spectra were 
Hanning-smoothed which divides the number of channels by two.}\tablefoottext{b}{From Ladd et al. (1998).} \tablefoottext{c}{From Pagani et al. (2009b).}\tablefoottext{d}{From the Jet Propulsion Laboratory (JPL; {\tt http://spec.jpl.nasa.gov/}) spectroscopic database (\cite{pickett1998}).}\tablefoottext{e}{From the Cologne Database for Molecular Spectroscopy (CDMS; {\tt http://www.astro.uni-koeln.de/cdms/catalog}) (\cite{muller2005}).}}
\end{minipage}
\end{table*}

\section{Observational results}

\subsection{SABOCA 350-$\mu$m emission}

\begin{figure*}
\begin{center}
\includegraphics[width=0.85\textwidth]{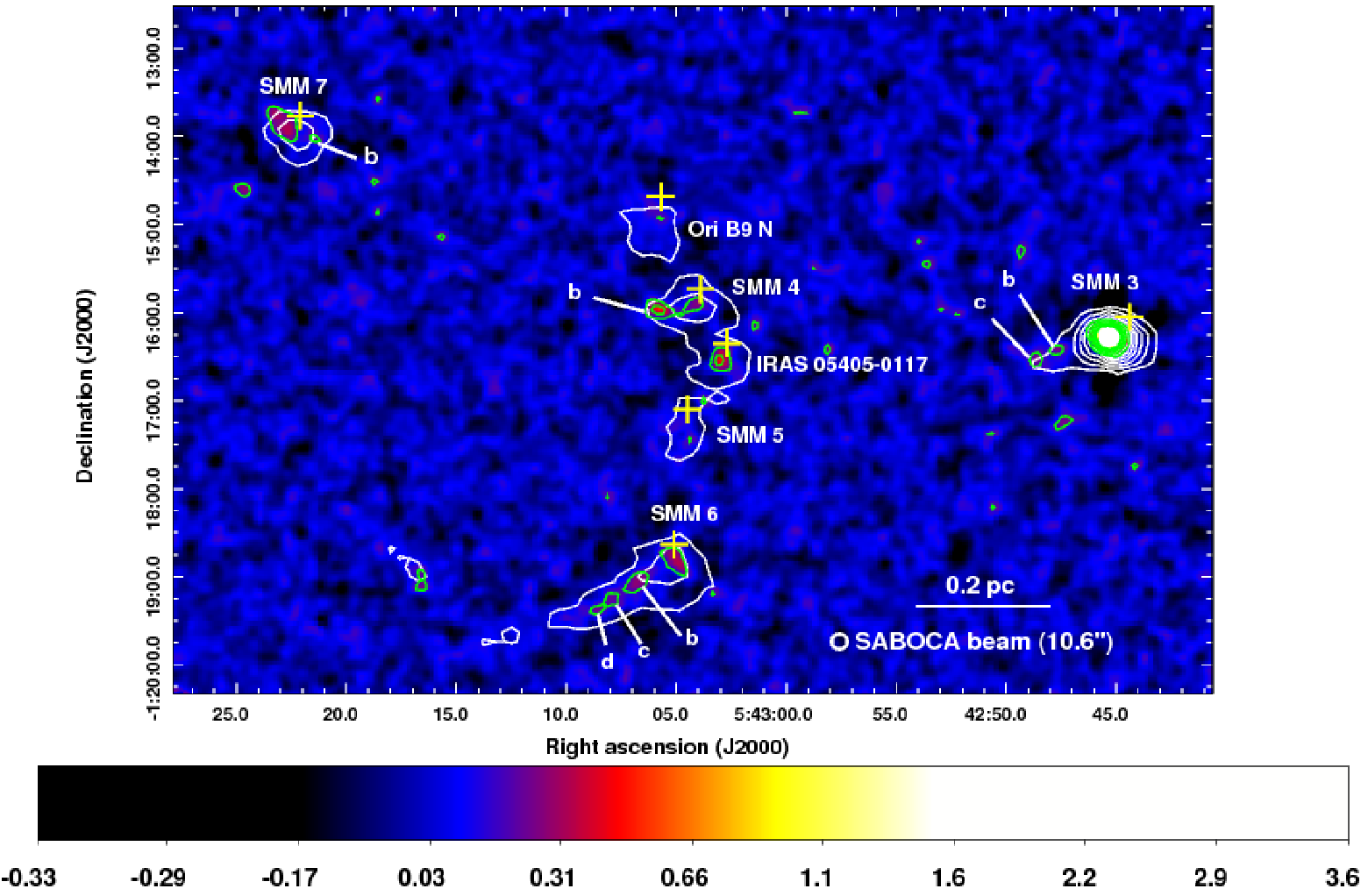}
\caption{SABOCA 350-$\mu$m image (smoothed to $10\farcs6$ resolution) of the 
central part of Orion B9 (colour scale and green contours). The image is shown 
with a square root scaling, and the colour bar indicates the flux density in 
units of Jy~beam$^{-1}$. The rms level is 0.06 Jy~beam$^{-1}$ ($1\sigma$). 
The first 350-$\mu$m contour and the separation between contours is 
$3\sigma$. The white contours show the LABOCA 870-$\mu$m dust continuum 
emission as in Fig.~\ref{figure:laboca}. The yellow plus signs show 
the target positions of our molecular-line observations. The 0.2-pc scale bar 
and beam HPBW are shown in the bottom right.}
\label{figure:saboca}
\end{center}
\end{figure*}

The 350-$\mu$m SABOCA map is shown in Fig.~\ref{figure:saboca}. 
Almost all the cores detected with LABOCA show also clear 350 $\mu$m 
emission. The exceptions are SMM 5 and Ori B9 N for which the 350-$\mu$m peak 
flux densities are at the levels $3.5\sigma$ (210 mJy) and $3\sigma$ 
(180 mJy), res\-pectively. With the peak flux density of 
$60.5\sigma$ (3630 mJy), SMM 3 is by far the strongest 350-$\mu$m 
source in the region. 
%At about two times better resolution, 
%the 350-$\mu$m peak positions are offset from those at 870 $\mu$m by 
%8\farcs6--18\farcs7 ($12\farcs9$ on average); the smallest and largest offsets 
%are found for SMM 4 and SMM 3, respectively. However, the difference in 
%angular resolution does not appear to be the cause of the offset in peak 
%positions; by smoothing the SABOCA map to correspond the resolution of the 
%LABOCA data, and regridding the maps onto the same grid, the above offsets 
%still remain (but are within the beam size). There is also a slight trend for 
%the 350 $\mu$m peaks to lie southeast from the 870 $\mu$m peaks. This suggests 
%that there is a shift between SABOCA and LABOCA absolute astrometric frames. 
%The positional offset can also (partly) reflect the source physical structure, 
%such as temperature gradient. 
Four of the cores are resolved into at least 
two emission peaks in the SABOCA image. In the case of SMM 3, there is a 
350-$\mu$m condensation, we call SMM 3b ($4.2\sigma$ or 250 mJy), 
at about $36\arcsec$ east of the ``main'' source. At about $17\arcsec$ from 
SMM 3b, there is a\-nother $4\sigma$ (240 mJy) emission peak, 
designated here as SMM 3c. The 870 $\mu$m emission of SMM 3 extends to the 
direction of the subcondensations SMM 3b and 3c, and both of them lie either 
within or at the borderline of the $3.3\sigma$ (99 mJy) 870-$\mu$m 
contour. SMM 4 is resolved into two condensations. The western condensation 
coincides with the LABOCA peak position, whereas the eastern one, SMM 4b, is 
coincident with a \textit{Spitzer} 24-$\mu$m source near SMM 4 
(see Sect.~5.7.2). The elongated core SMM 6 is resolved into at least three 
subcondensations. The northwesternmost condensation corresponds to LABOCA peak 
of SMM 6. Finally, SMM 7 shows a hint of substructure 
in the western part of the core, where the 870 $\mu$m emission is 
above $3.3\sigma$ (99 mJy) level. SMM 3, 4, 6, 7, and their 
substructure will be discussed further in Sect.~5.7. There are also two 
$4-4.5\sigma$ (240 -- 270 mJy) emission 
peaks at $\alpha_{2000.0}=05^{\rm h}42^{\rm m}47.4^{\rm s}$, 
$\delta_{2000.0}=-01\degr17\arcmin15\arcsec$ (south of SMM 3b) and 
$\alpha_{2000.0}=05^{\rm h}43^{\rm m}24.7^{\rm s}$, 
$\delta_{2000.0}=-01\degr14\arcmin35\arcsec$ (southeast of SMM 7), 
and a few $3.5-4\sigma$ (210 -- 240 mJy) peaks to the east of SMM 6 at 
$\alpha_{2000.0}=05^{\rm h}43^{\rm m}16.5^{\rm s}$, 
$\delta_{2000.0}\simeq-01\degr19\arcmin00\arcsec$. There is some 
weak LABOCA emission just slightly north of the latter peaks. However, because 
these ``additional'' sources are below $5\sigma$ (300 mJy), were not 
detected by LABOCA at 870 $\mu$m, and were not identified by the SIMBAD 
Astronomical Database\footnote{\texttt{http://simbad.u-strasbg.fr/simbad/}}, 
they may well be unreal and are not discussed further in this paper. 
 
To identify and extract the cores from the SABOCA map, we employed the 
commonly used two-dimensional \texttt{clumpfind} algorithm, 
\texttt{clfind2d}, developed by Williams et al. (1994). 
The \texttt{clfind2d} routine determines the peak position, the 
FWHM size (not corrected for beam size), and the peak and total integrated 
flux density of the source based on specified contour levels. The algorithm 
requires two configuration parameters: \textit{i}) the intensity threshold, 
i.e., the lowest contour level, $T_{\rm low}$, which determines the minimum 
emission to be included into the source; and \textit{ii}) the contour level 
spacing, $\Delta T$, which determines the required ``contrast'' between two 
sources to be considered as different objects. We set both parameters to 
$3\sigma$ (180 mJy). The selected $3\sigma$ contour levels 
turned out to give the best agreement with the identification by eye. 
With these parameter settings, the SMM 6c and 6d condensations are treated 
as a single source by \texttt{clumpfind}.

The J2000.0 coordinates of the peak 350 $\mu$m emission, source effective 
radius ($R_{\rm eff}=\sqrt{A/\pi}$, where $A$ is the projected area within the 
$3\sigma$ contour), and peak and integrated flux densities are listed in 
Cols.~(2)--(6) of Table~\ref{table:cores}. In Col.~(7), we also list the 
flux densities measured in a $40\arcsec$ diameter aperture from the SABOCA map 
smoothed to the resolution of our LABOCA data. These flux densities were 
determined using the CRUSH-2 programme. The effective radius is only 
given for sources which are larger than the beam size. The total flux density 
uncertainty was derived from 
$\sigma(S_{\lambda})=\sqrt{\sigma_{\rm cal}^2+\sigma_{\rm S}^2}$, where 
$\sigma_{\rm cal}$ is the absolute calibration error (adopted to be 
30\% of flux density), and $\sigma_{\rm S}$ is the uncertainty in the 
flux-density determination based on the rms noise near the source area.
The $1\sigma$ rms noise in the smoothed SABOCA map, $\sigma=0.08$ 
Jy~beam$^{-1}$, is slightly higher than in the original map. We note that 
the data reduction produces ne\-gative artefacts (``holes'') around regions of 
bright emission, most notably around SMM 3. This decreases the source's peak 
intensity, introducing an additional uncertainty in the flux density. 
The uncertainties due to negative bowls are neglected in the subsequent 
analysis.

\begin{table*}
\caption{The 350-$\mu$m core properties.}
\begin{minipage}{2\columnwidth}
\centering
\renewcommand{\footnoterule}{}
\label{table:cores}
\begin{tabular}{c c c c c c c}
\hline\hline 
 & \multicolumn{2}{c}{Peak position} & $R_{\rm eff}$ & $S_{350}^{\rm peak}$ & $S_{350}$ & $S_{350}^{40\arcsec}$\tablefootmark{a}\\
Name & $\alpha_{2000.0}$ [h:m:s] & $\delta_{2000.0}$ [$\degr$:$\arcmin$:$\arcsec$] & [\arcsec] & [Jy beam$^{-1}$] & [Jy] & [Jy] \\
\hline
SMM 3 & 05 42 45.3 & -01 16 16 & 13.3 & 3.63 & $5.4\pm1.6$ & $3.9\pm1.2$ \\
SMM 3b & 05 42 47.6 & -01 16 24 & \ldots & 0.25 & \ldots & \ldots \\
SMM 3c & 05 42 48.6 & -01 16 32 & \ldots & 0.24 & \ldots & \ldots \\
IRAS 05405-0117 & 05 43 03.0 & -01 16 31 & 7.6 & 0.47 & $0.5\pm0.2$ & $0.6\pm0.2$ \\
SMM 4 & 05 43 04.0 & -01 15 52 & 5.8 & 0.25 & $0.2\pm0.1$ & $0.7\pm0.3$\\
SMM 4b & 05 43 05.7 & -01 15 57 & 7.5 & 0.49 & $0.5\pm0.2$ & \ldots \\
SMM 5 & 05 43 03.8 & -01 16 59 & \ldots & 0.21 & \ldots & $0.5\pm0.2$ \\
SMM 6 & 05 43 04.9 & -01 18 49 & 8.2 & 0.33 & $0.5\pm0.2$ & $0.7\pm0.3$ \\
SMM 6b & 05 43 06.7 & -01 19 02 & 7.0 & 0.27 & $0.3\pm0.1$ & \ldots \\
SMM 6c & 05 43 07.9 & -01 19 13 & 6.2\tablefootmark{b} & 0.21 & $0.2\pm0.1$\tablefootmark{b} & \ldots \\
SMM 6d & 05 43 08.6 & -01 19 22 & \ldots & 0.23 & \ldots & \ldots \\
Ori B9 N & 05 43 05.6 & -01 14 55 & \ldots & 0.18 & \ldots & $0.5\pm0.2$ \\
SMM 7 & 05 43 22.7 & -01 13 55 & 10.1 & 0.32 & $0.7\pm0.2$ & $0.8\pm0.3$ \\
SMM 7b & 05 43 21.3 & -01 14 02 & \ldots & 0.21 & \ldots & \ldots \\
\hline 
\end{tabular} 
\tablefoot{\tablefoottext{a}{Flux density measured in a $40\arcsec$ diameter 
aperture from the SABOCA map smoothed to the resolution of the LABOCA map.} \tablefoottext{b} These values refer to the combined size and flux density of SMM 6c and 6d.}
\end{minipage}
\end{table*}

\subsection{Spectra}

The Hanning-smoothed spectra are shown in Fig.~\ref{figure:spectra}. 
The C$^{17}$O$(2-1)$ line is clearly detected towards all sources except SMM 4 
where only a low-velocity component is observed at 
${\rm v_{\rm LSR}}\simeq1.7$ km~s$^{-1}$ (hereafter, SMM 4-LVC, where 
LVC stands for low-velocity component). We note that SMM 4 shows strong 
NH$_3(1,\,1)$ emission at about 9.1 km~s$^{-1}$ (Paper II). IRAS 05405-0117 
(henceforth, IRAS05405 etc.) shows two additional velocity components at 
$\sim1.3$ and 3.0 km~s$^{-1}$, and Ori B9 N shows an additional and wide 
C$^{17}$O$(2-1)$ line at $\sim1.9$ km~s$^{-1}$. There is also a hint of 
C$^{17}$O$(2-1)$ emission at the systemic velocity of the region 
($\sim9$ km~s$^{-1}$) in the spectrum towards SMM 7, although the radial 
velocity of SMM 7, as determined from NH$_3(1,\,1)$ measurements in Paper II, 
is 3.6 km~s$^{-1}$. The hf-structure of the C$^{17}$O$(2-1)$ line is 
partially resolved in IRAS05399, SMM 1 and 3, and Ori B9 N.

The H$^{13}$CO$^+(4-3)$ observations were carried 
out only towards three sources (IRAS05399, SMM 1, and SMM 4). The line 
is only detected in IRAS05399 and SMM 4-LVC. The DCO$^+(4-3)$ line is 
detected towards all the other sources except IRAS05405, SMM 4, and SMM 5. 
The line is also quite weak ($4\sigma$) in Ori B9 N. Again, clear emission 
from an additional velocity component can be seen towards SMM 4 and Ori B9 N.

As already presented in Paper II, N$_2$H$^+(3-2)$ emission is clearly seen 
in all sources with the exception of target position near Ori B9 N 
where the line is very weak. It should be noted, however, that towards SMM 4 
only the LVC is detected. Ori B9 N shows an additional velocity component at 
about 1.8 km~s$^{-1}$. Finally, the N$_2$D$^+(3-2)$ line was detected in 
IRAS05399, SMM 1, SMM 3, IRAS05405, SMM 4-LVC, and SMM 6. In addition, the 
line appears at $\sim3$ km~s$^{-1}$ towards IRAS05405.

\begin{figure*}
\begin{center}
\includegraphics[width=3.1cm, angle=-90]{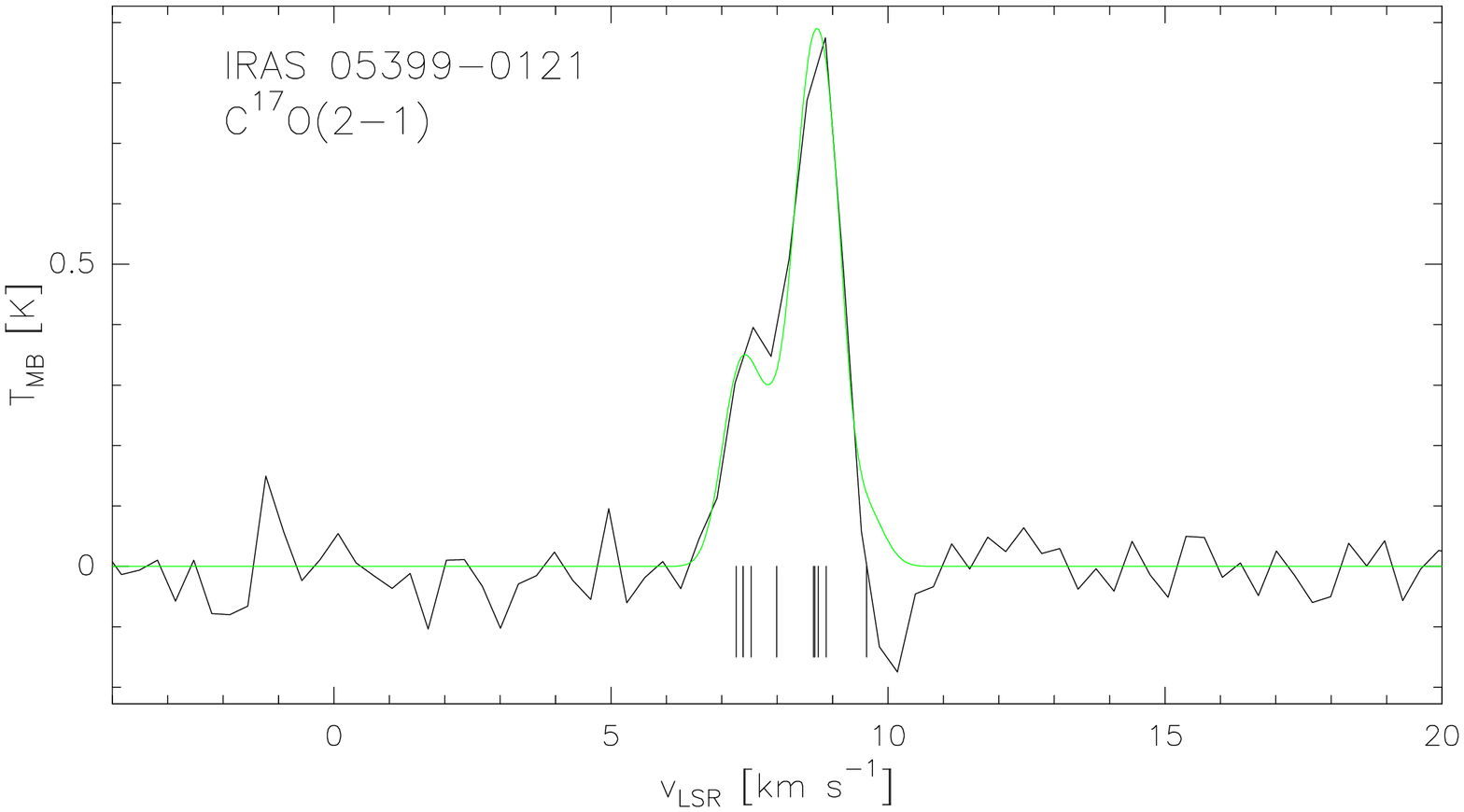}
\includegraphics[width=3.1cm, angle=-90]{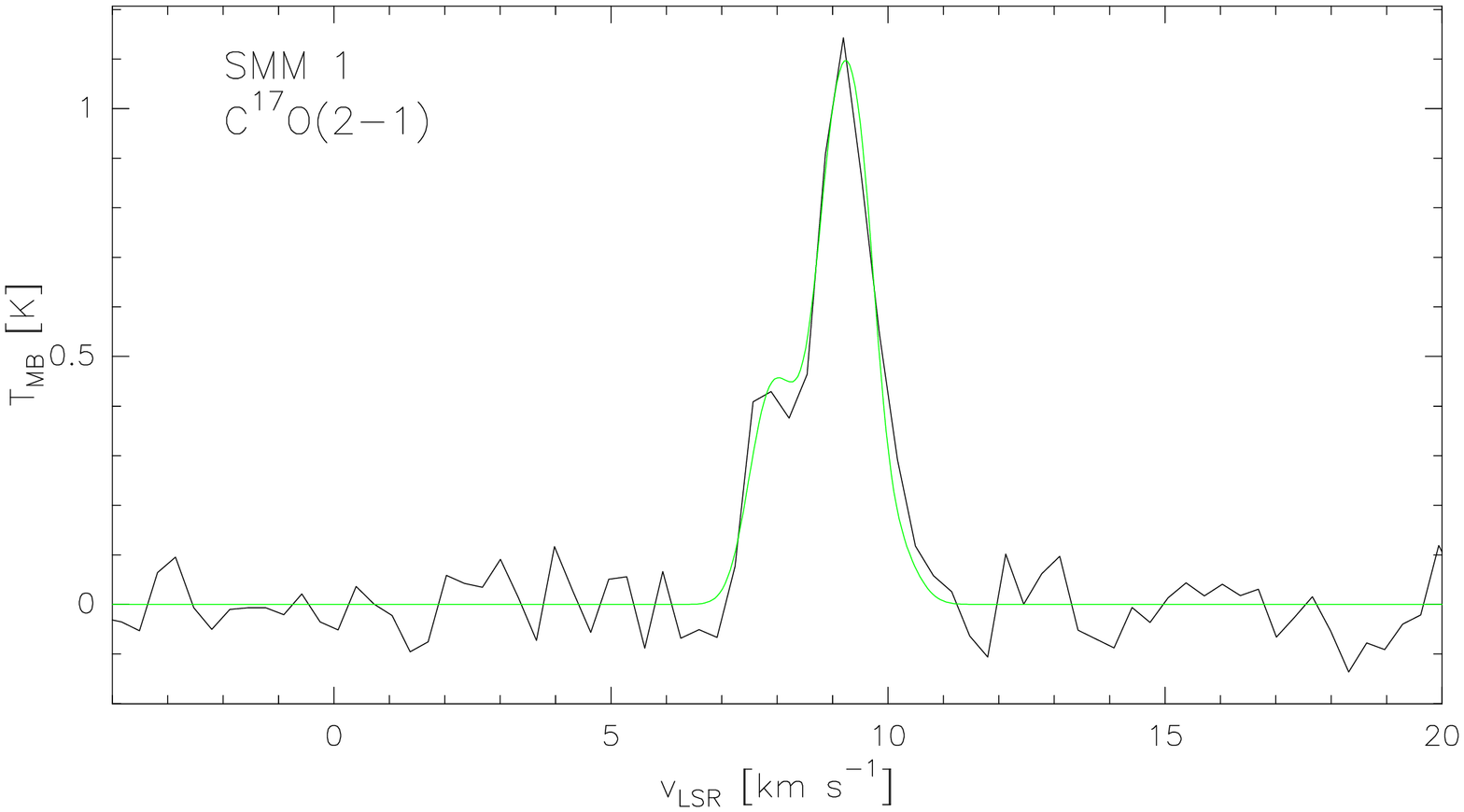}
\includegraphics[width=3.1cm, angle=-90]{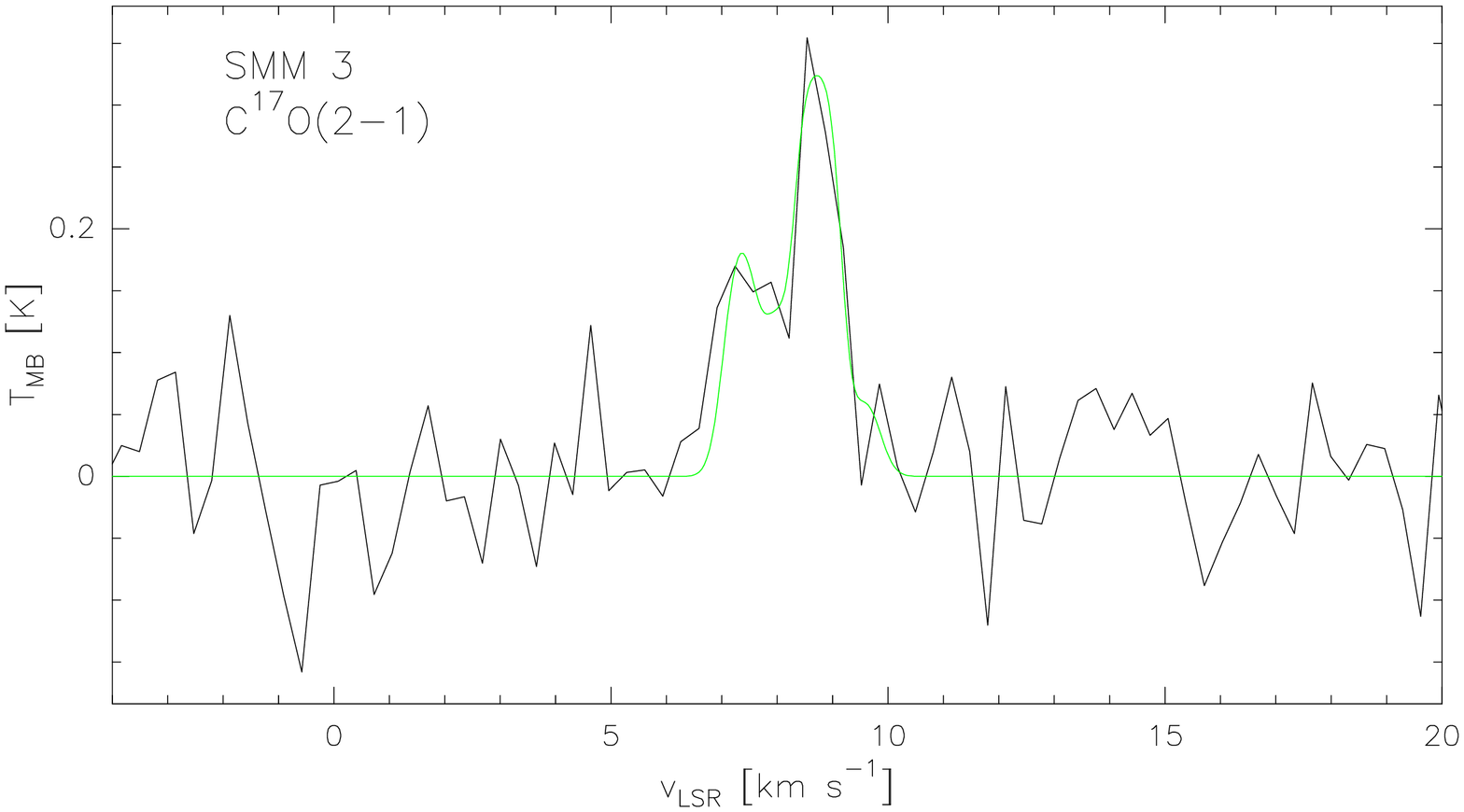}
\includegraphics[width=3.1cm, angle=-90]{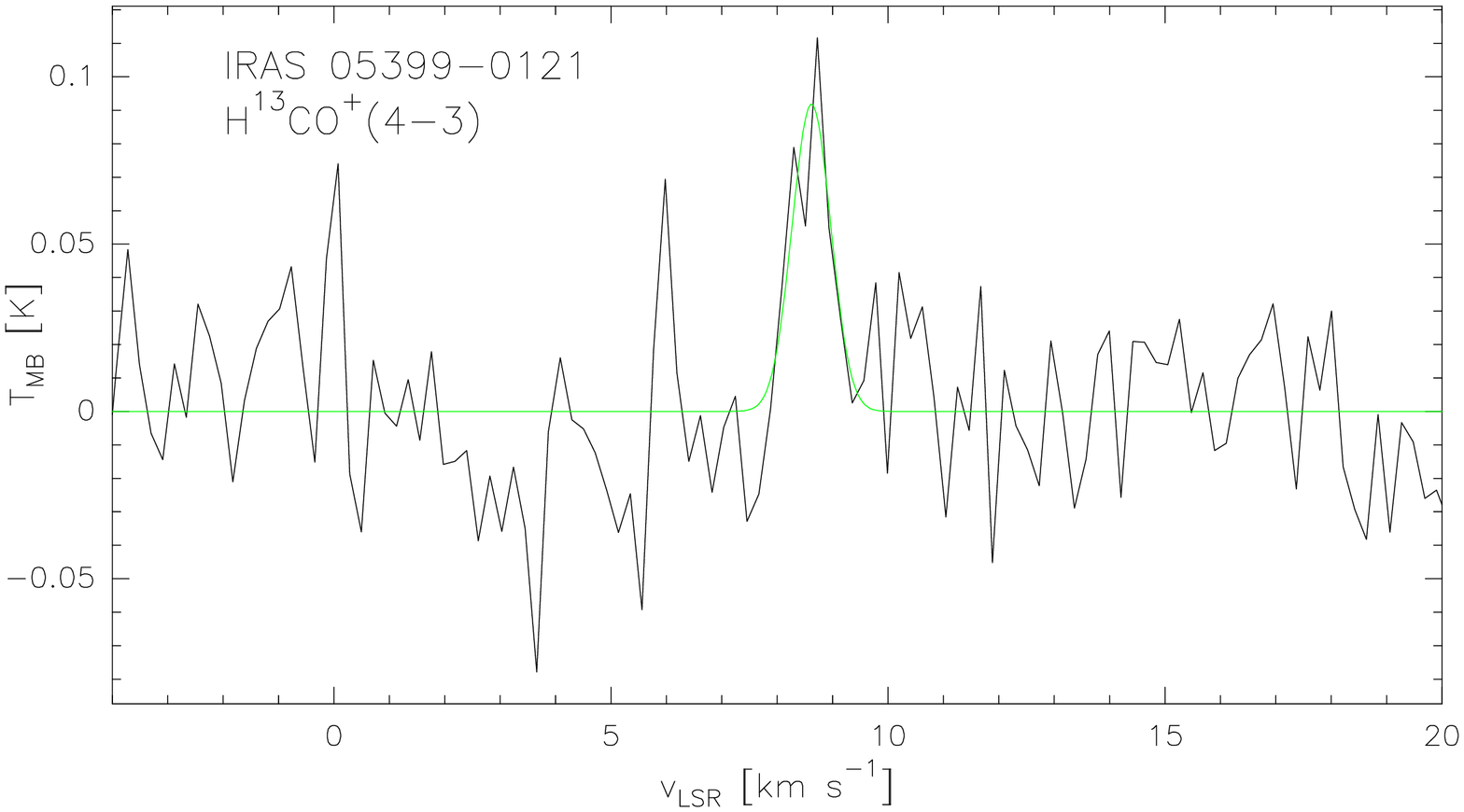}
\includegraphics[width=3.1cm, angle=-90]{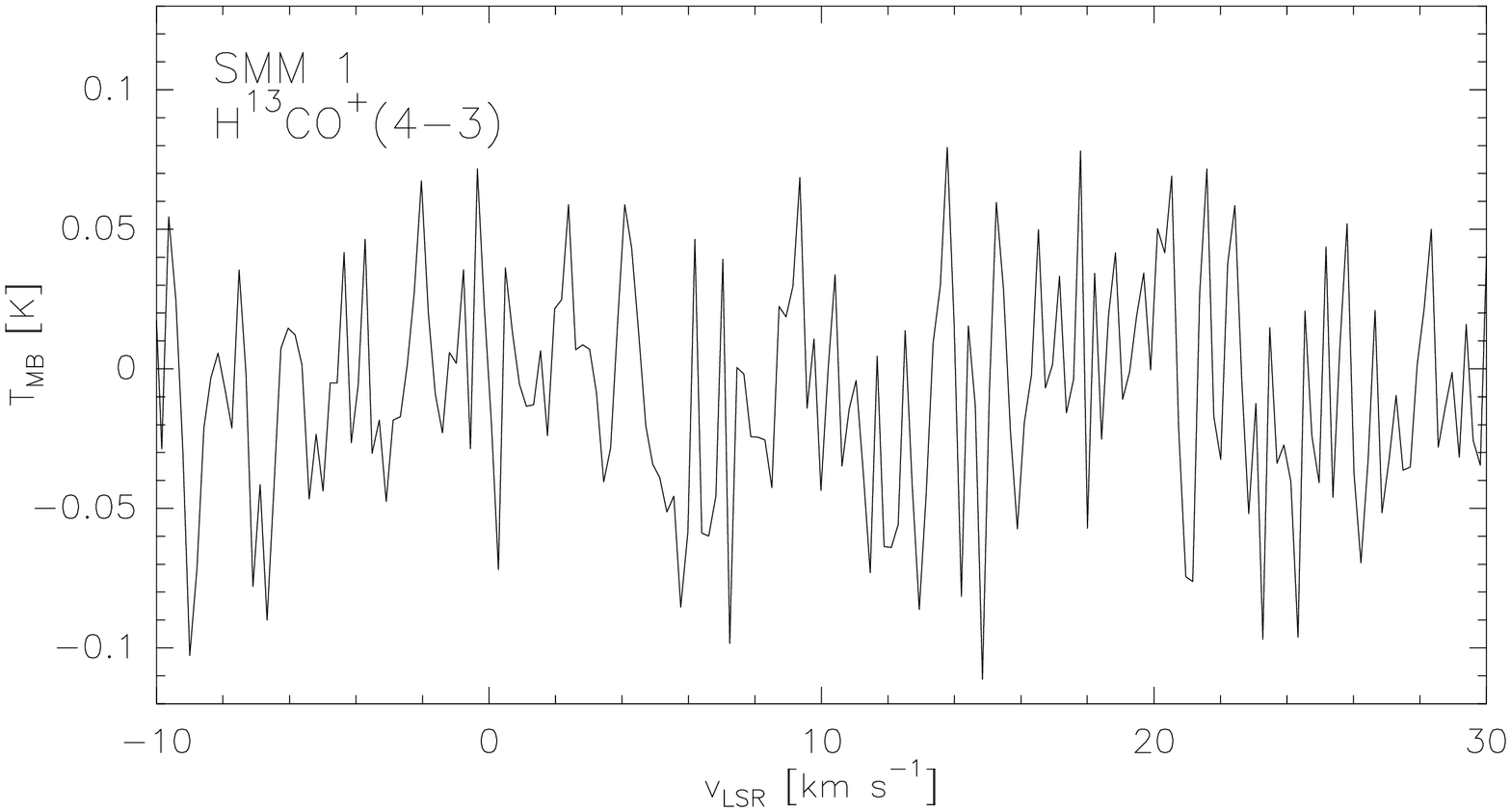}
\includegraphics[width=3.1cm, angle=-90]{blank.eps}
\includegraphics[width=3.1cm, angle=-90]{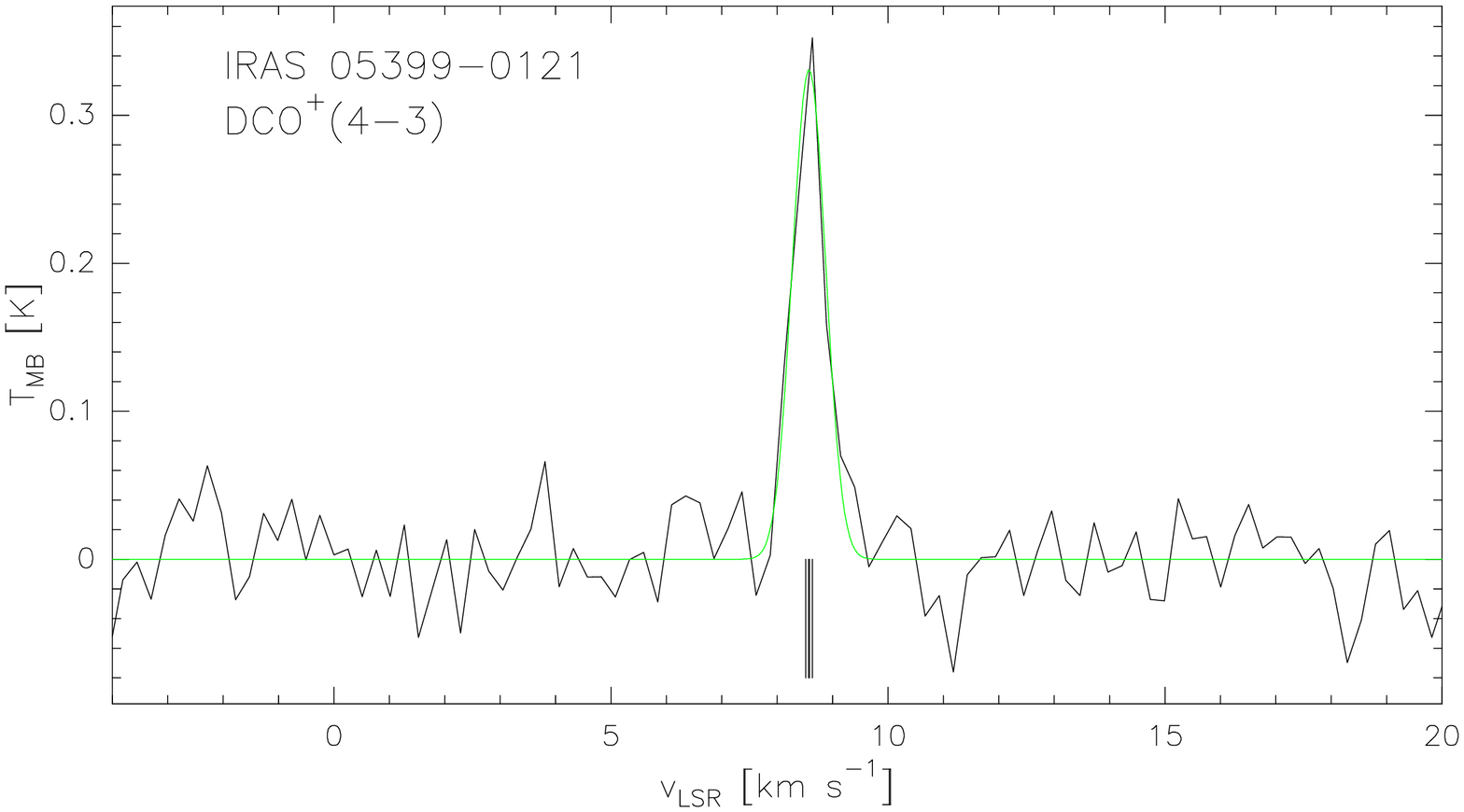}
\includegraphics[width=3.1cm, angle=-90]{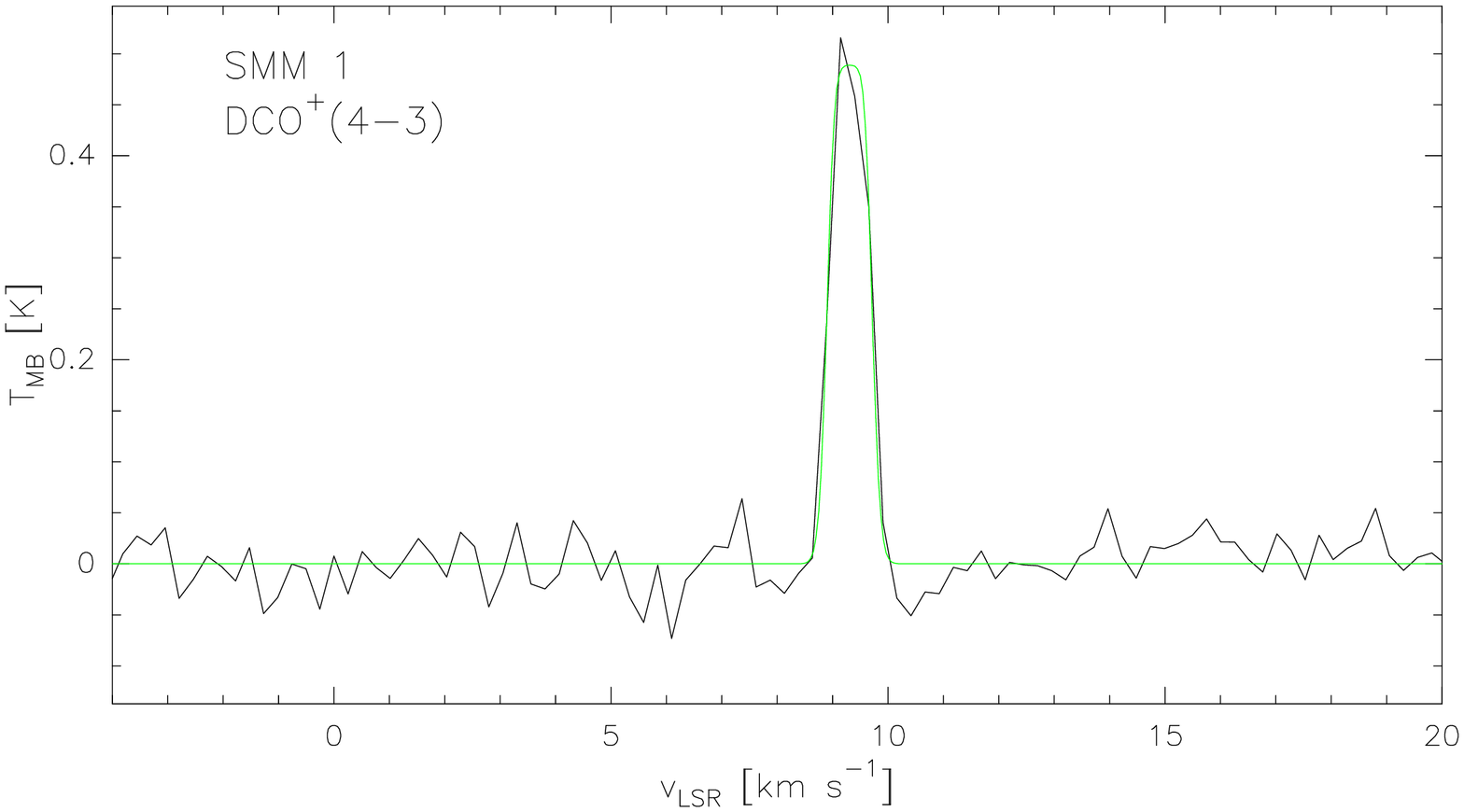}
\includegraphics[width=3.1cm, angle=-90]{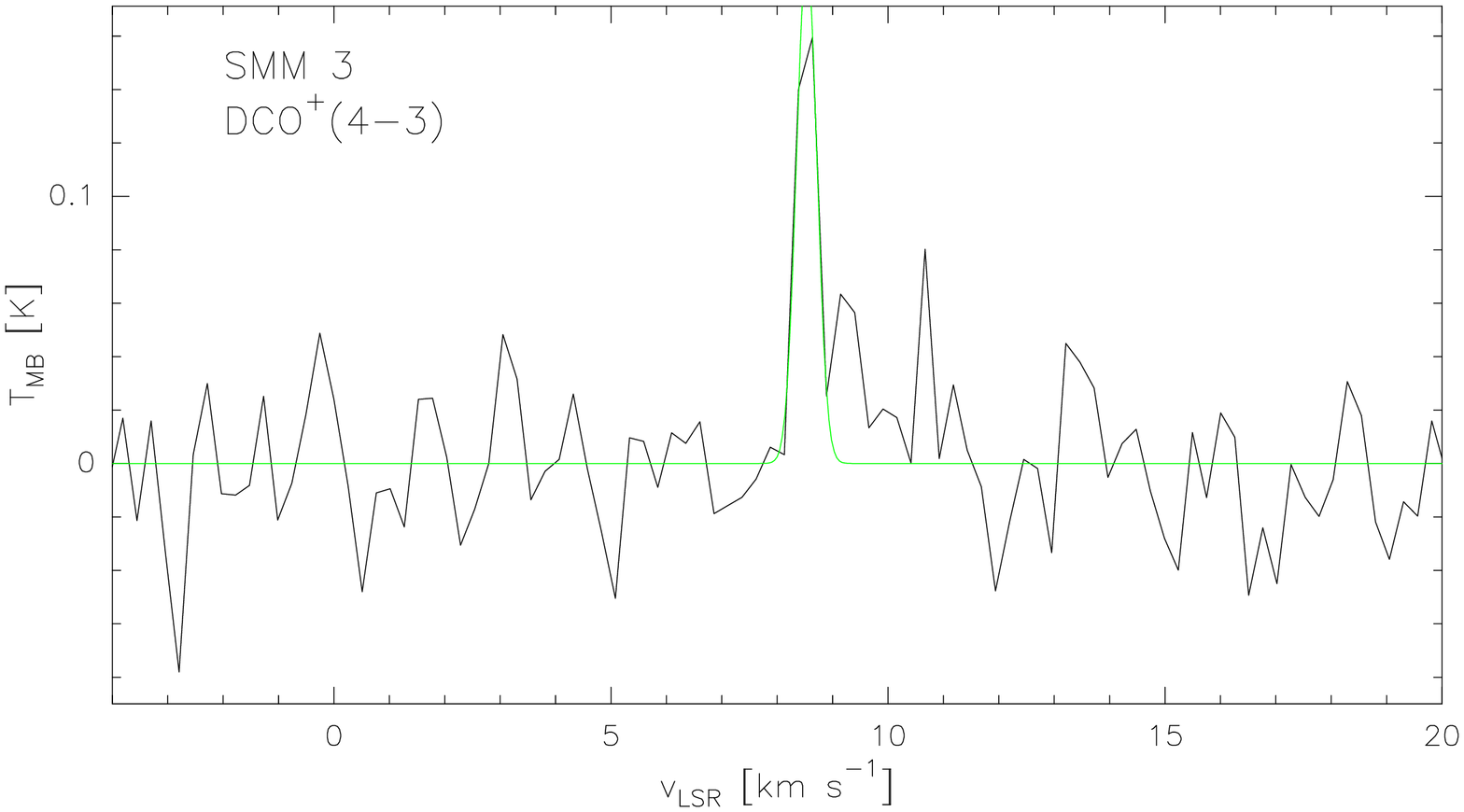}
\includegraphics[width=3.1cm, angle=-90]{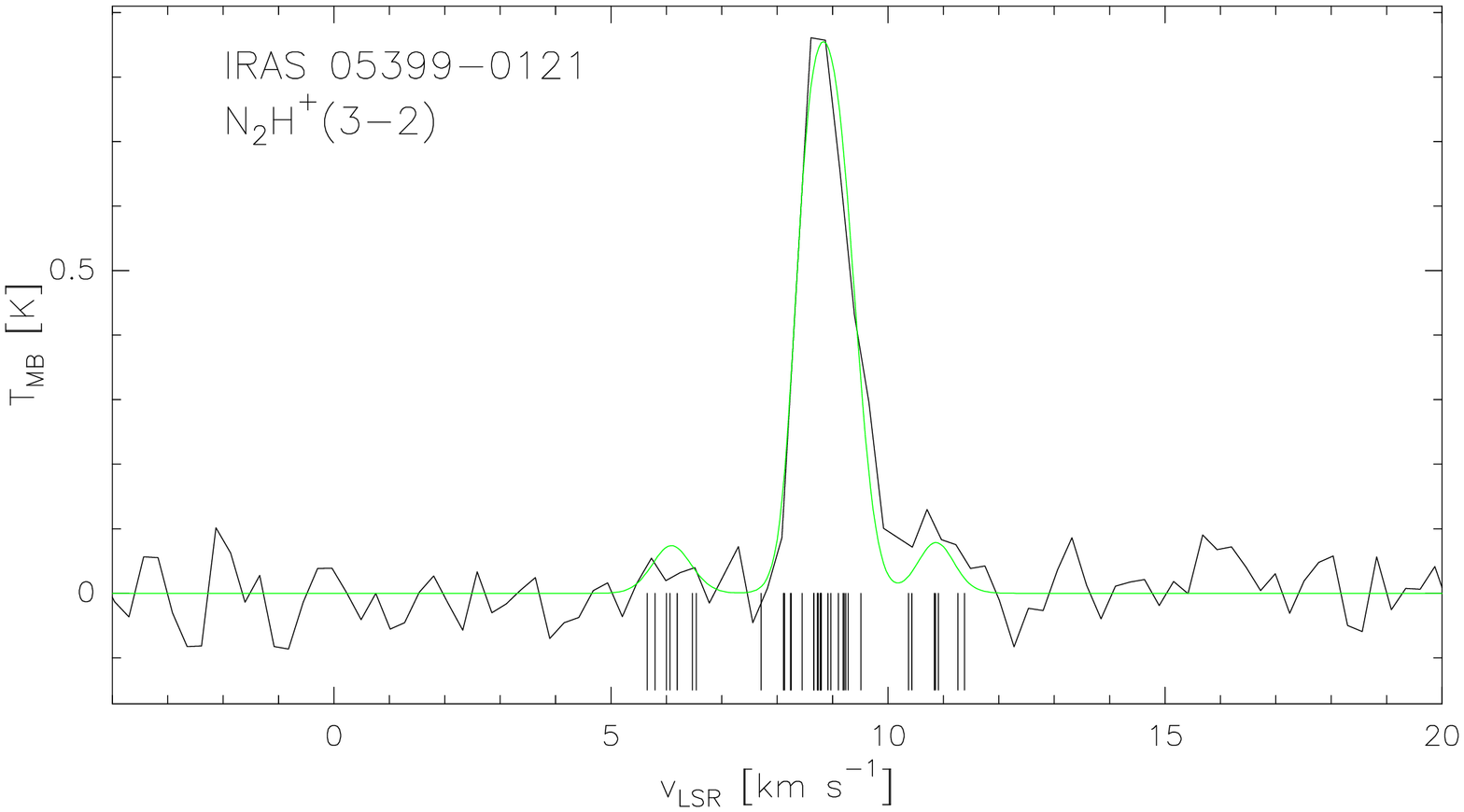}
\includegraphics[width=3.1cm, angle=-90]{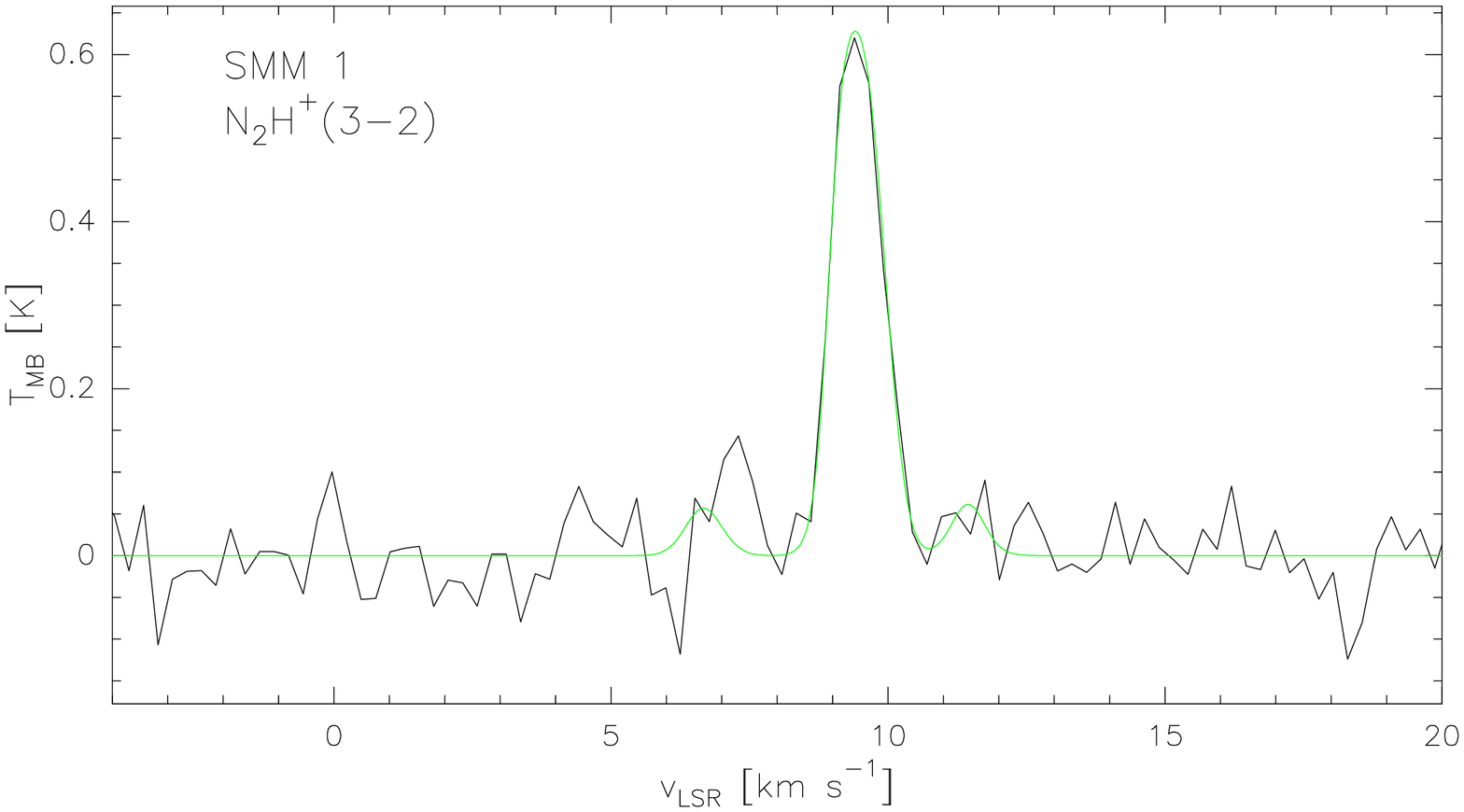}
\includegraphics[width=3.1cm, angle=-90]{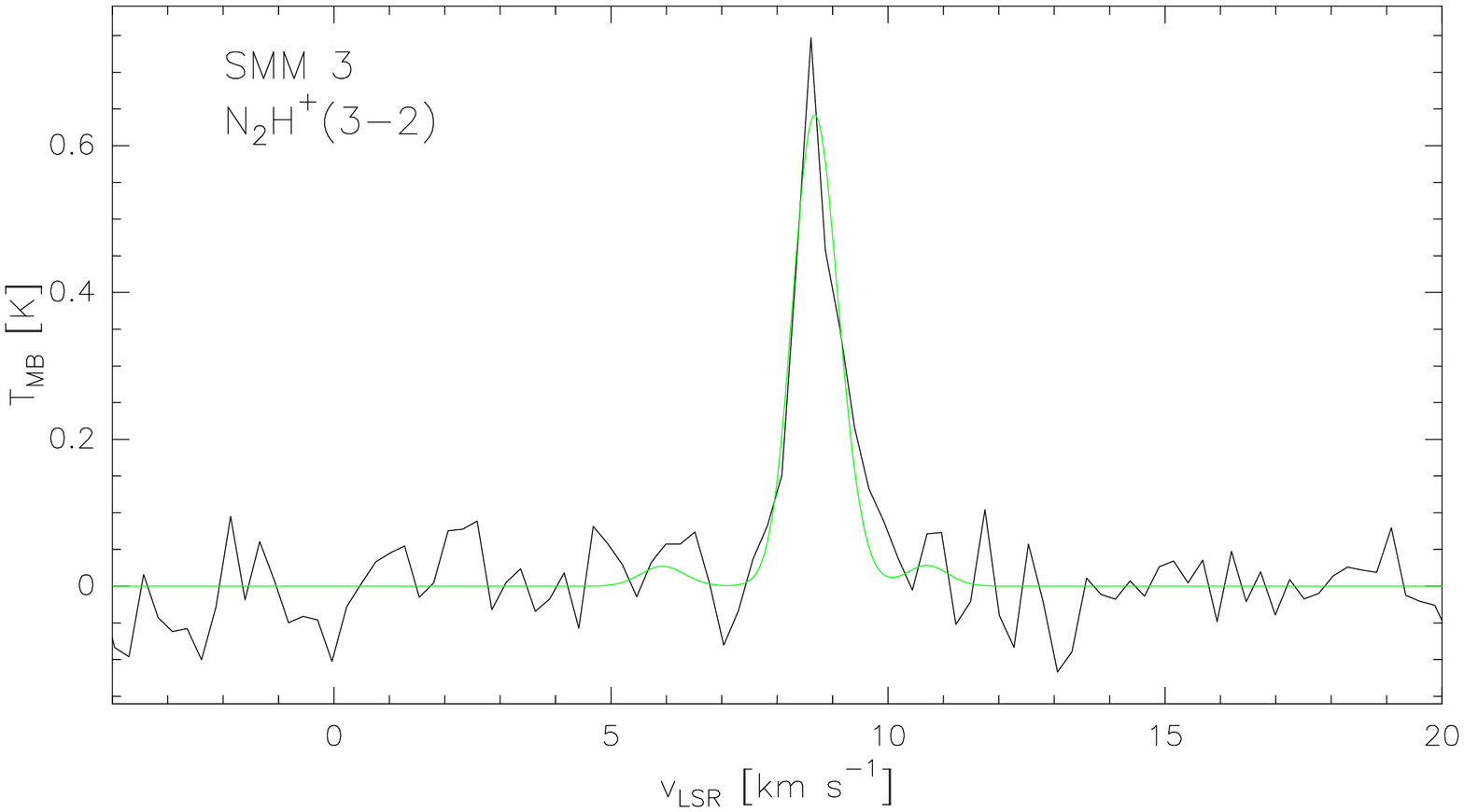}
\includegraphics[width=3.1cm, angle=-90]{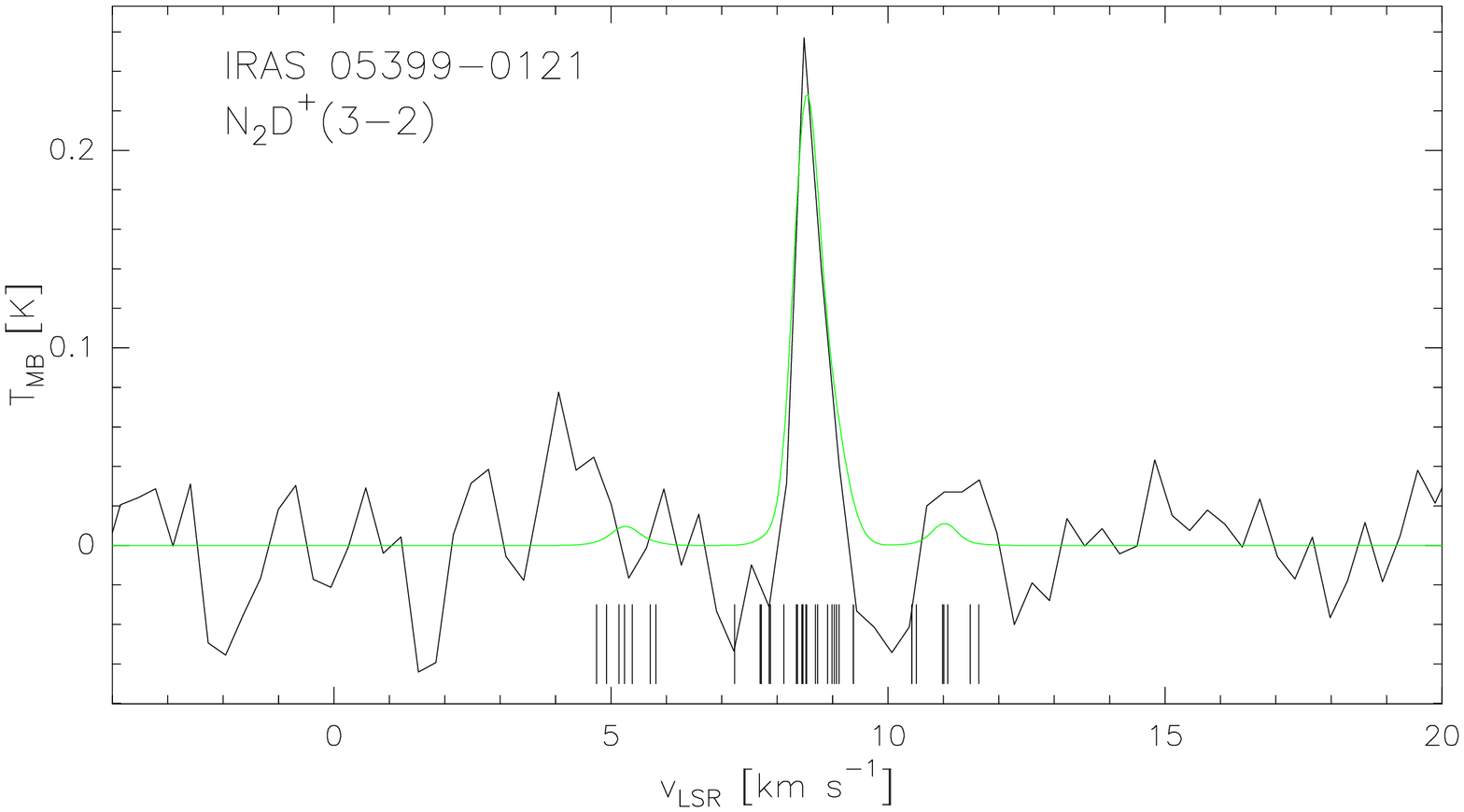}
\includegraphics[width=3.1cm, angle=-90]{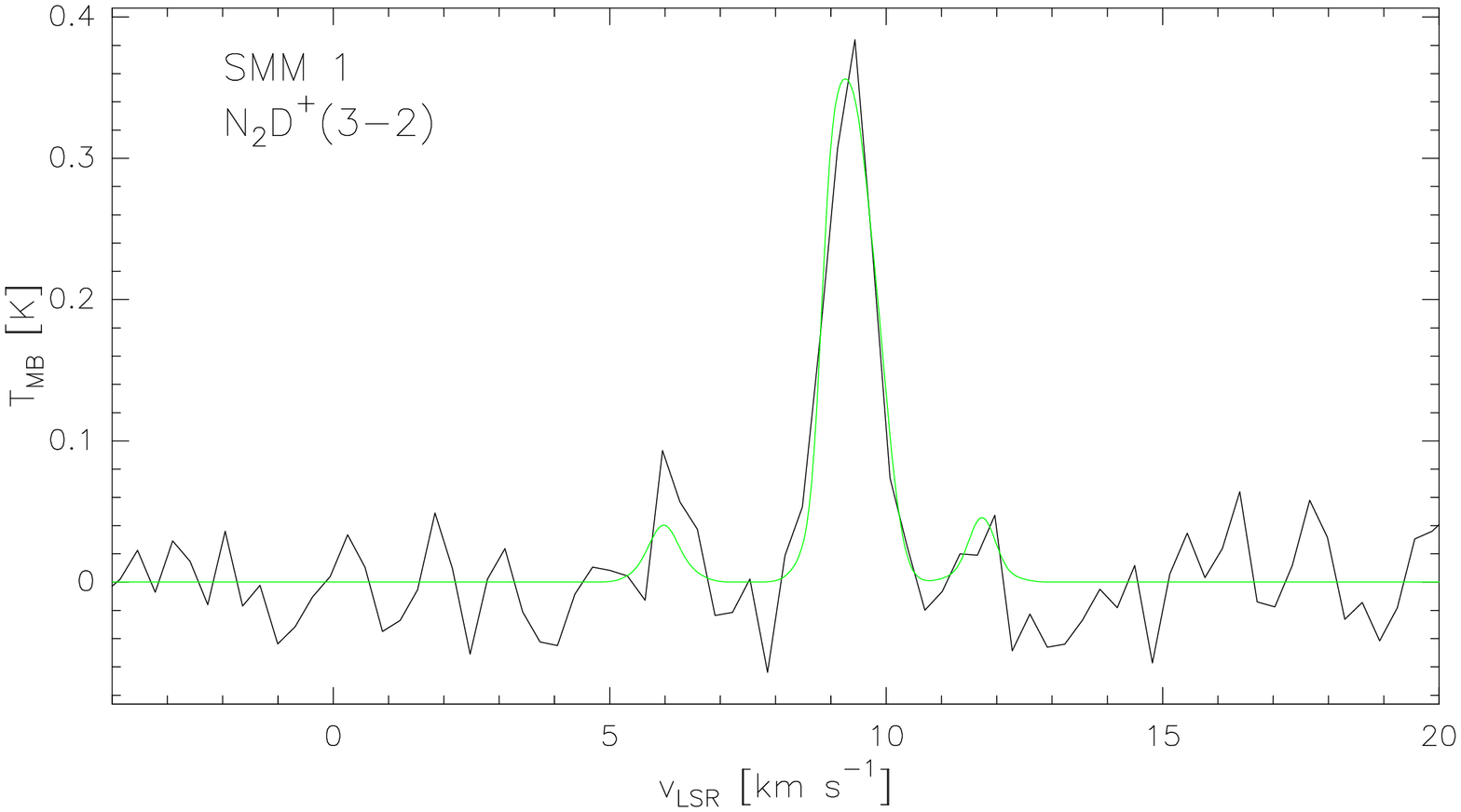}
\includegraphics[width=3.1cm, angle=-90]{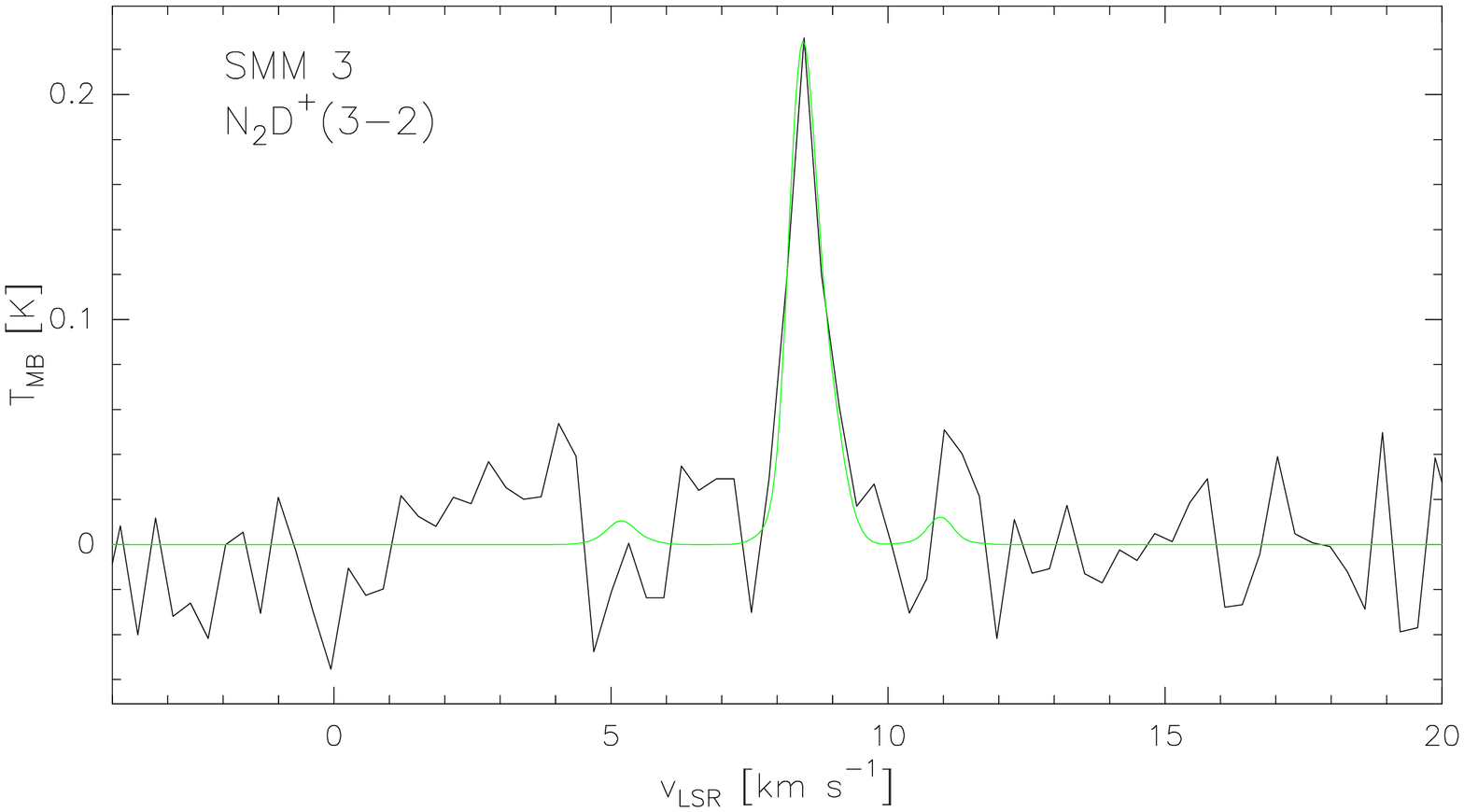}
\caption{Smoothed C$^{17}$O$(2-1)$, H$^{13}$CO$^+(4-3)$, DCO$^+(4-3)$, 
N$_2$H$^+(3-2)$, and N$_2$D$^+(3-2)$ spectra. Overlaid on the C$^{17}$O, 
DCO$^+$, N$_2$H$^+$, and N$_2$D$^+$ spectra are the hf-structure fits. 
The relative velocities of individual hf components in these transitions 
are labelled with a short bar on the spectra towards IRAS05399. The 
H$^{13}$CO$^+$ spectra are overlaid with a single Gaussian fit. 
The fits to the lines at the systemic velocity $\sim9$ km~s$^{-1}$ are shown 
as green lines, whereas the red lines show fits to the lines at lower 
velocities. Note that H$^{13}$CO$^+(4-3)$ observations were carried 
out only towards three sources and the line is detected only towards 
IRAS05399 and SMM 4 (at $\sim1.5$ km~s$^{-1}$).}
\label{figure:spectra}
\end{center}
\end{figure*}

\begin{figure*}
\begin{center}
\includegraphics[width=3.1cm, angle=-90]{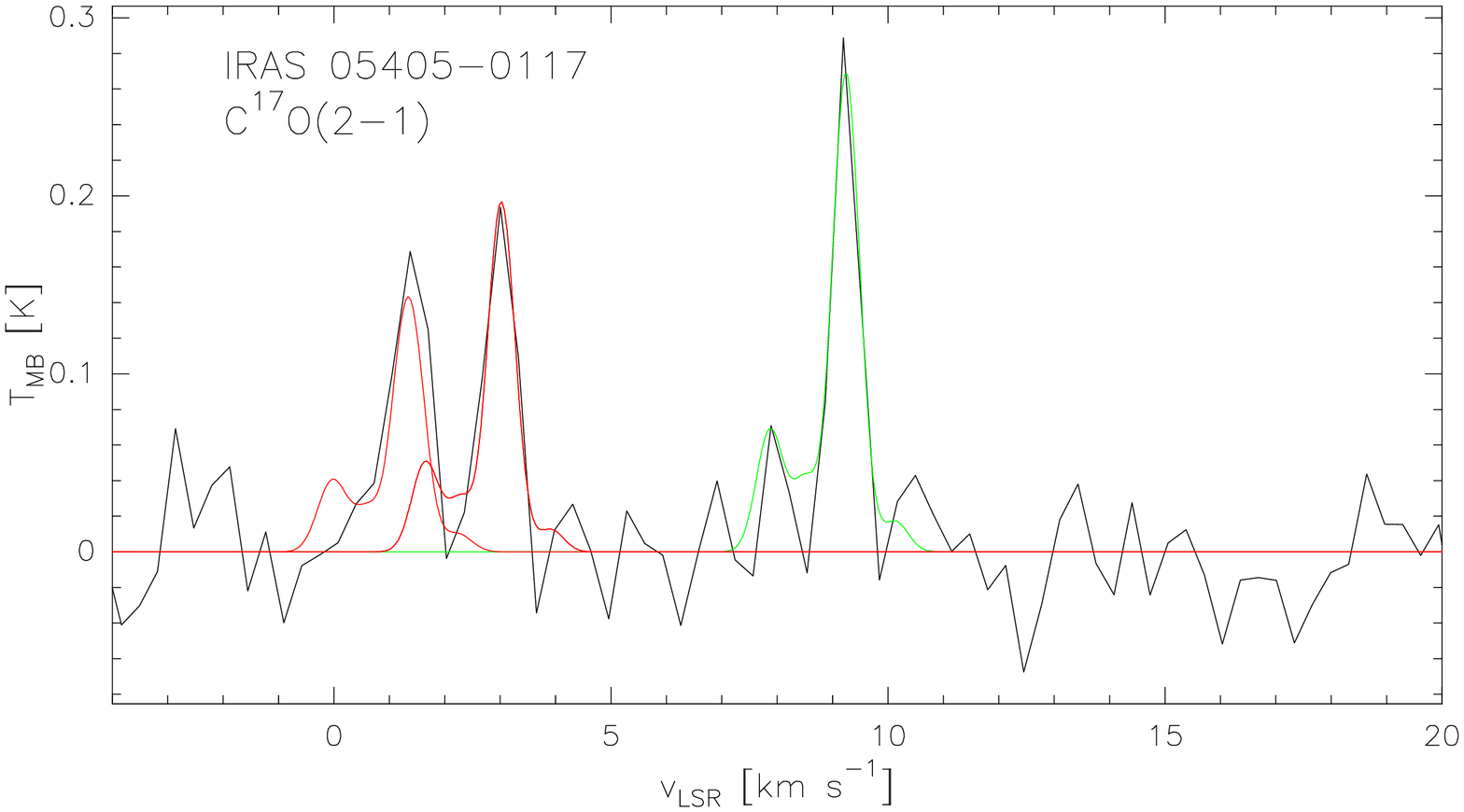}
\includegraphics[width=3.1cm, angle=-90]{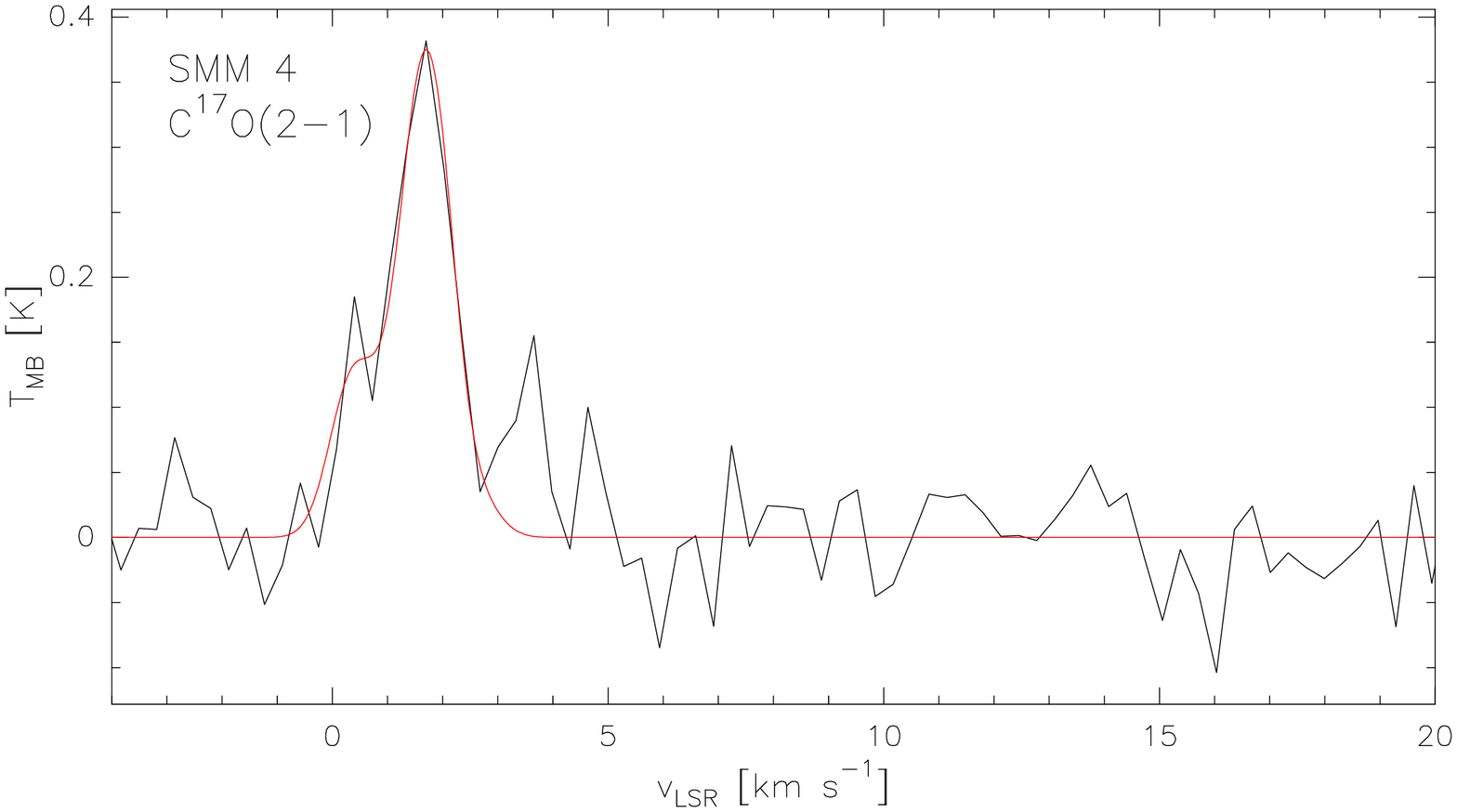}
\includegraphics[width=3.1cm, angle=-90]{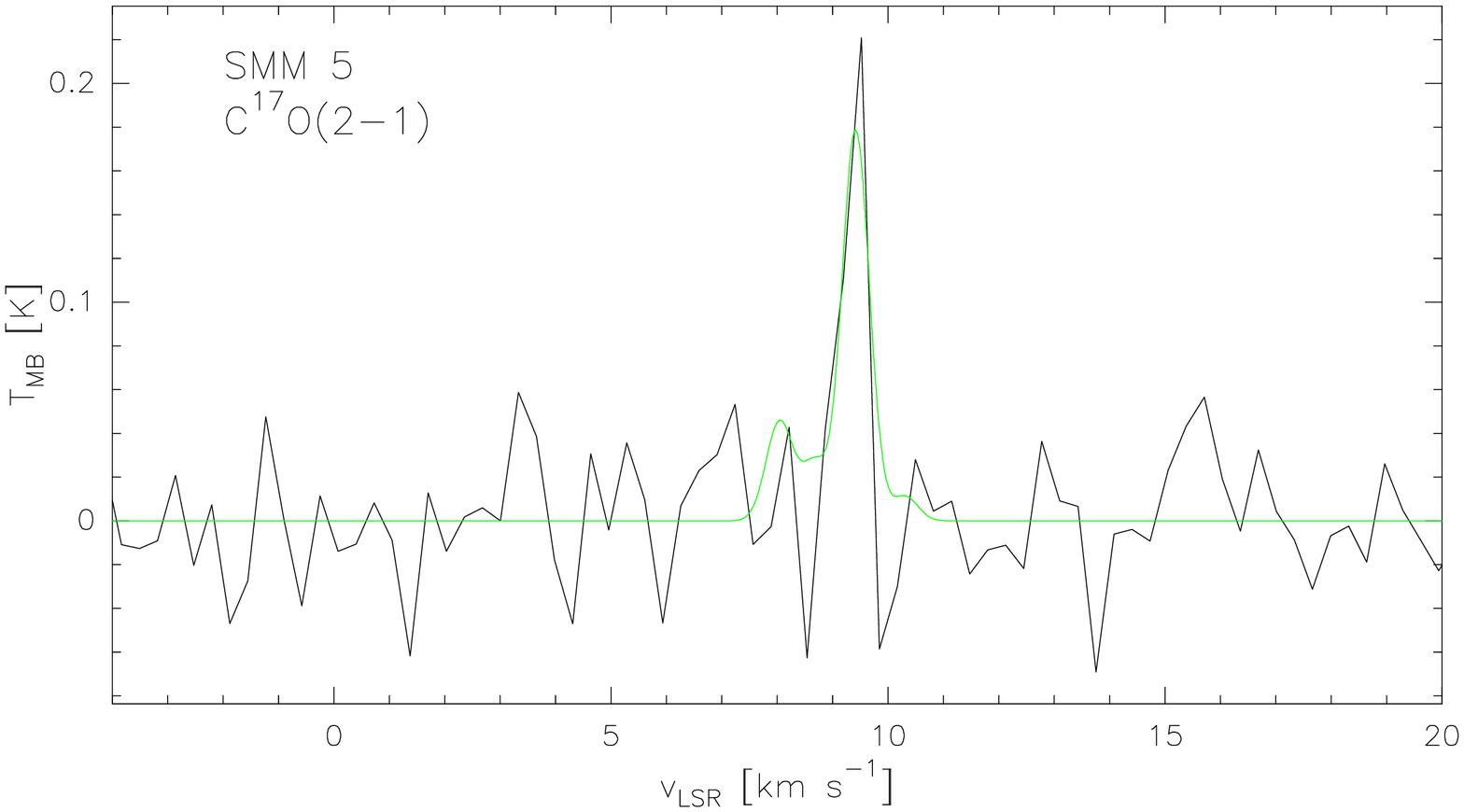}
\includegraphics[width=3.1cm, angle=-90]{blank.eps}
\includegraphics[width=3.1cm, angle=-90]{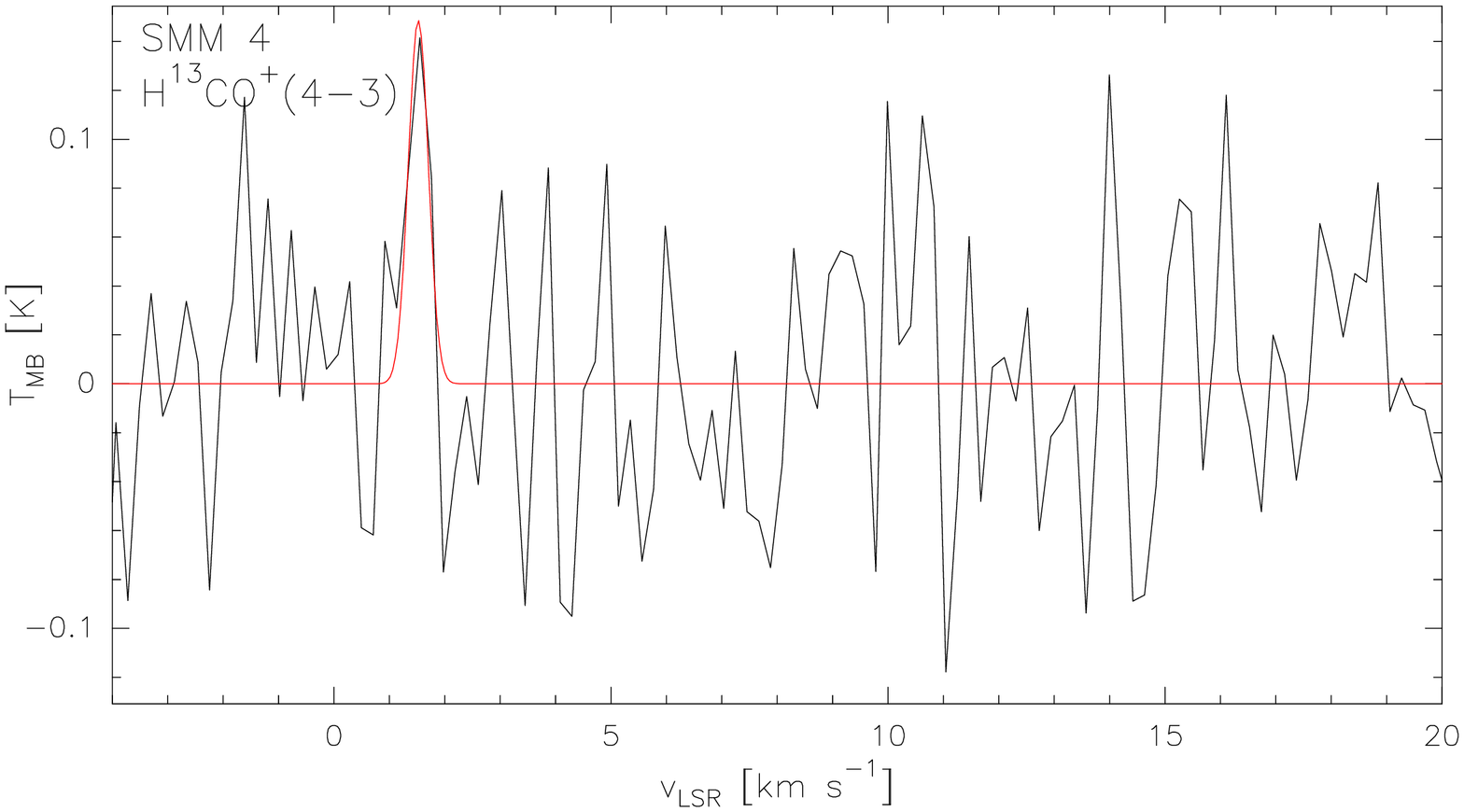}
\includegraphics[width=3.1cm, angle=-90]{blank.eps}
\includegraphics[width=3.1cm, angle=-90]{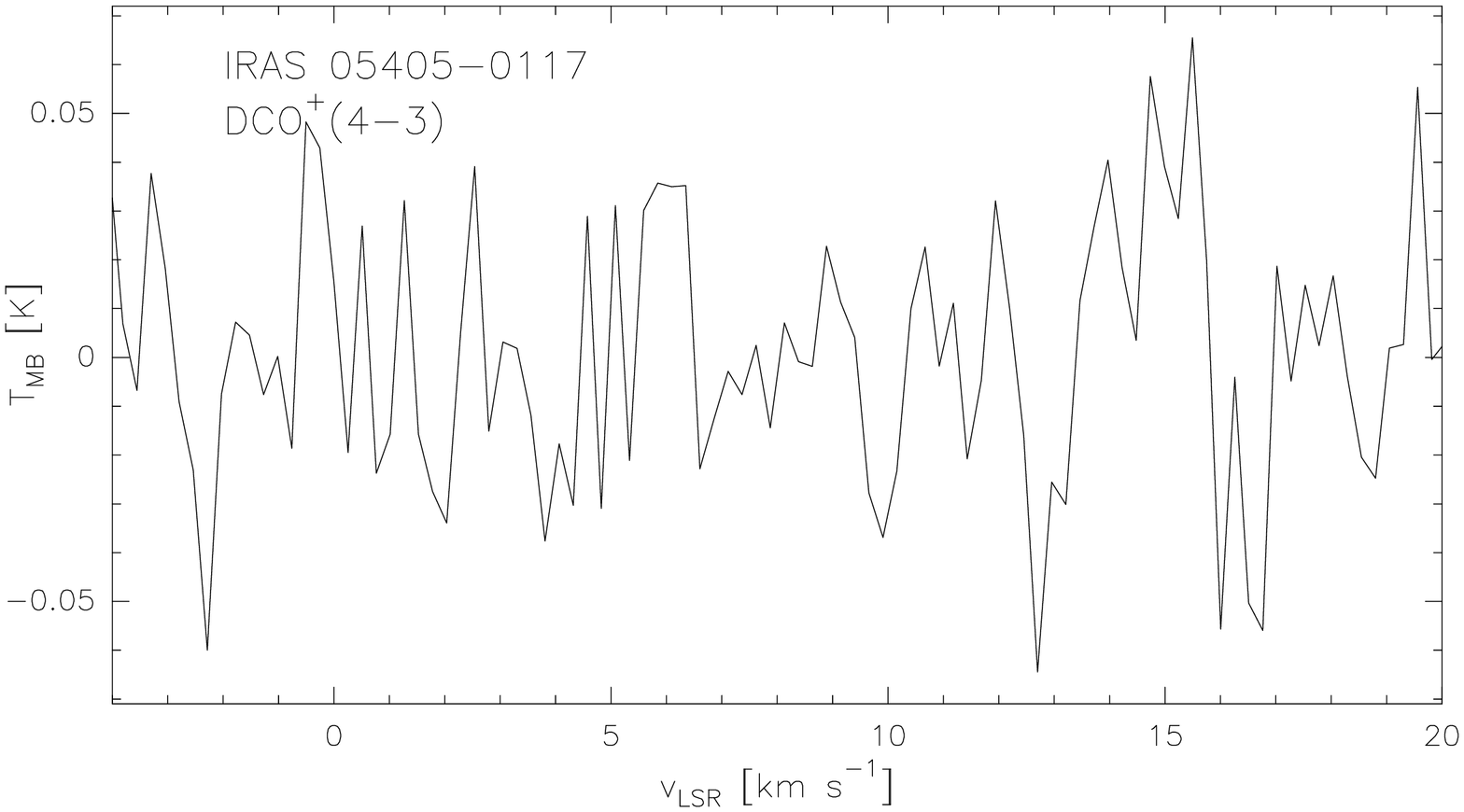}
\includegraphics[width=3.1cm, angle=-90]{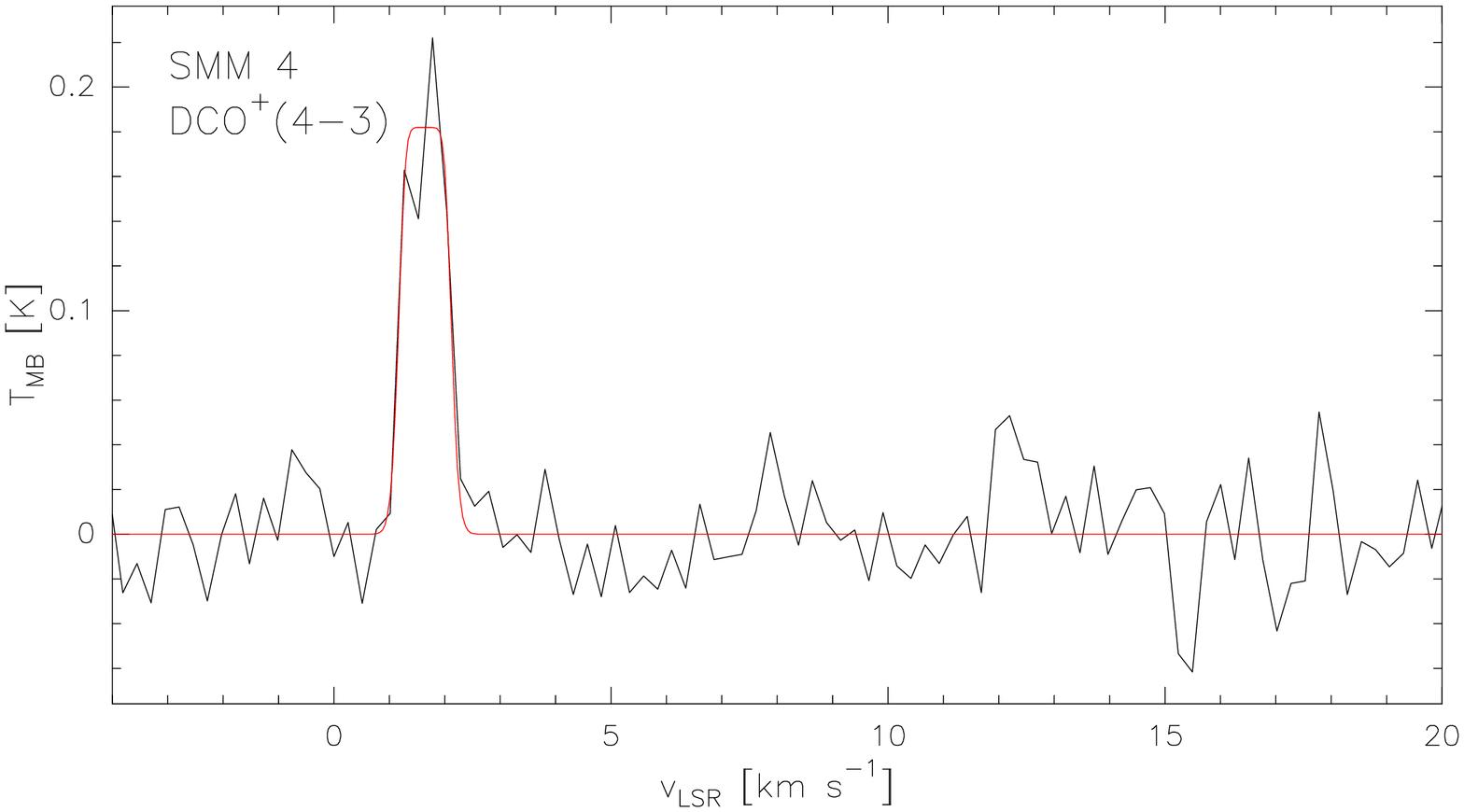}
\includegraphics[width=3.1cm, angle=-90]{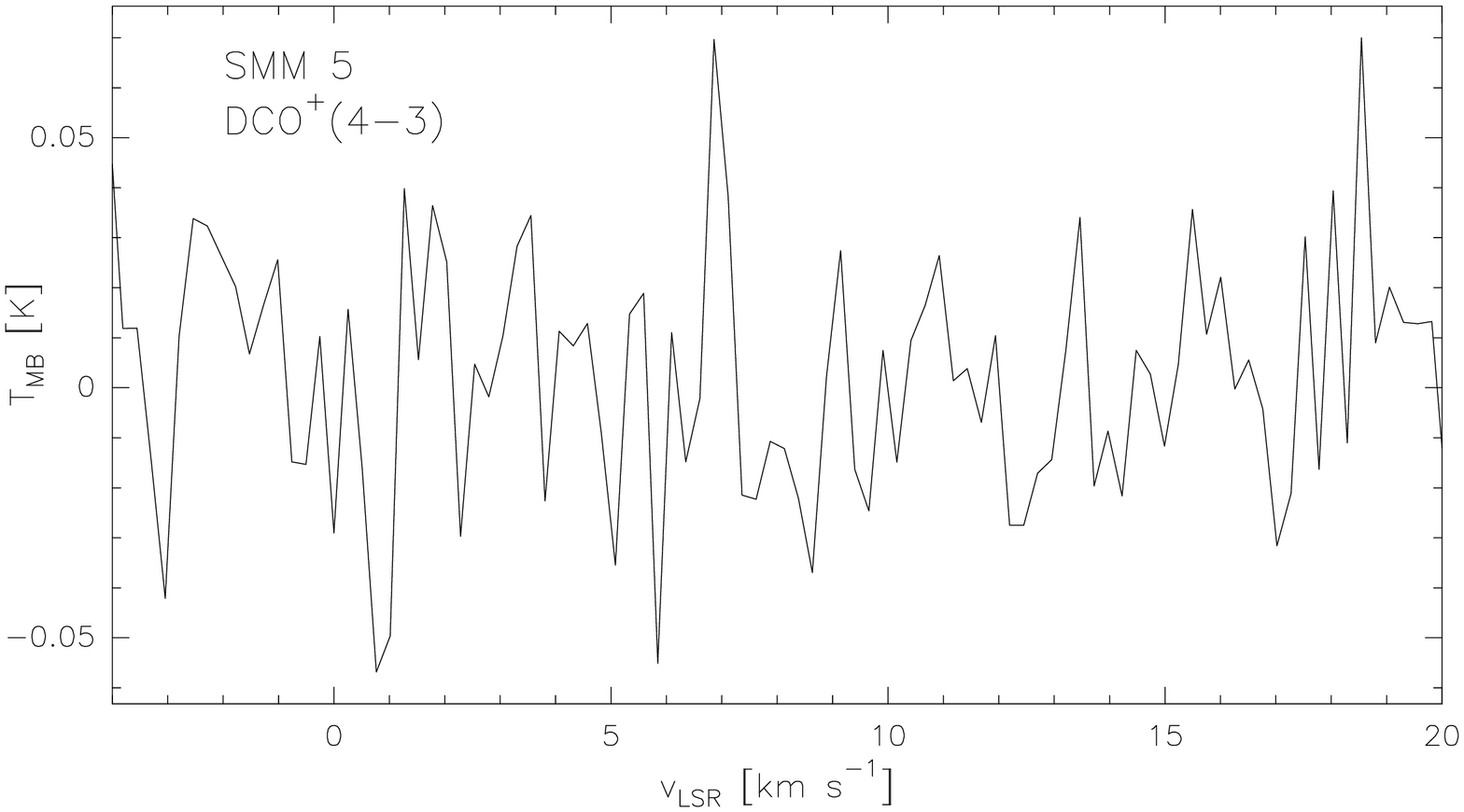}
\includegraphics[width=3.1cm, angle=-90]{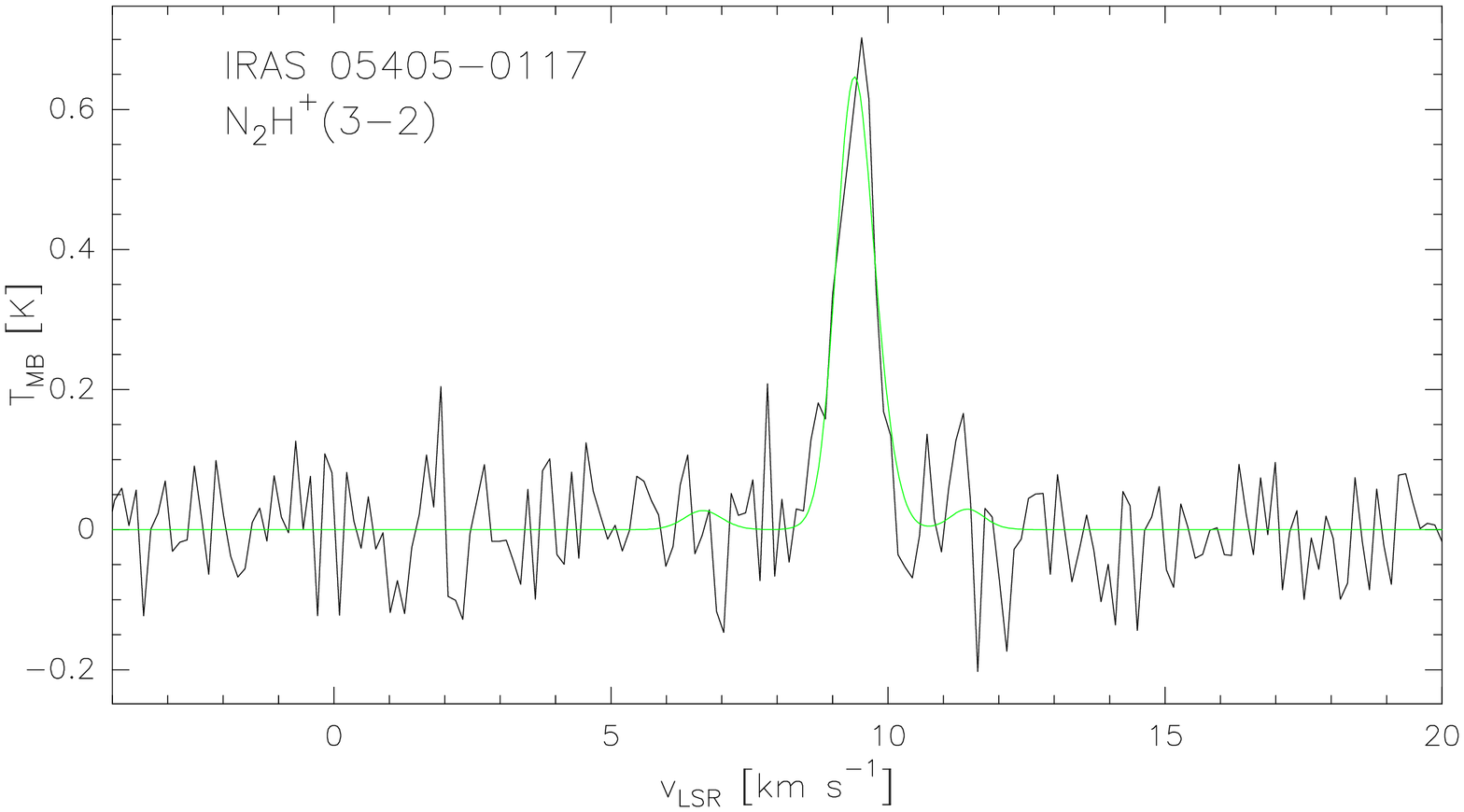}
\includegraphics[width=3.1cm, angle=-90]{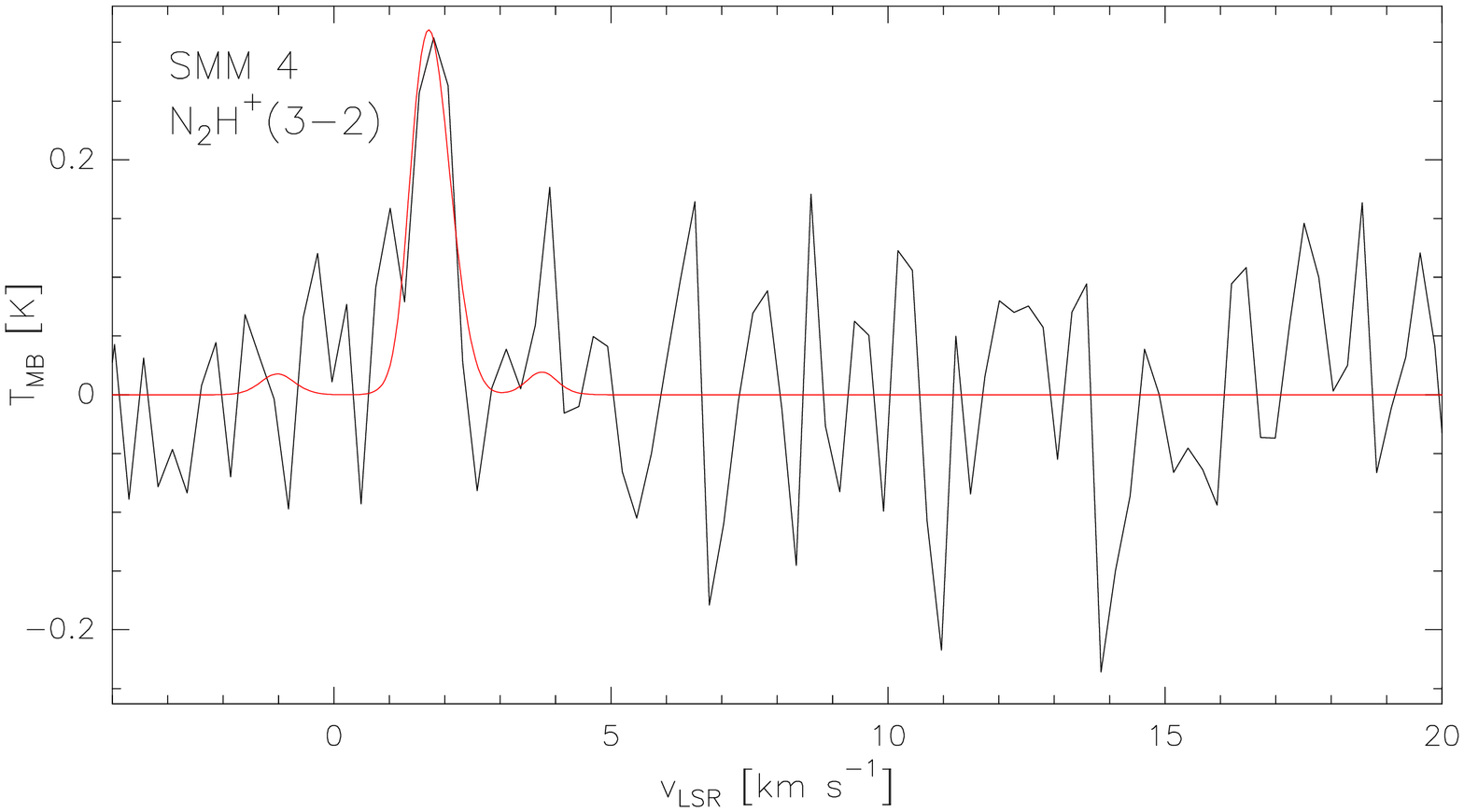}
\includegraphics[width=3.1cm, angle=-90]{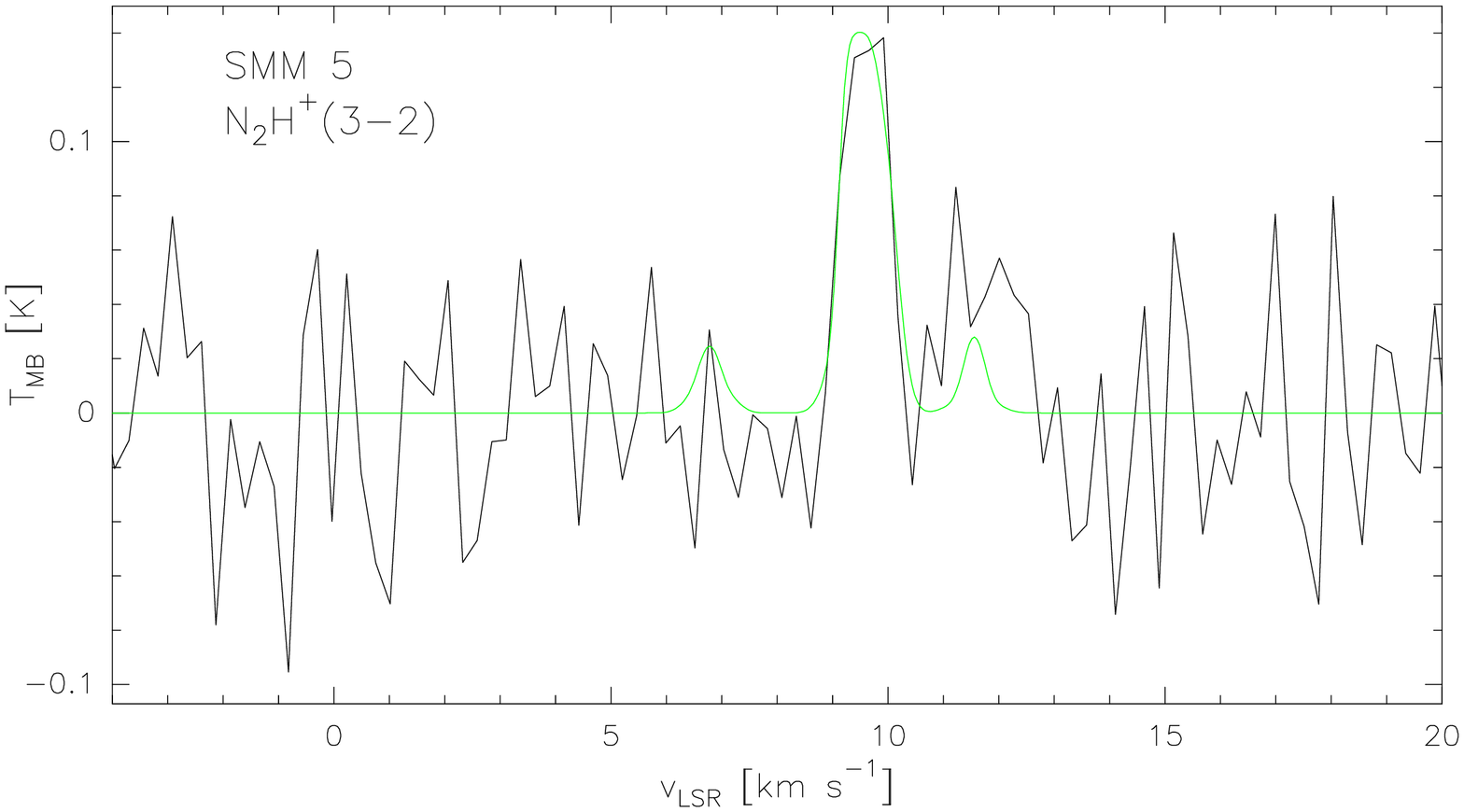}
\includegraphics[width=3.1cm, angle=-90]{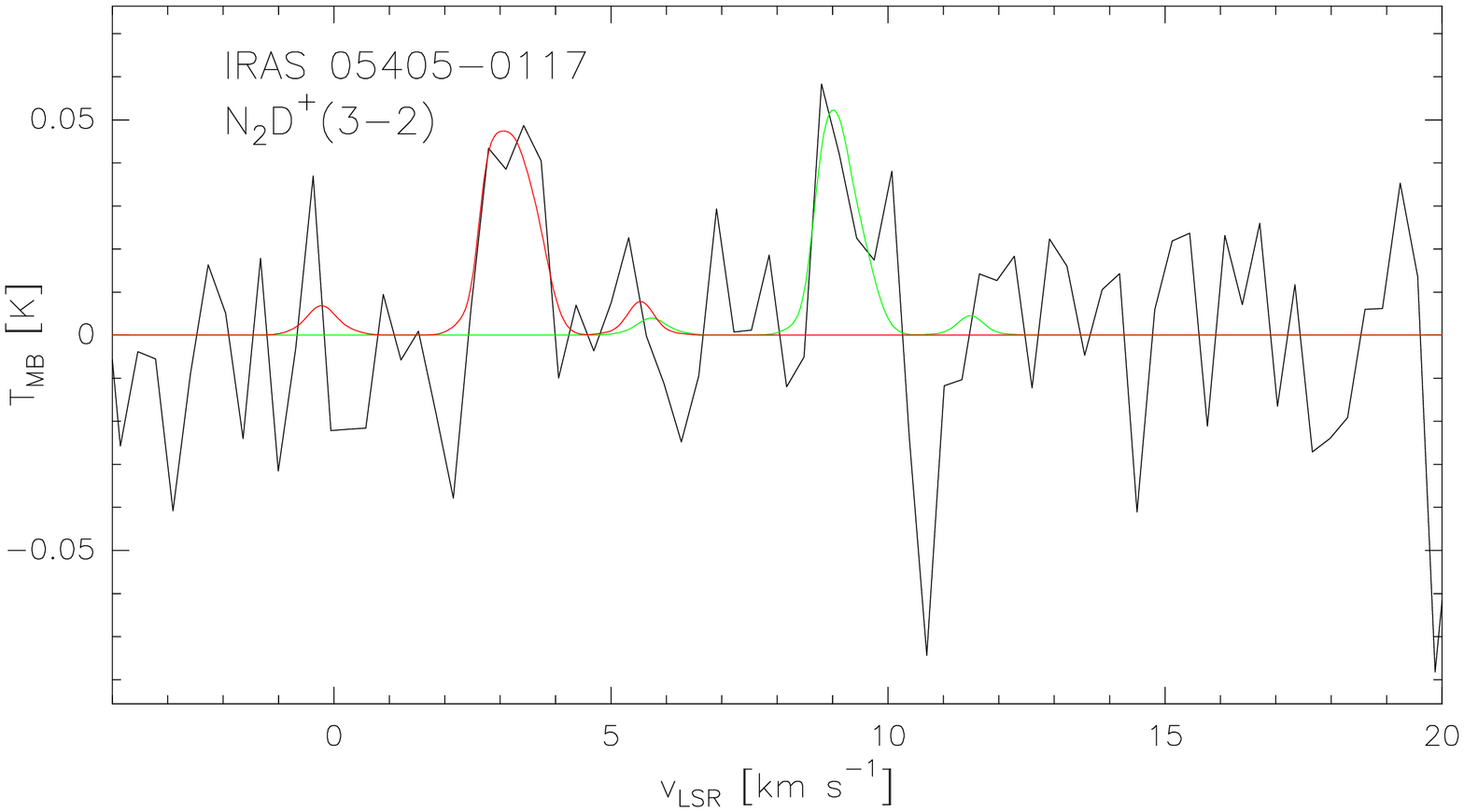}
\includegraphics[width=3.1cm, angle=-90]{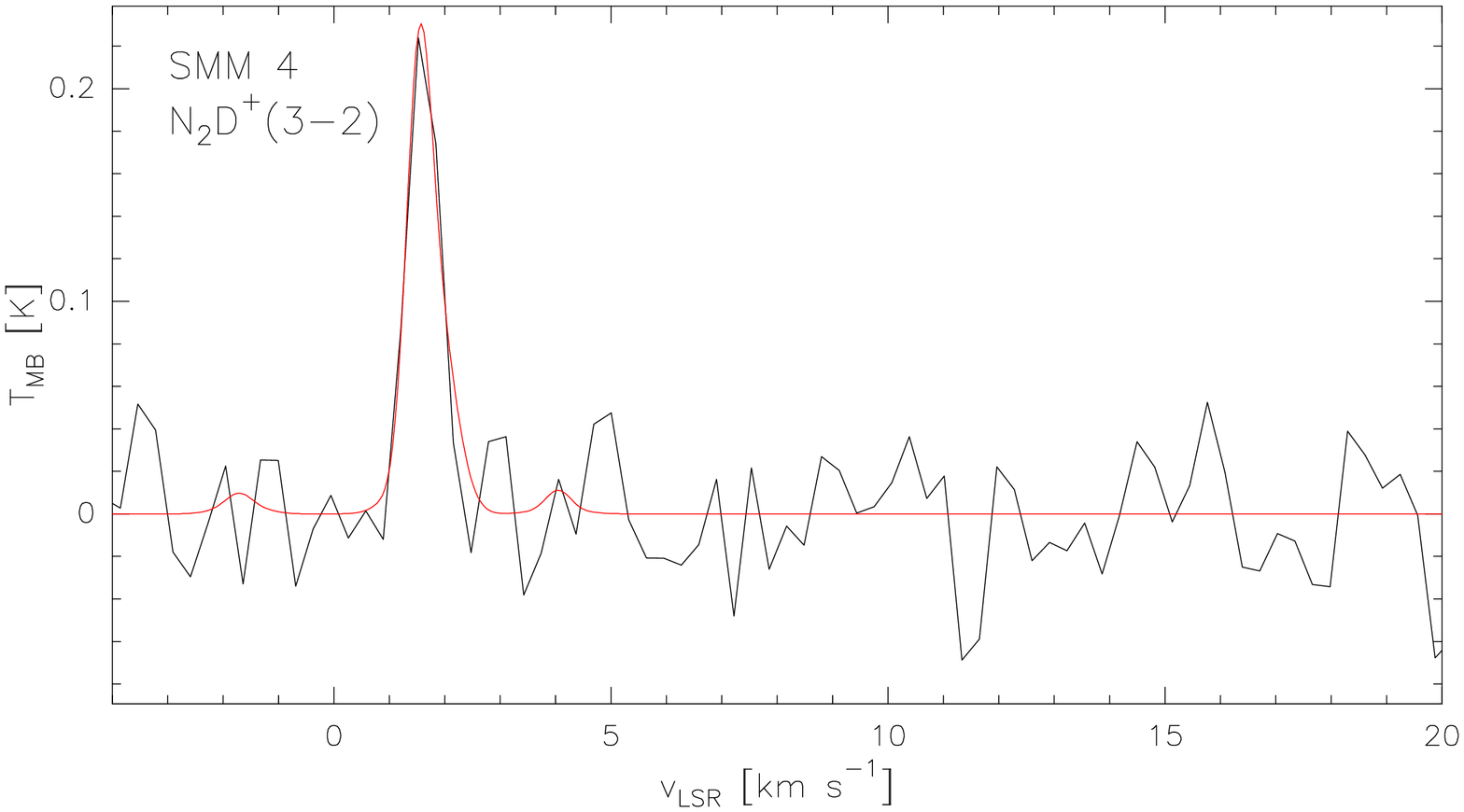}
\includegraphics[width=3.1cm, angle=-90]{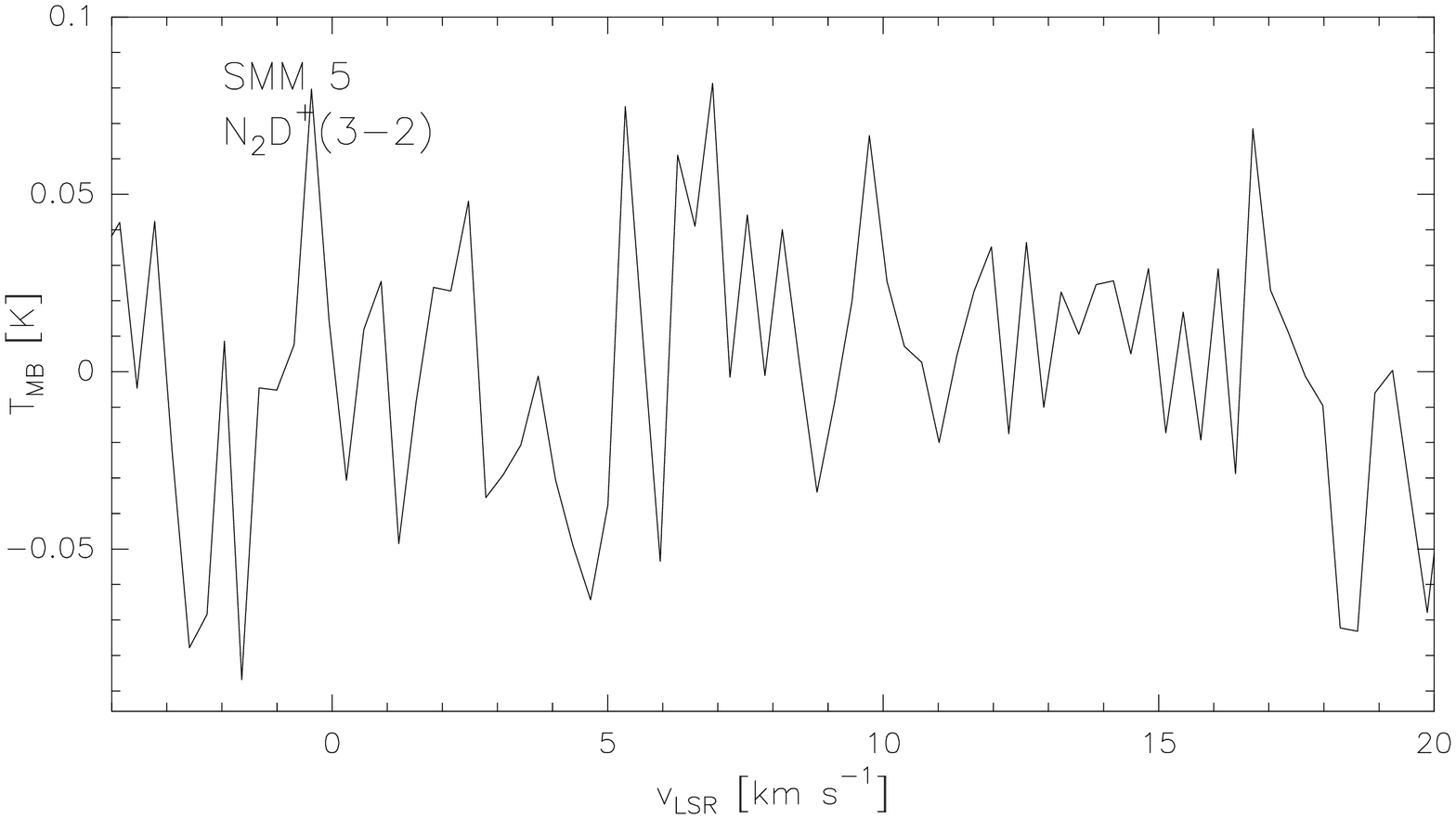}
\addtocounter{figure}{-1}
\caption{continued.}
\label{figure:spectra}
\end{center}
\end{figure*}

\begin{figure*}
\begin{center}
\includegraphics[width=3.1cm, angle=-90]{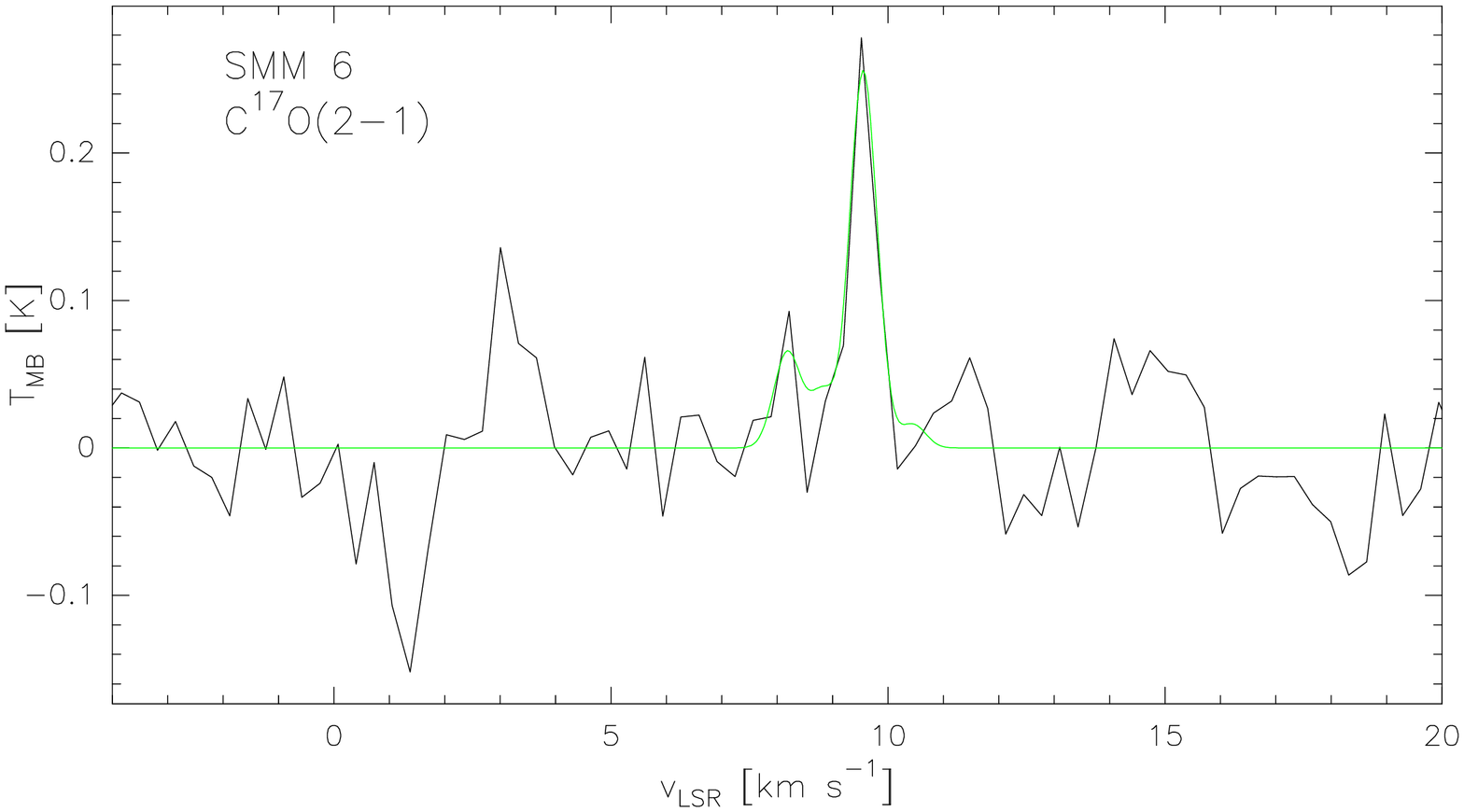}
\includegraphics[width=3.1cm, angle=-90]{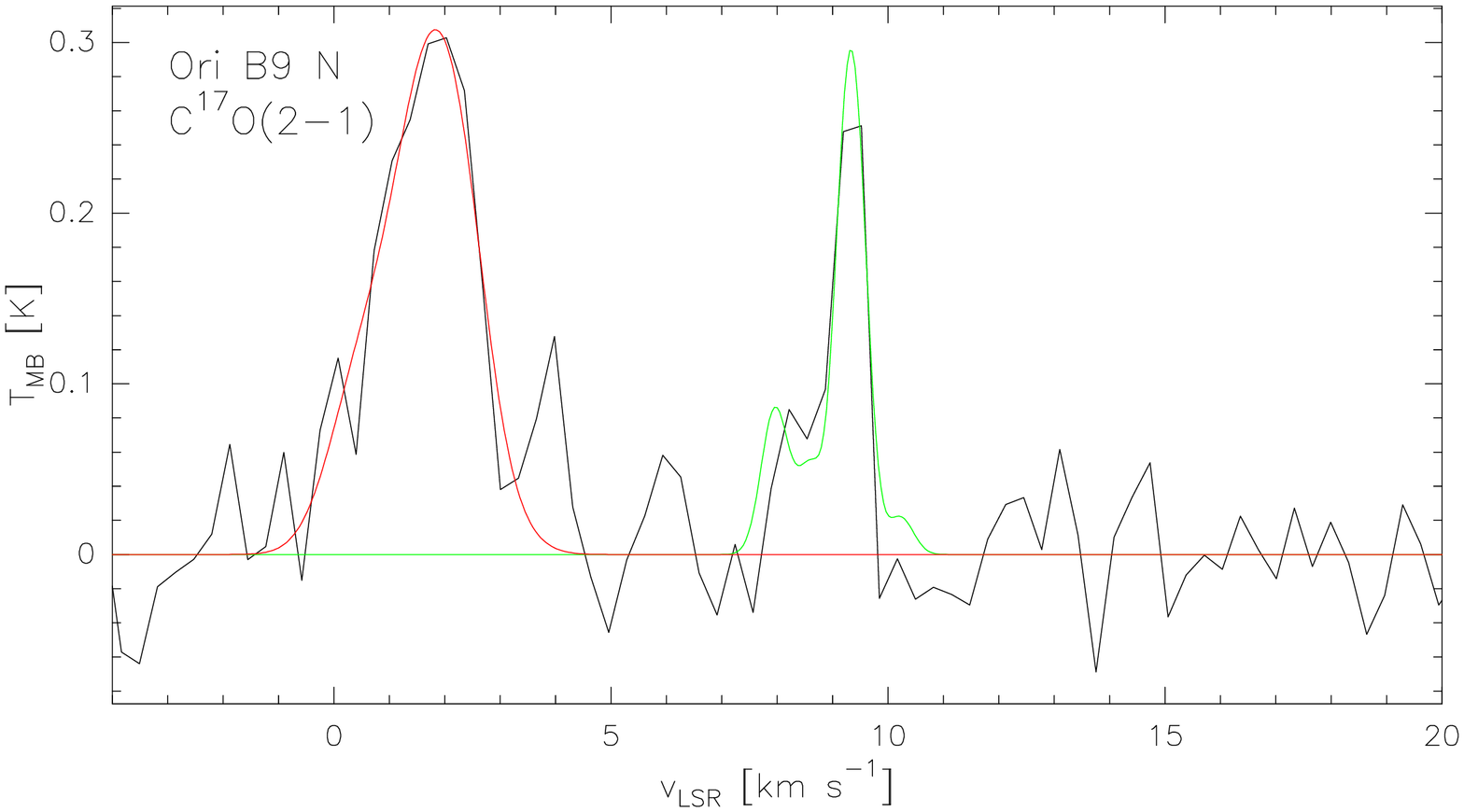}
\includegraphics[width=3.1cm, angle=-90]{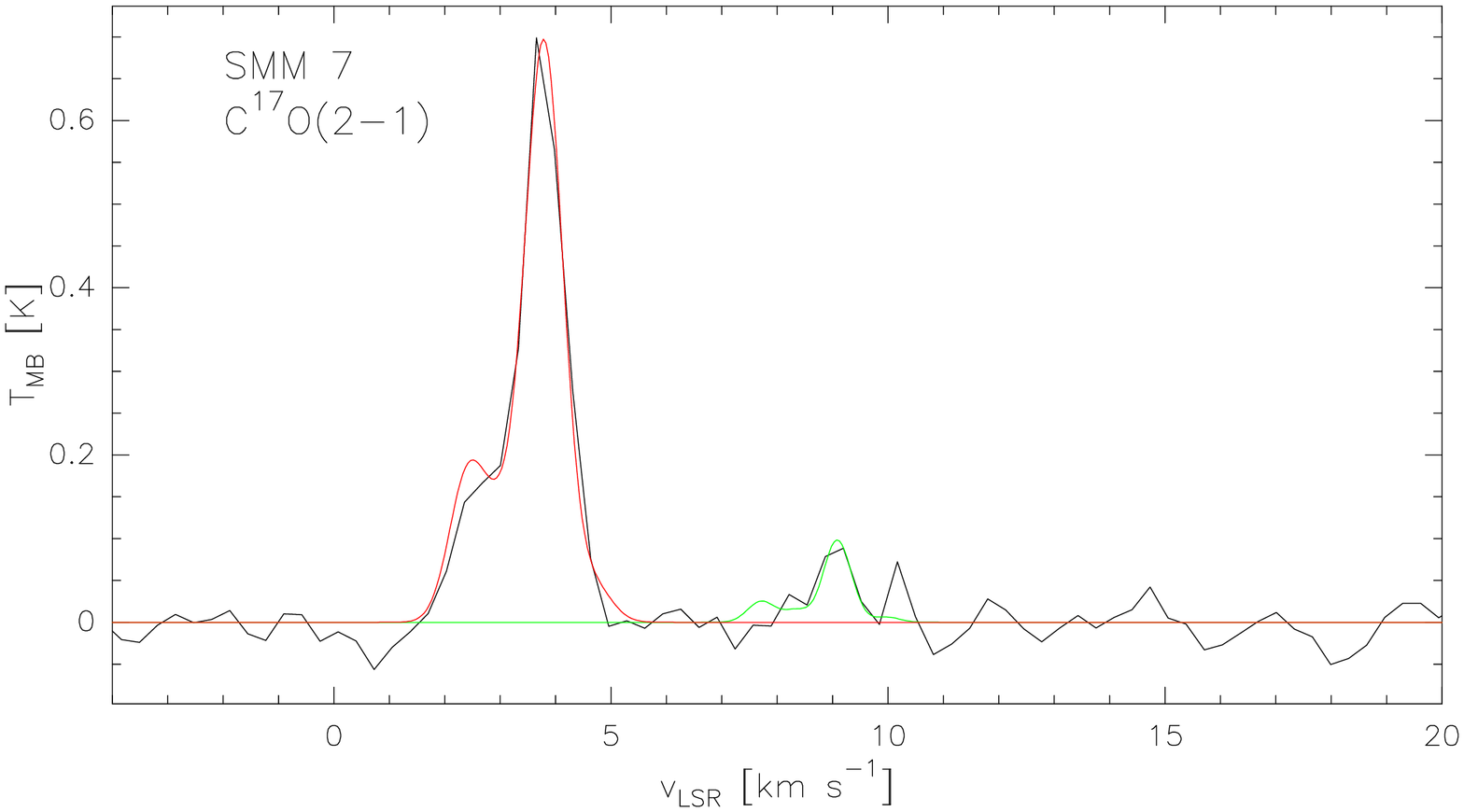}
\includegraphics[width=3.1cm, angle=-90]{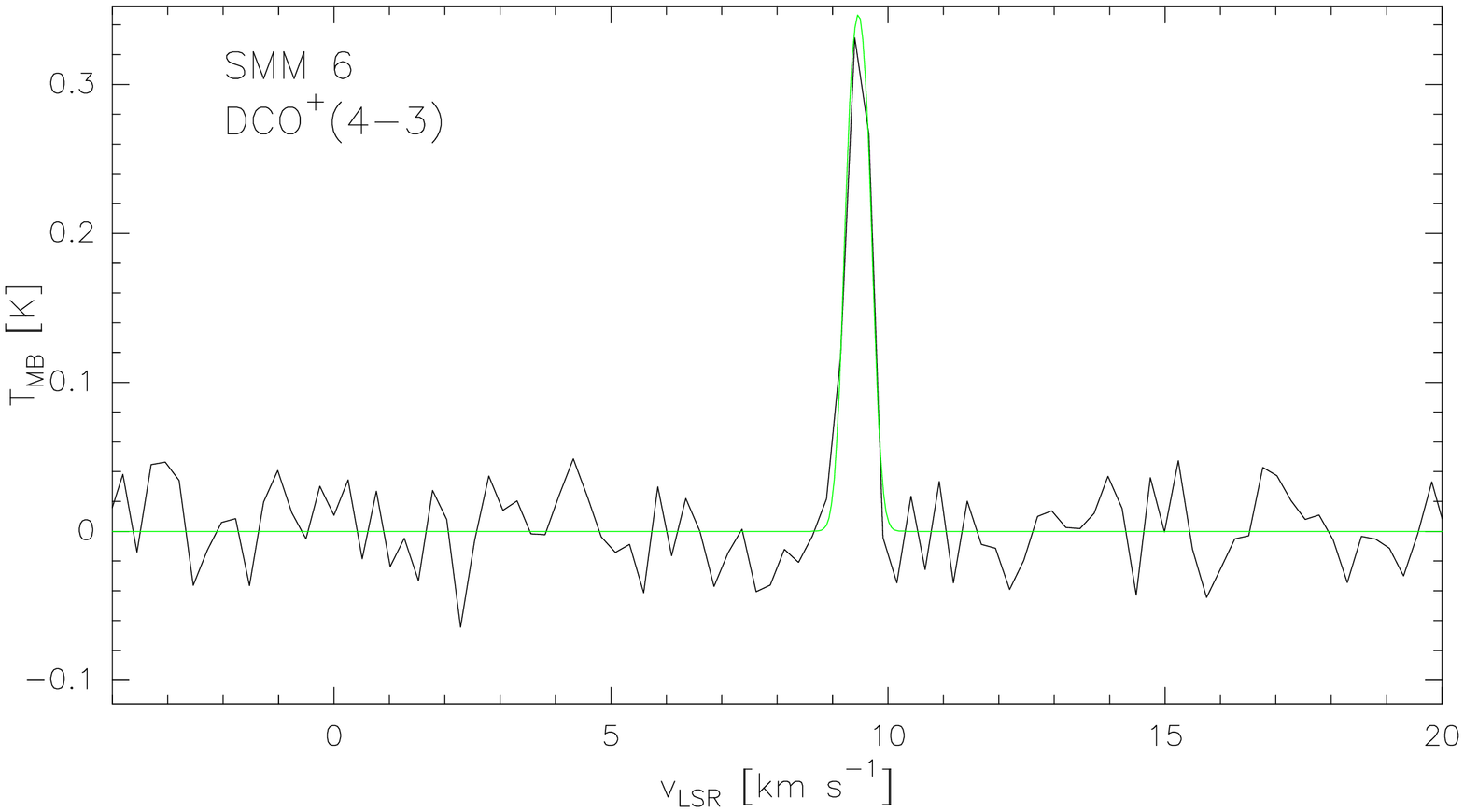}
\includegraphics[width=3.1cm, angle=-90]{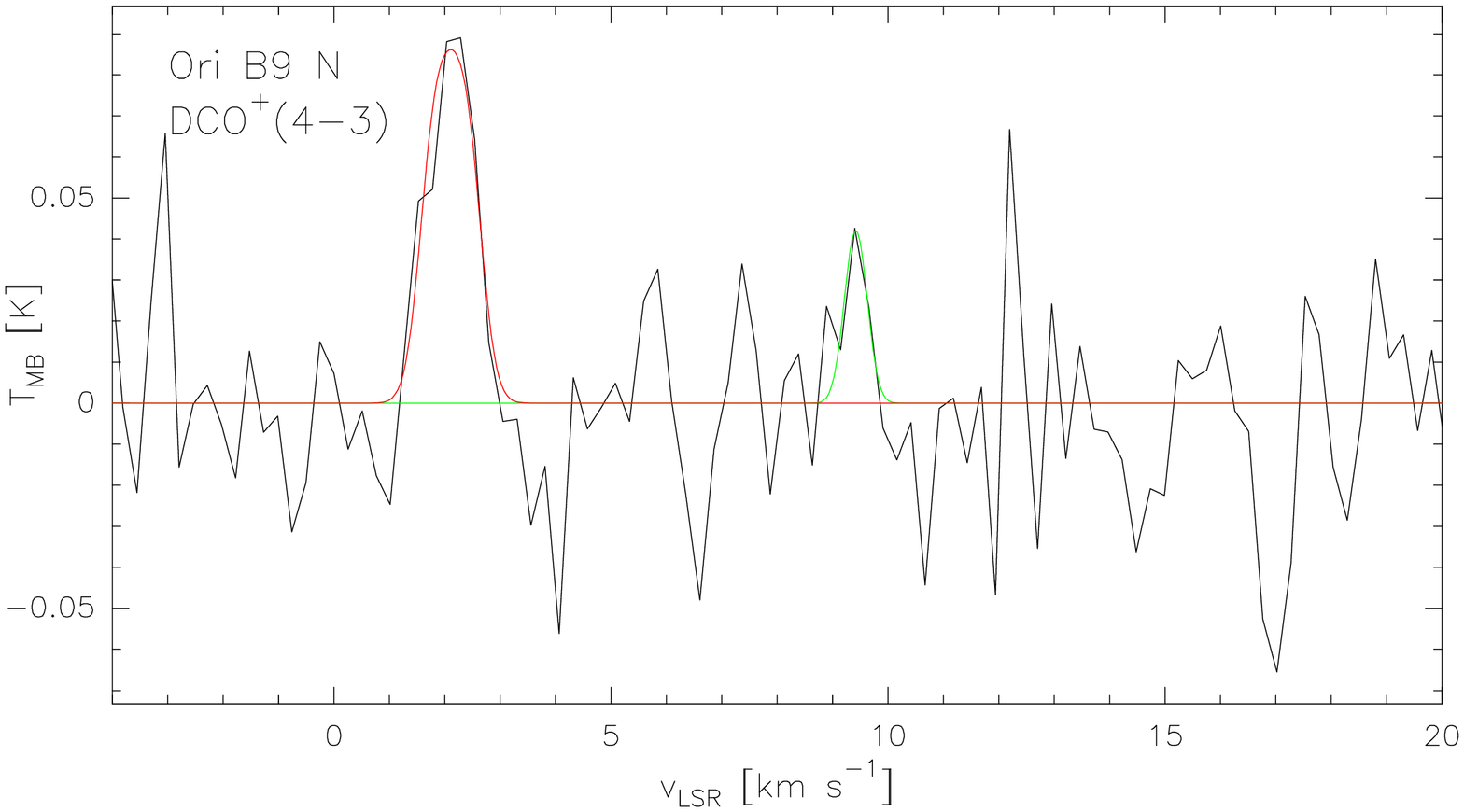}
\includegraphics[width=3.1cm, angle=-90]{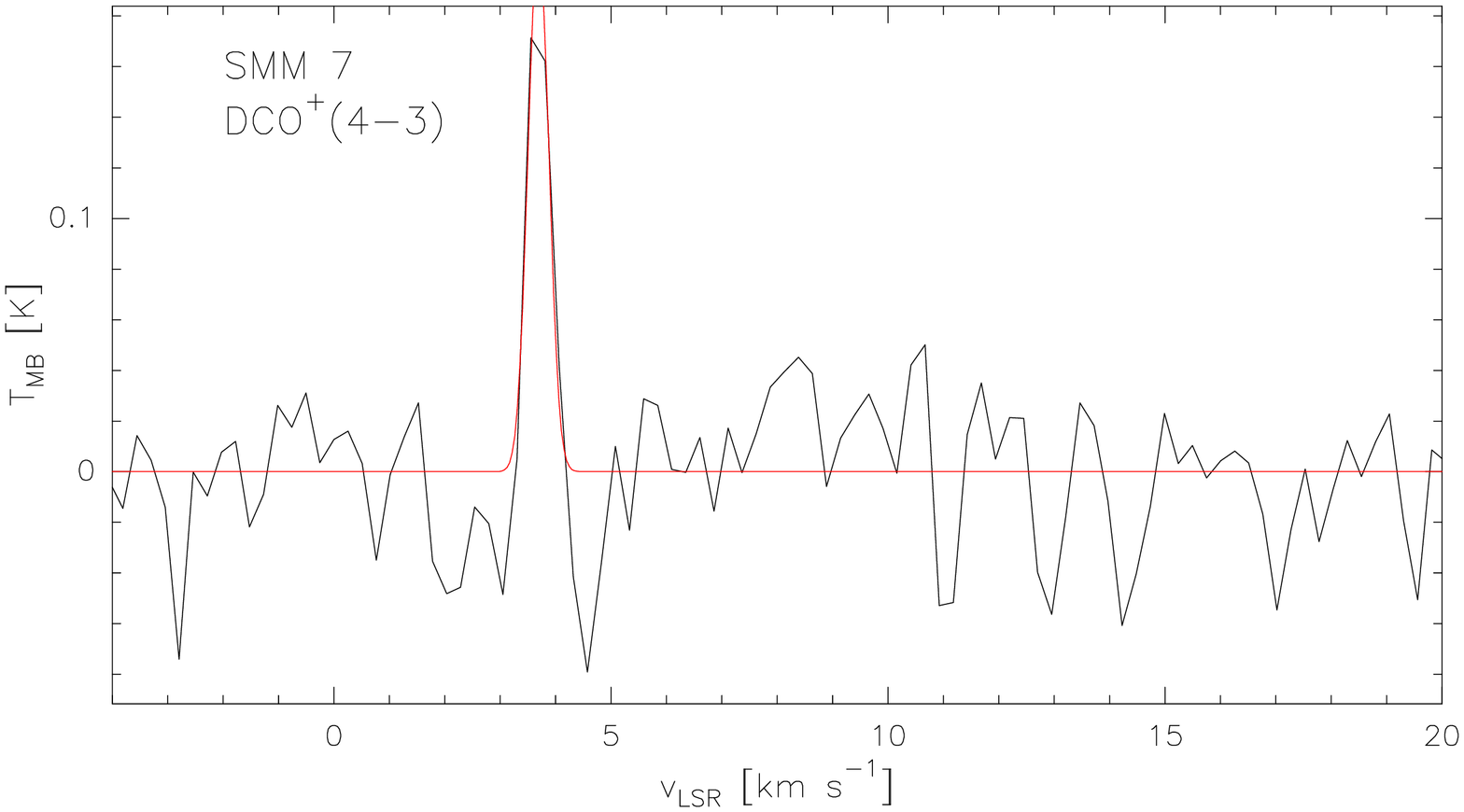}
\includegraphics[width=3.1cm, angle=-90]{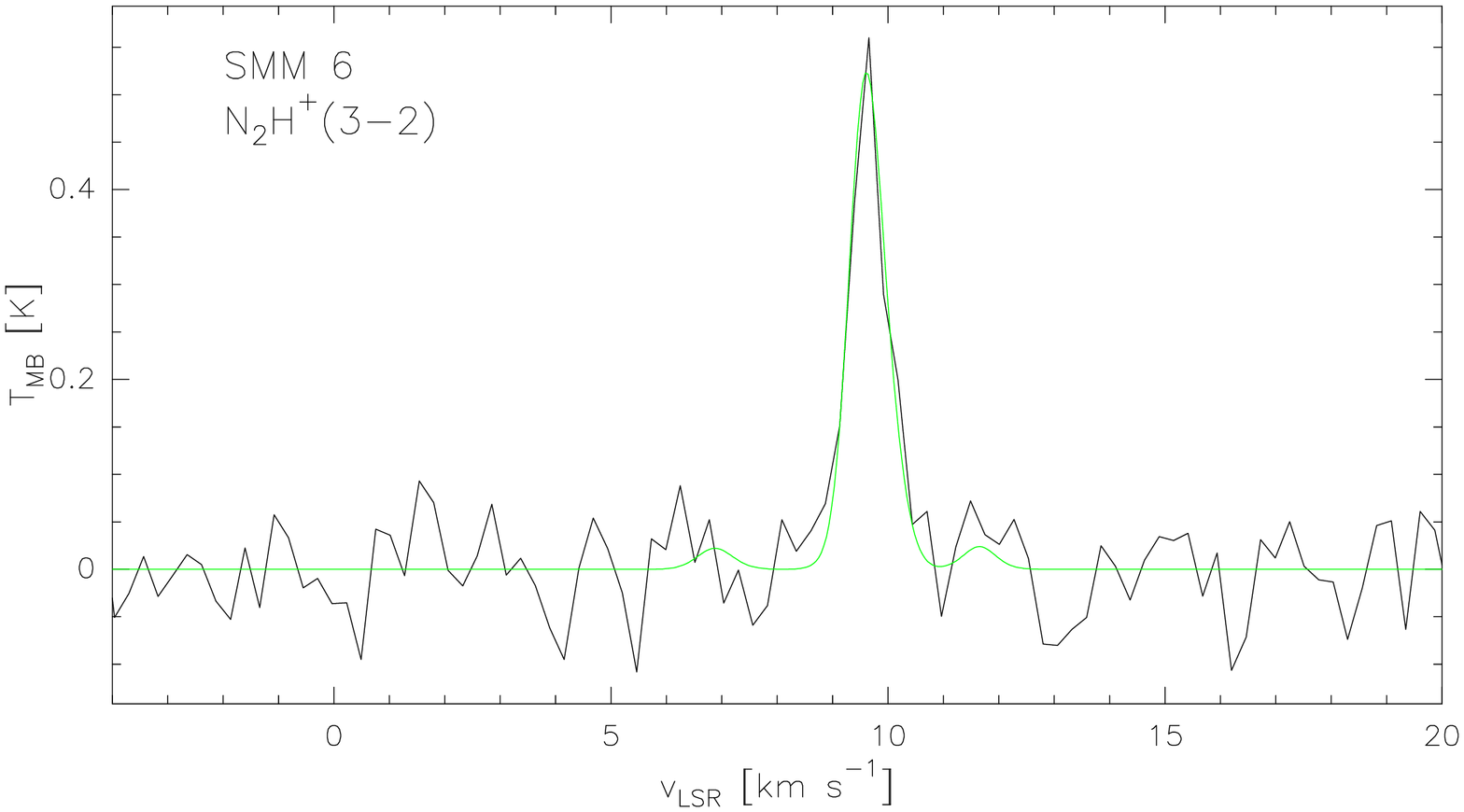}
\includegraphics[width=3.1cm, angle=-90]{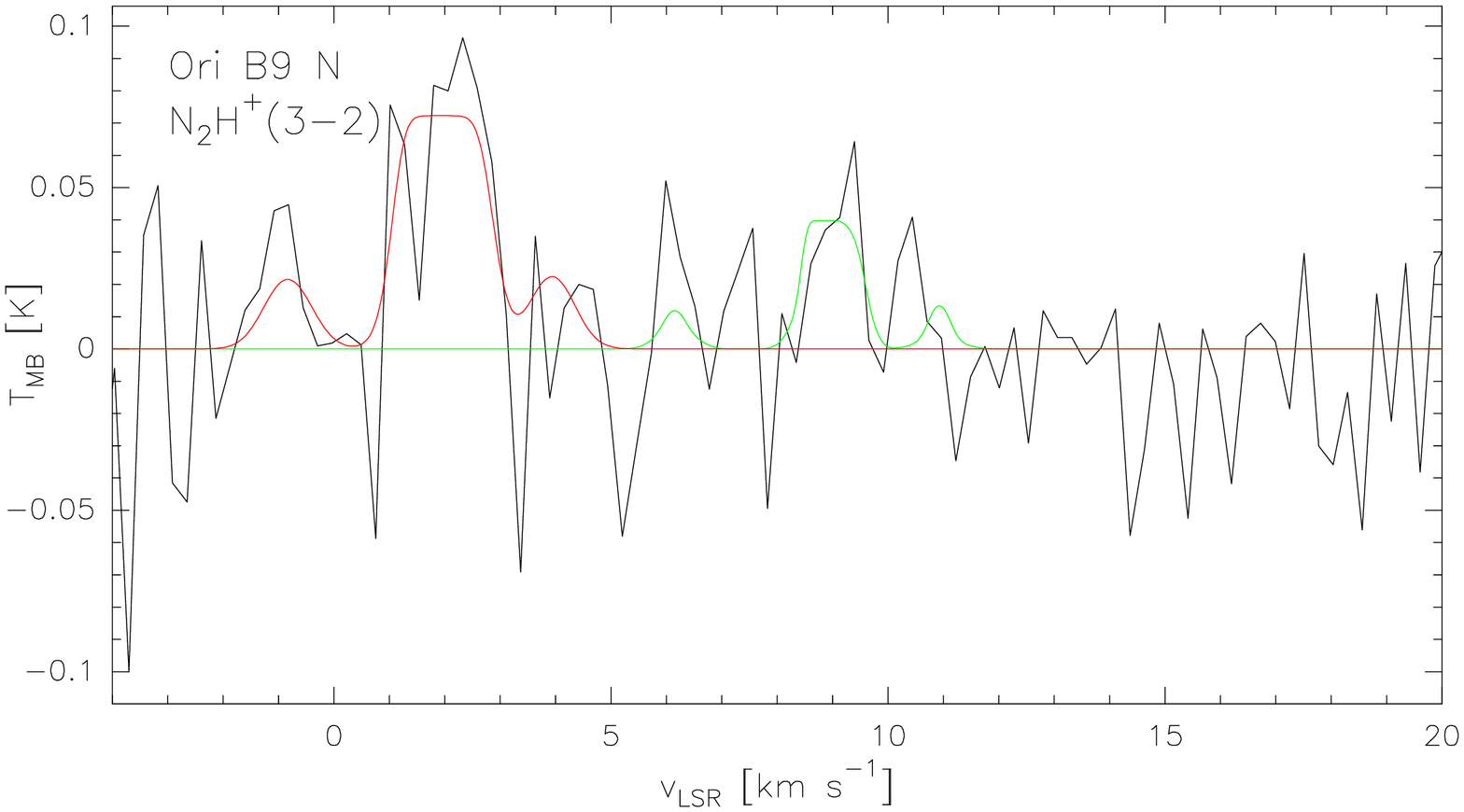}
\includegraphics[width=3.1cm, angle=-90]{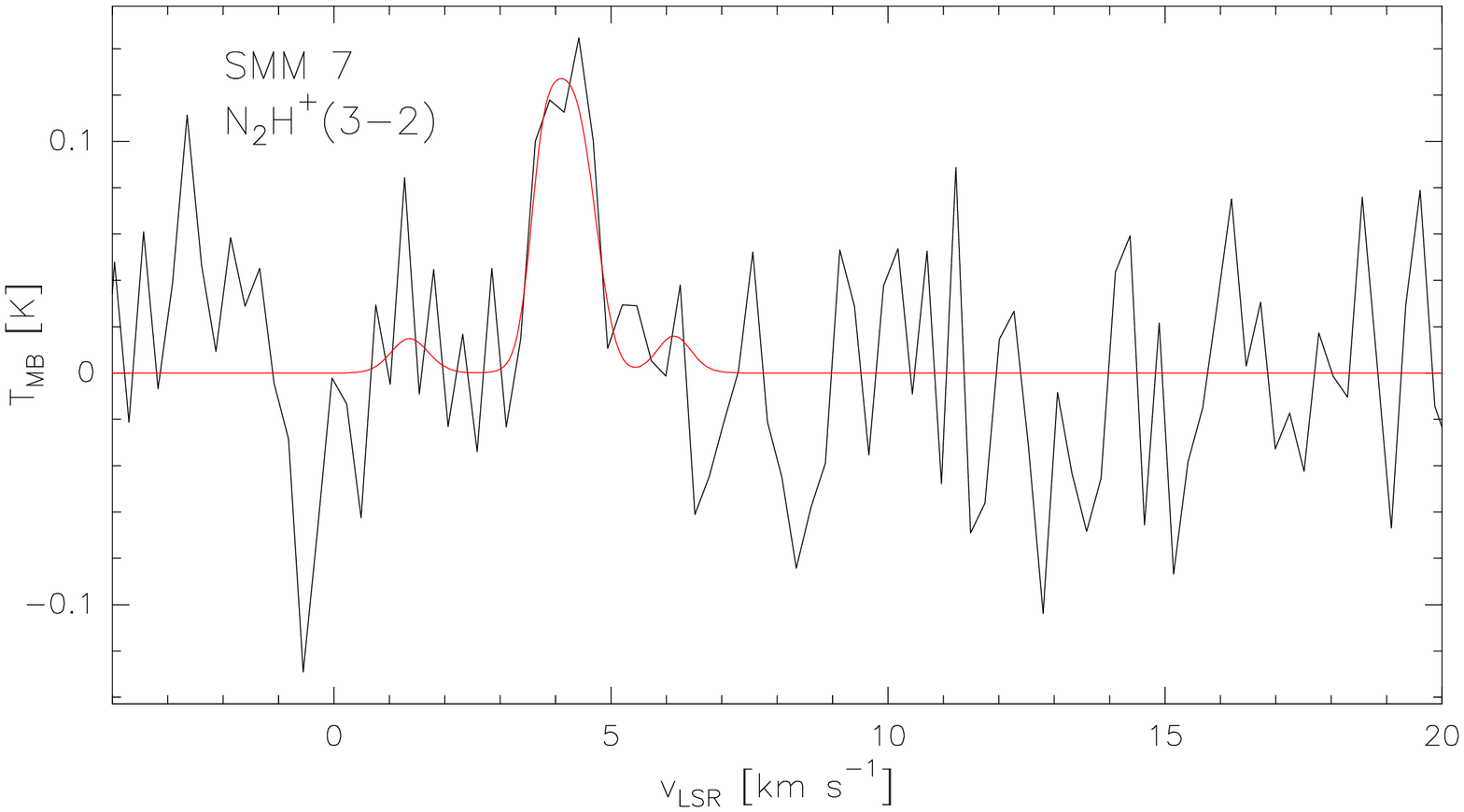}
\includegraphics[width=3.1cm, angle=-90]{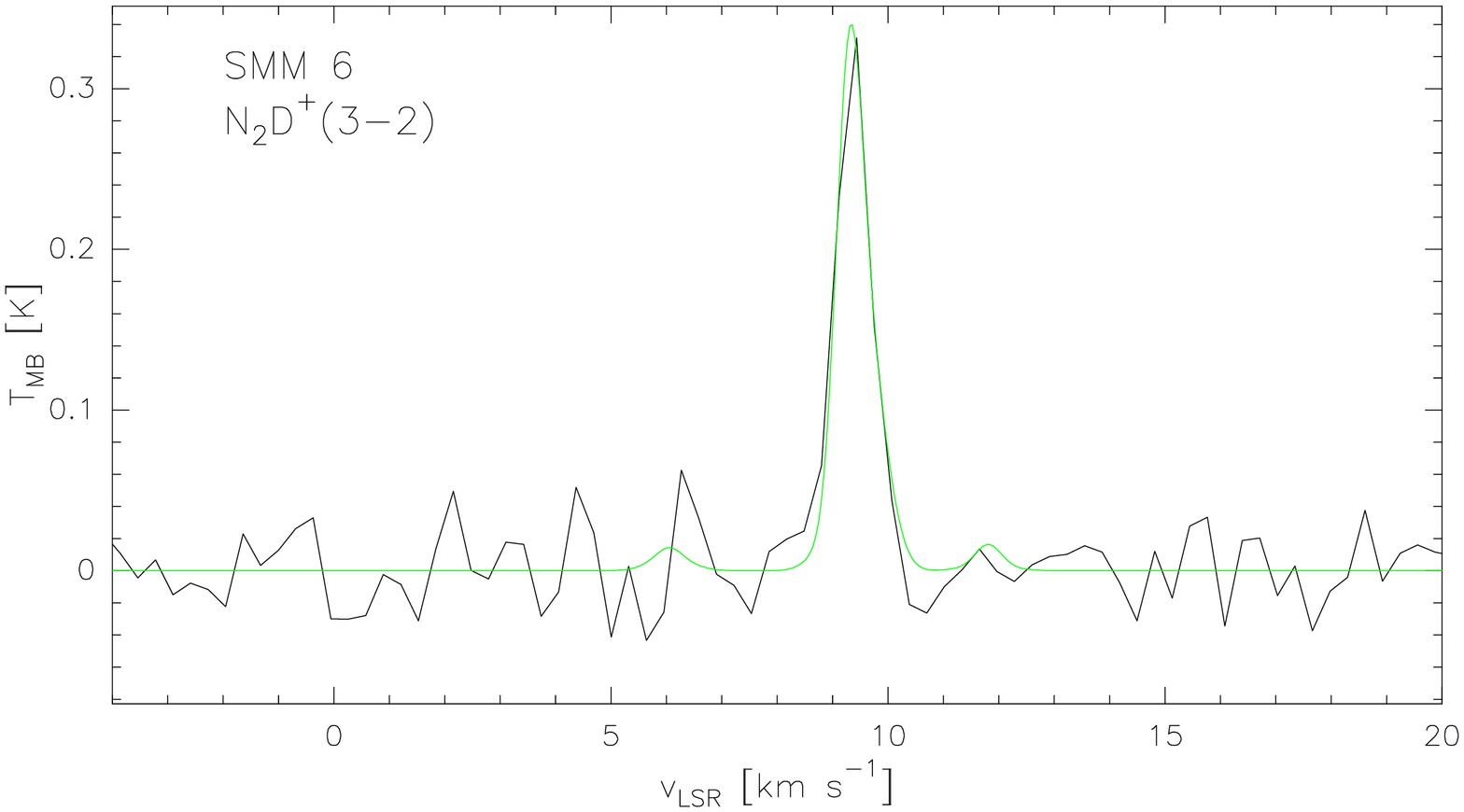}
\includegraphics[width=3.1cm, angle=-90]{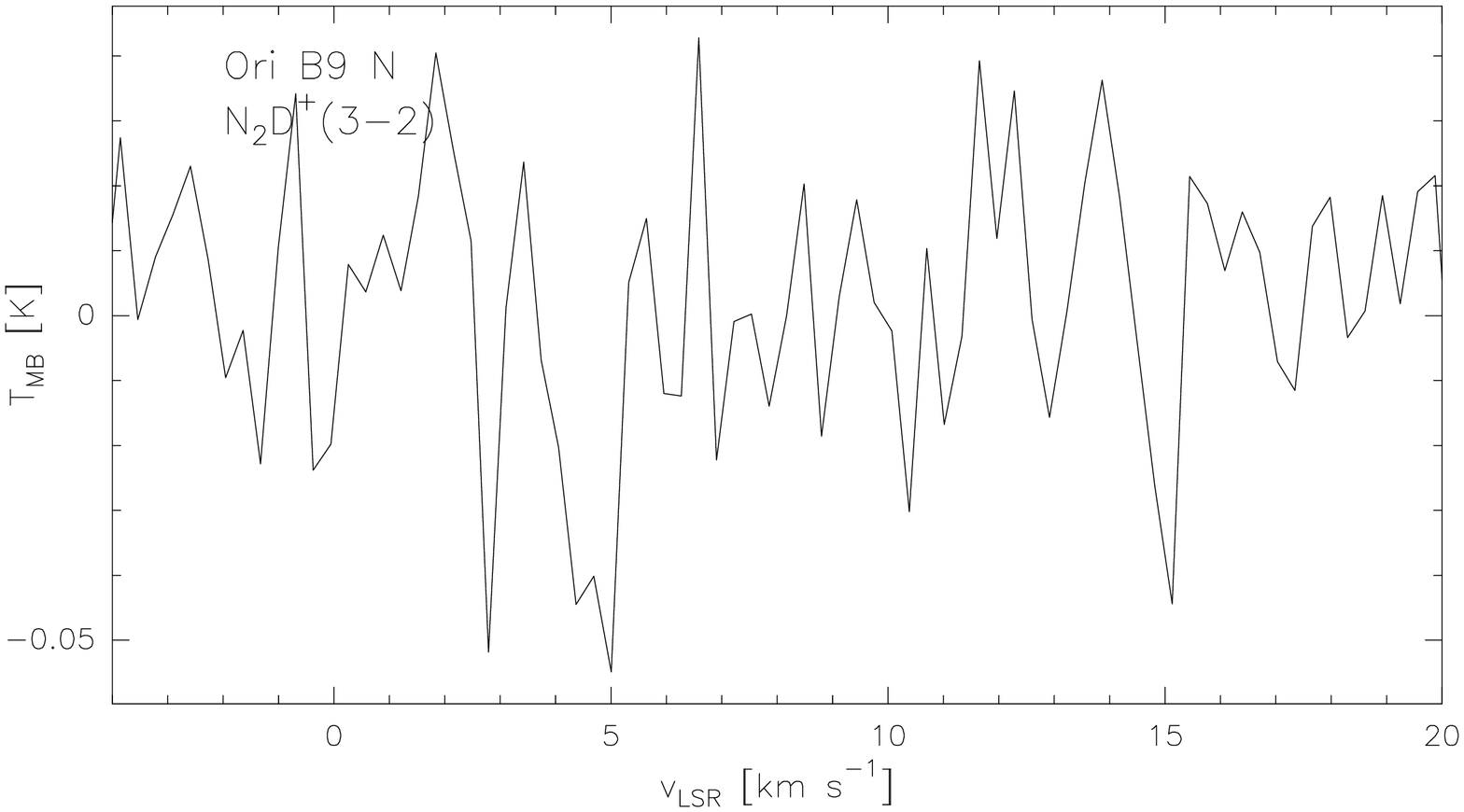}
\includegraphics[width=3.1cm, angle=-90]{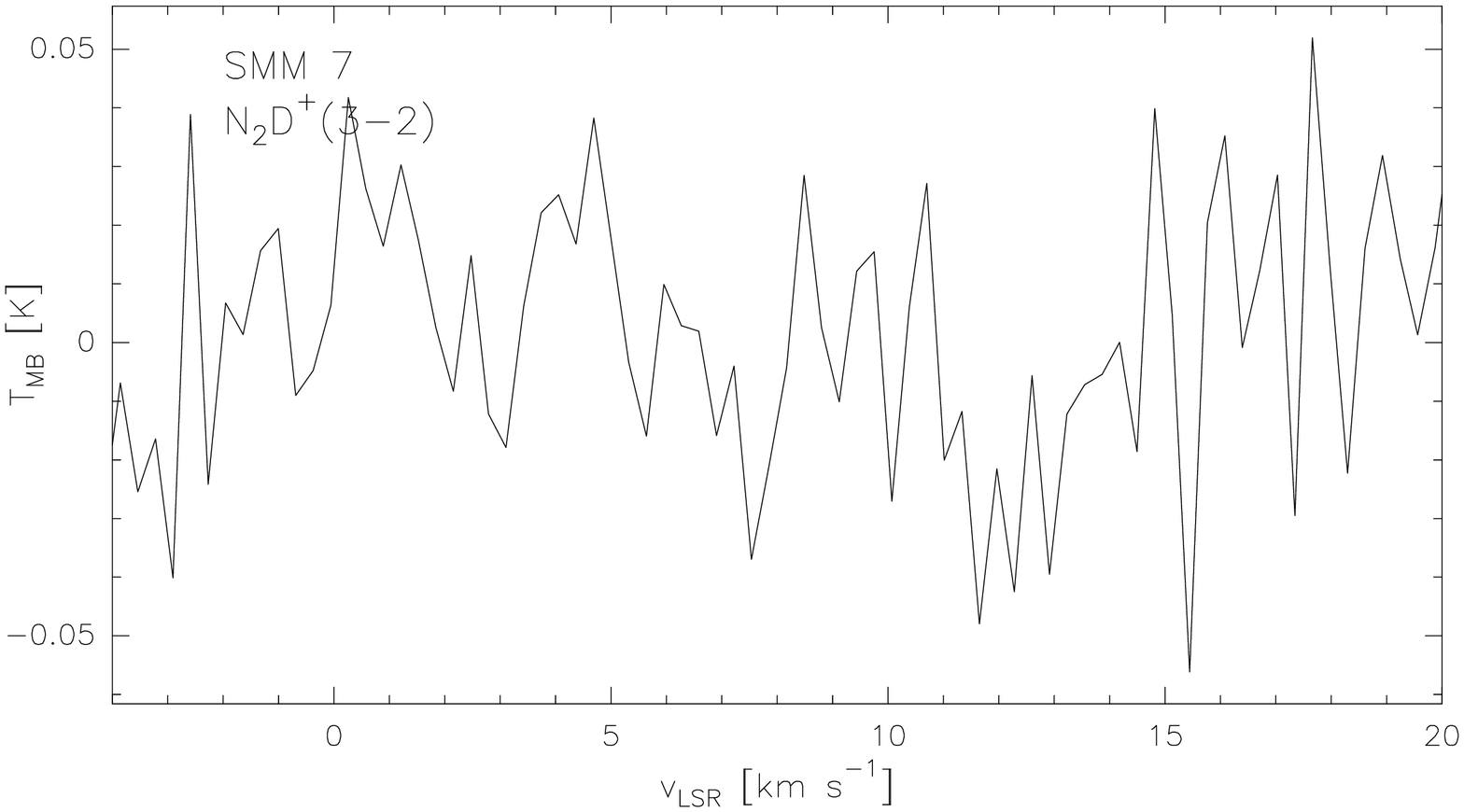}
\addtocounter{figure}{-1}
\caption{continued.}
\label{figure:spectra}
\end{center}
\end{figure*}

\subsubsection{Other line detections}

All the observed sources except SMM 5 show additional spectral lines in the 
frequency band covering the N$_2$D$^+(3-2)$ transition. The line 
identification was done by using Weeds, which is an extension of CLASS 
(\cite{maret2011}), and the JPL and CDMS spectroscopic databases. 
We used the LTE modelling application of Weeds to check if all predicted lines 
of a candidate molecule are present in the observed spectrum. In some cases, we 
were able to reject some line candidates on the basis of non-detection of 
other transitions expected at nearby frequencies.

\begin{figure}[!h]
\centering
\resizebox{0.6\hsize}{!}{\includegraphics[angle=-90]{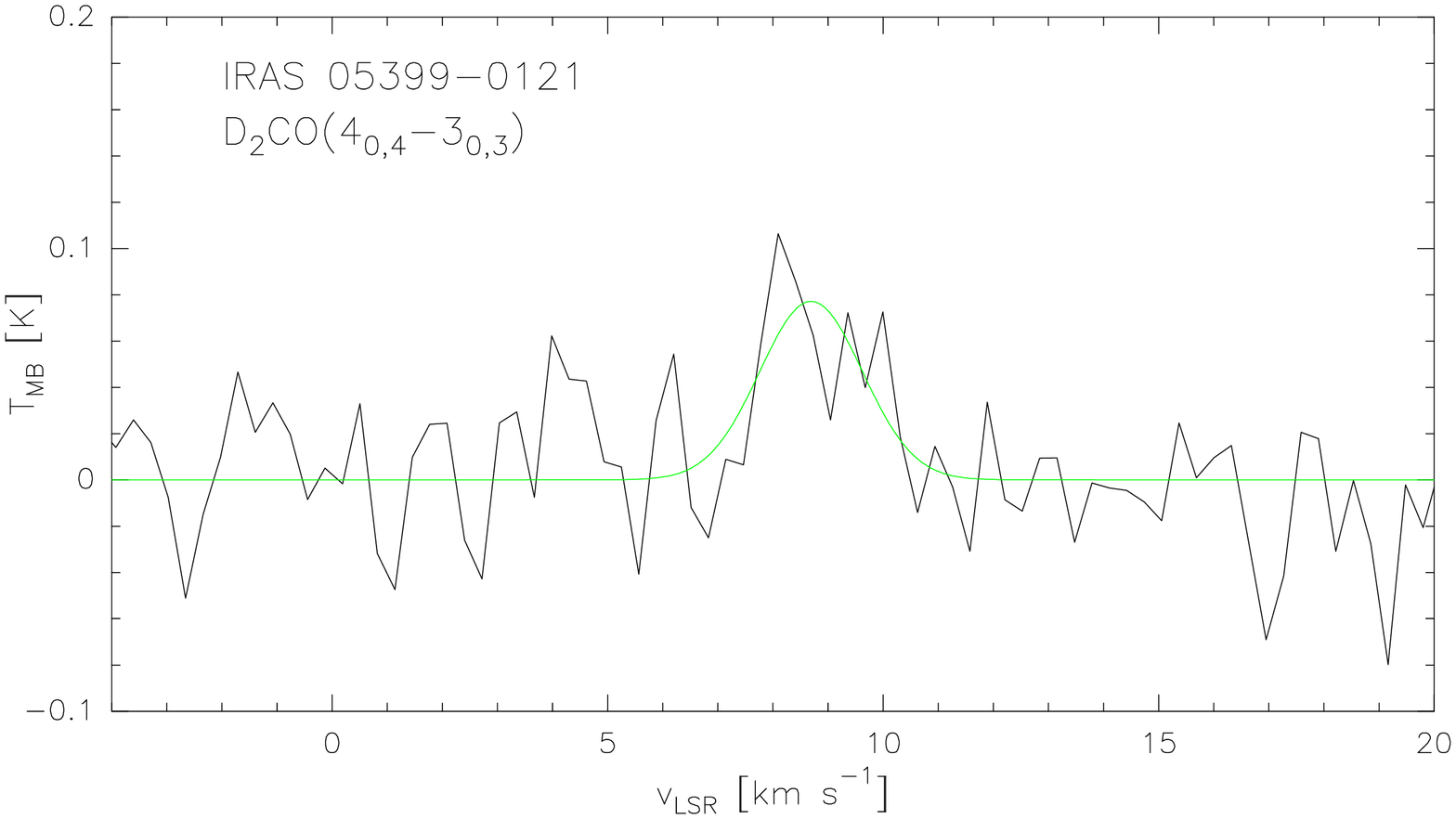}}
\resizebox{0.6\hsize}{!}{\includegraphics[angle=-90]{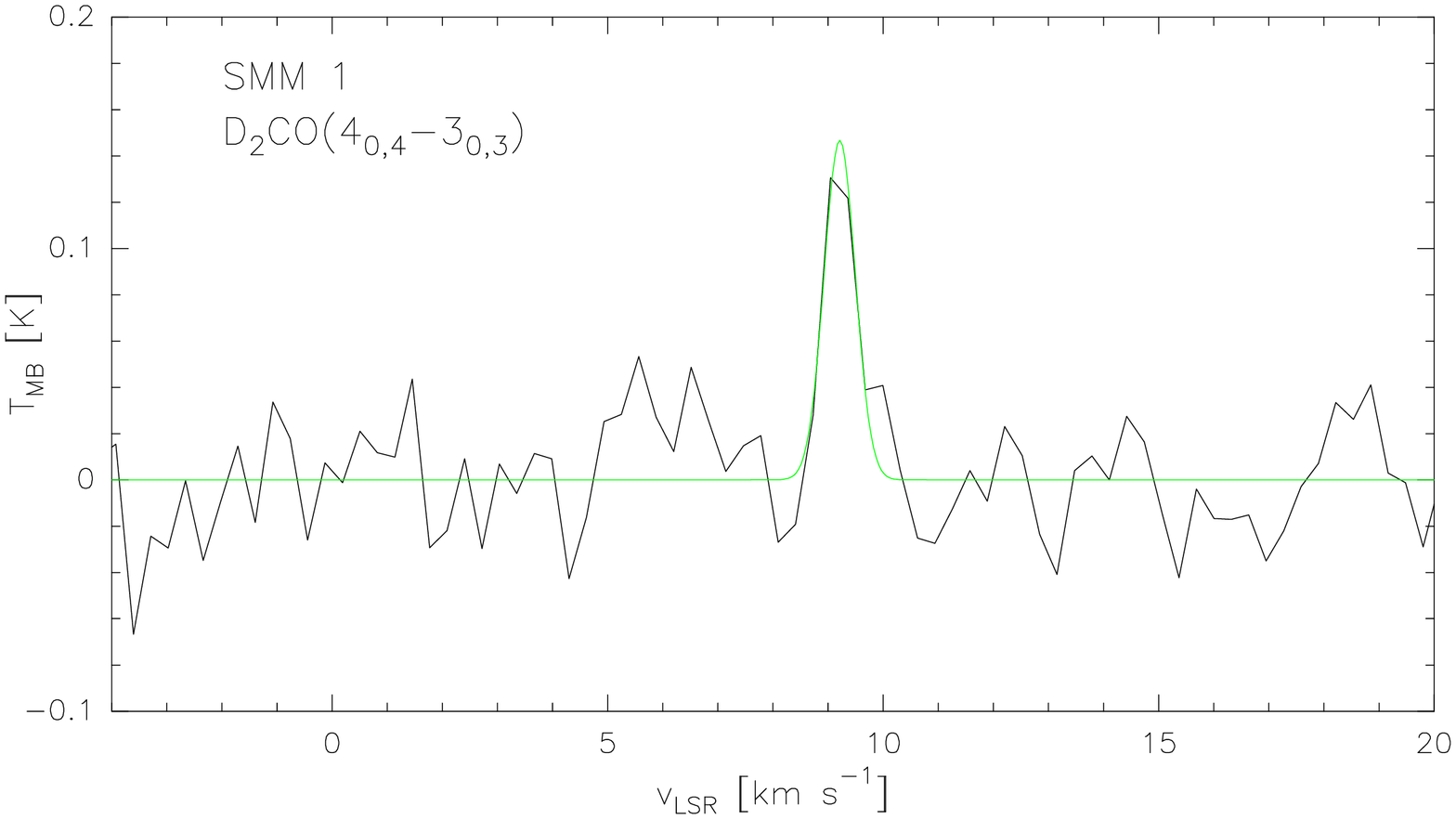}}
\caption{\textit{ortho}-D$_2$CO$(4_{0,4}-3_{0,3})$ spectra towards IRAS05399 
and SMM 1.}
\label{figure:D2CO}
\end{figure}

\begin{figure*}
\begin{center}
\includegraphics[width=2.5cm,angle=-90]{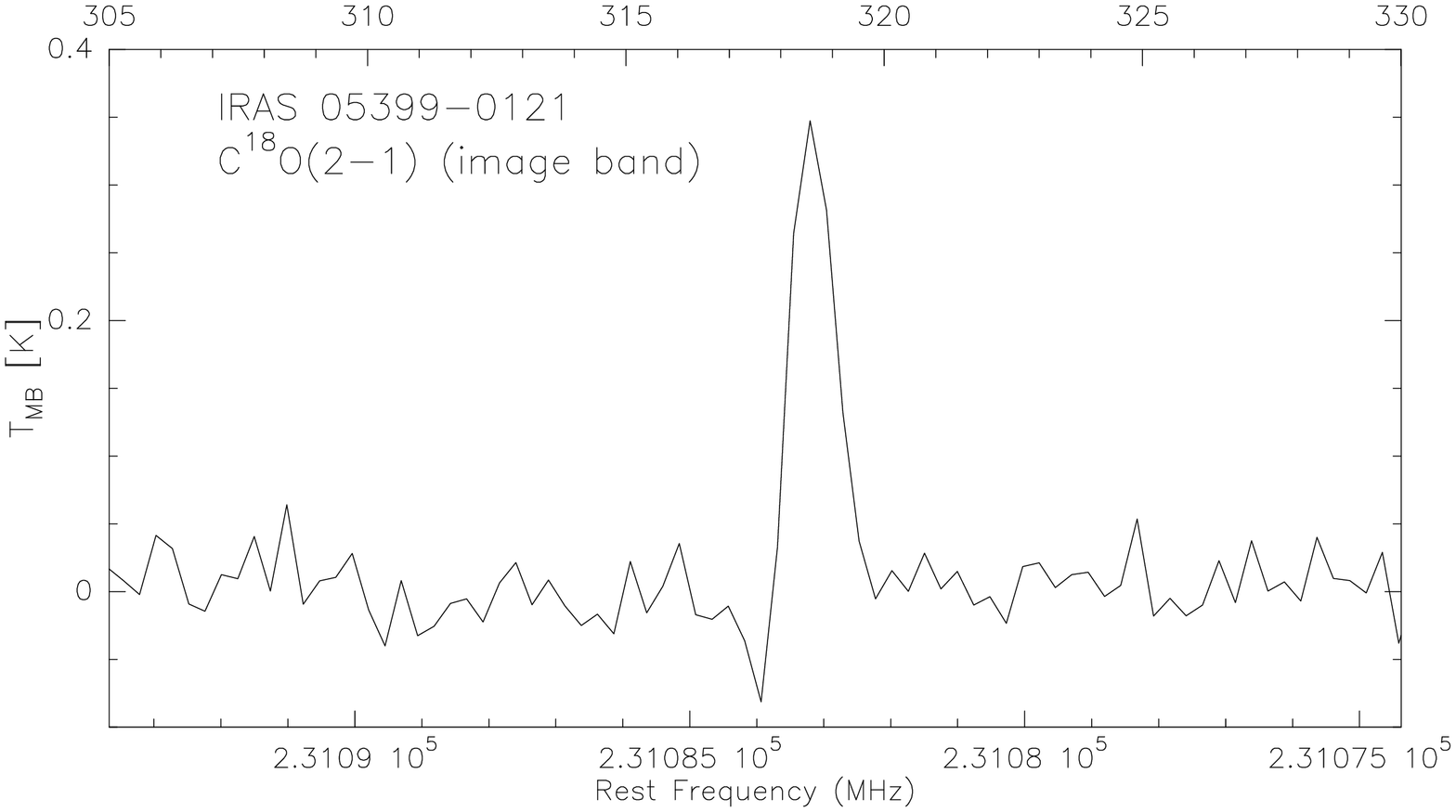}
\includegraphics[width=2.5cm,angle=-90]{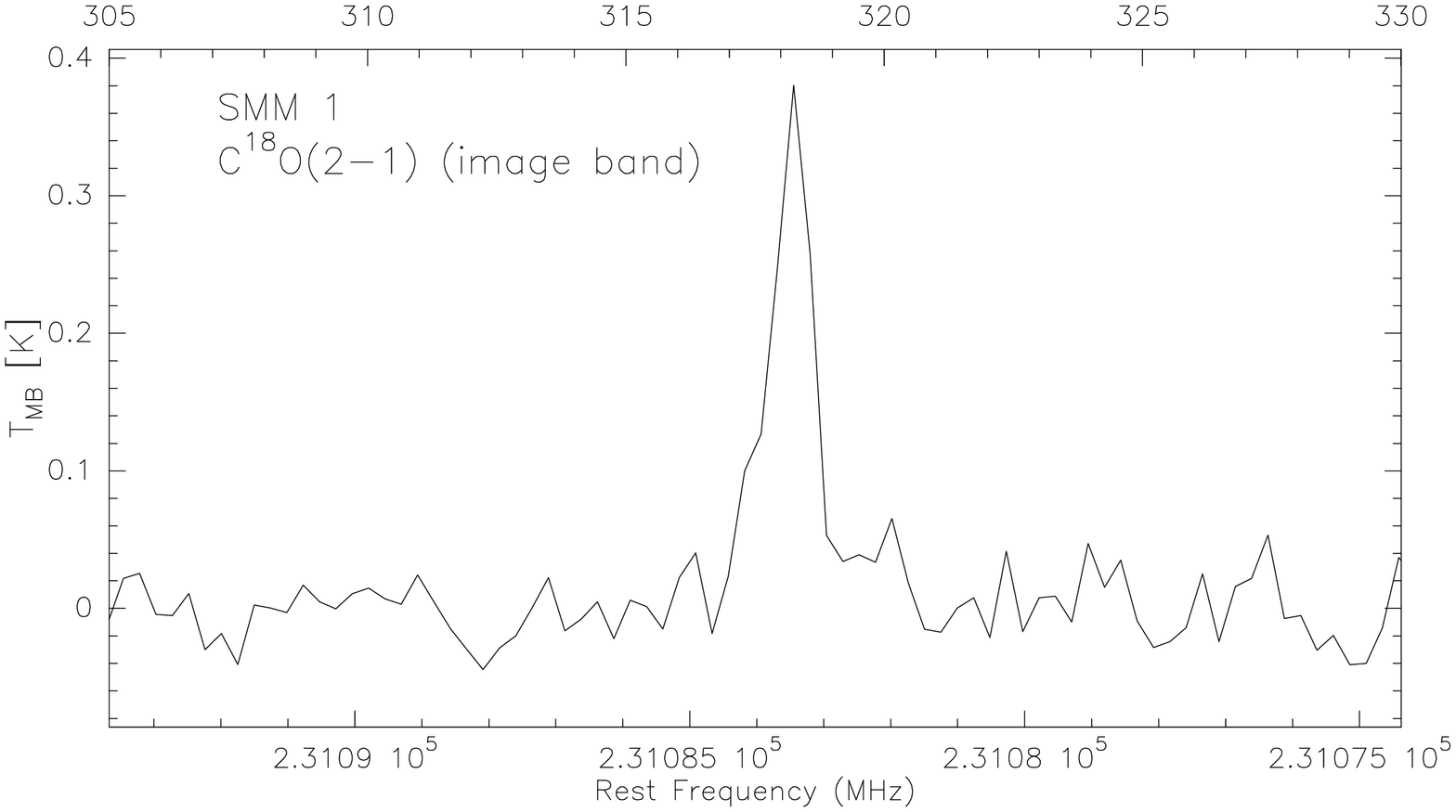}
\includegraphics[width=2.5cm,angle=-90]{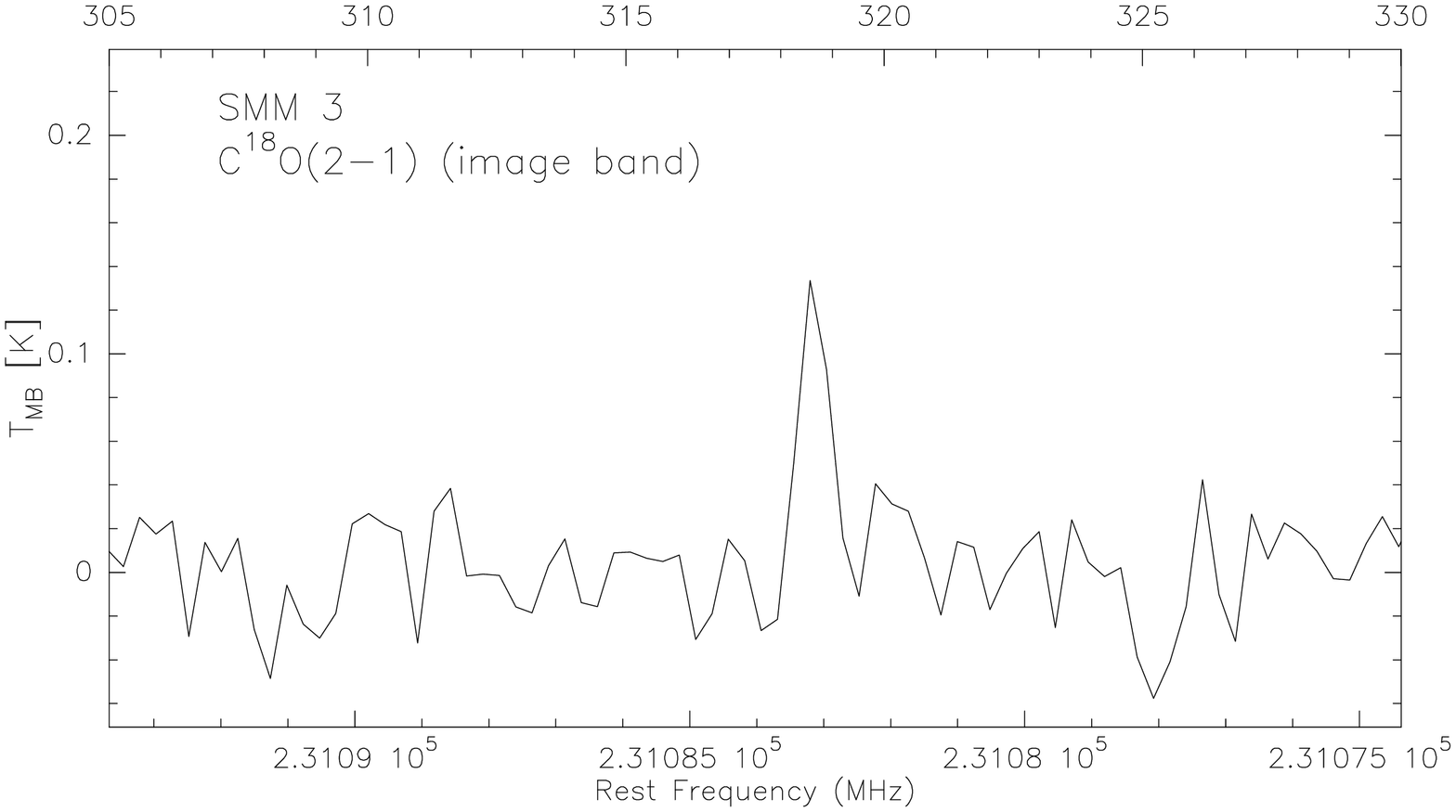}
\includegraphics[width=2.5cm,angle=-90]{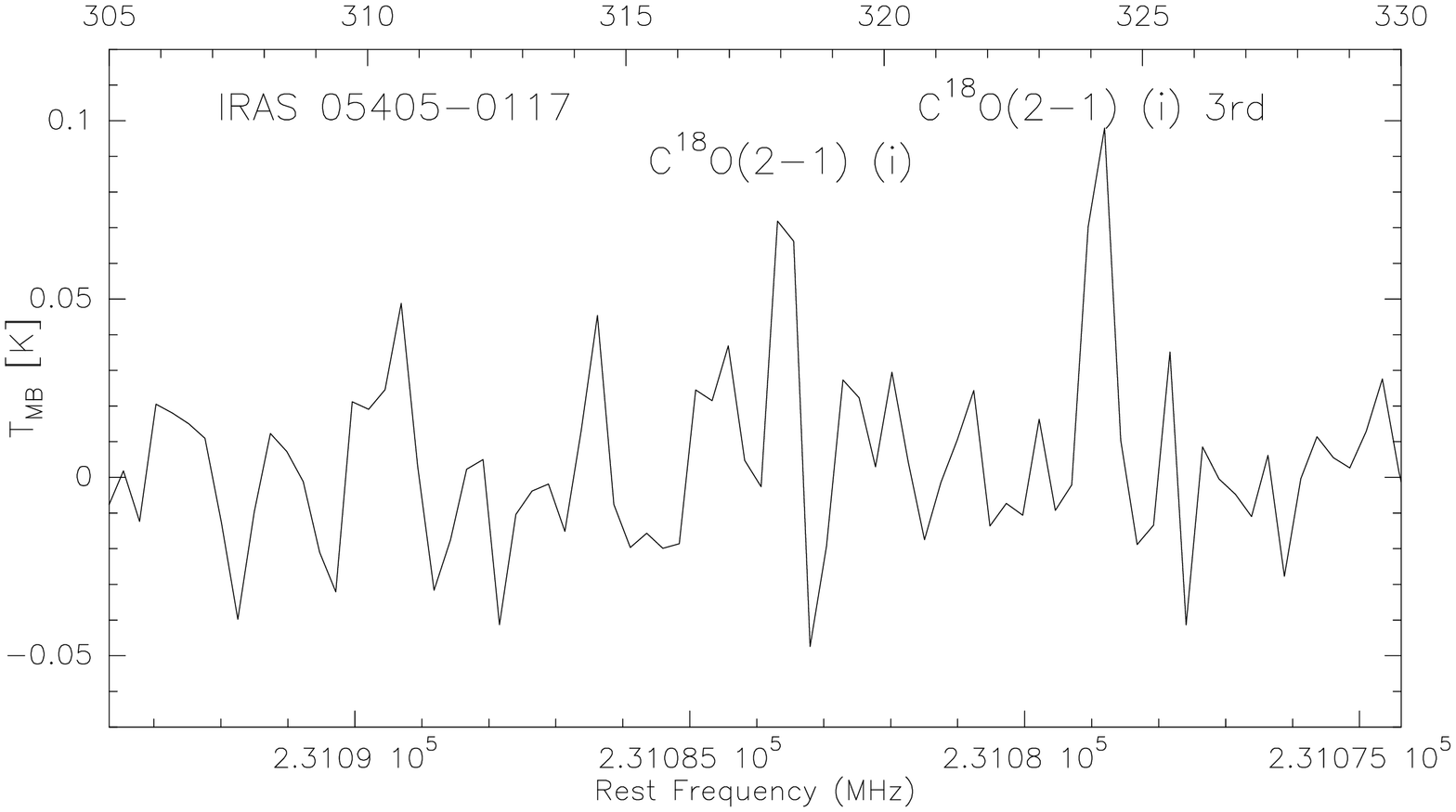}
\includegraphics[width=2.5cm,angle=-90]{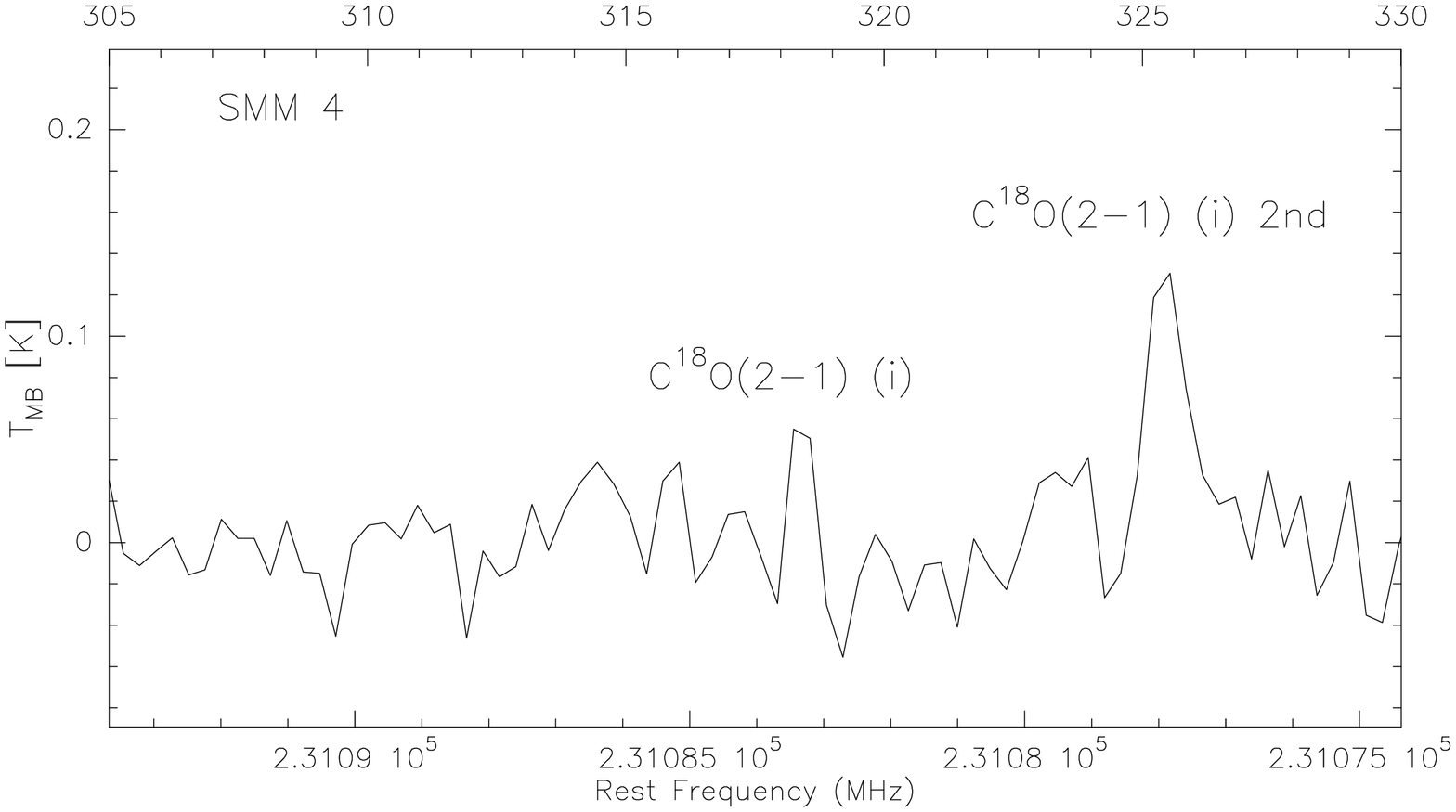}
\includegraphics[width=2.5cm,angle=-90]{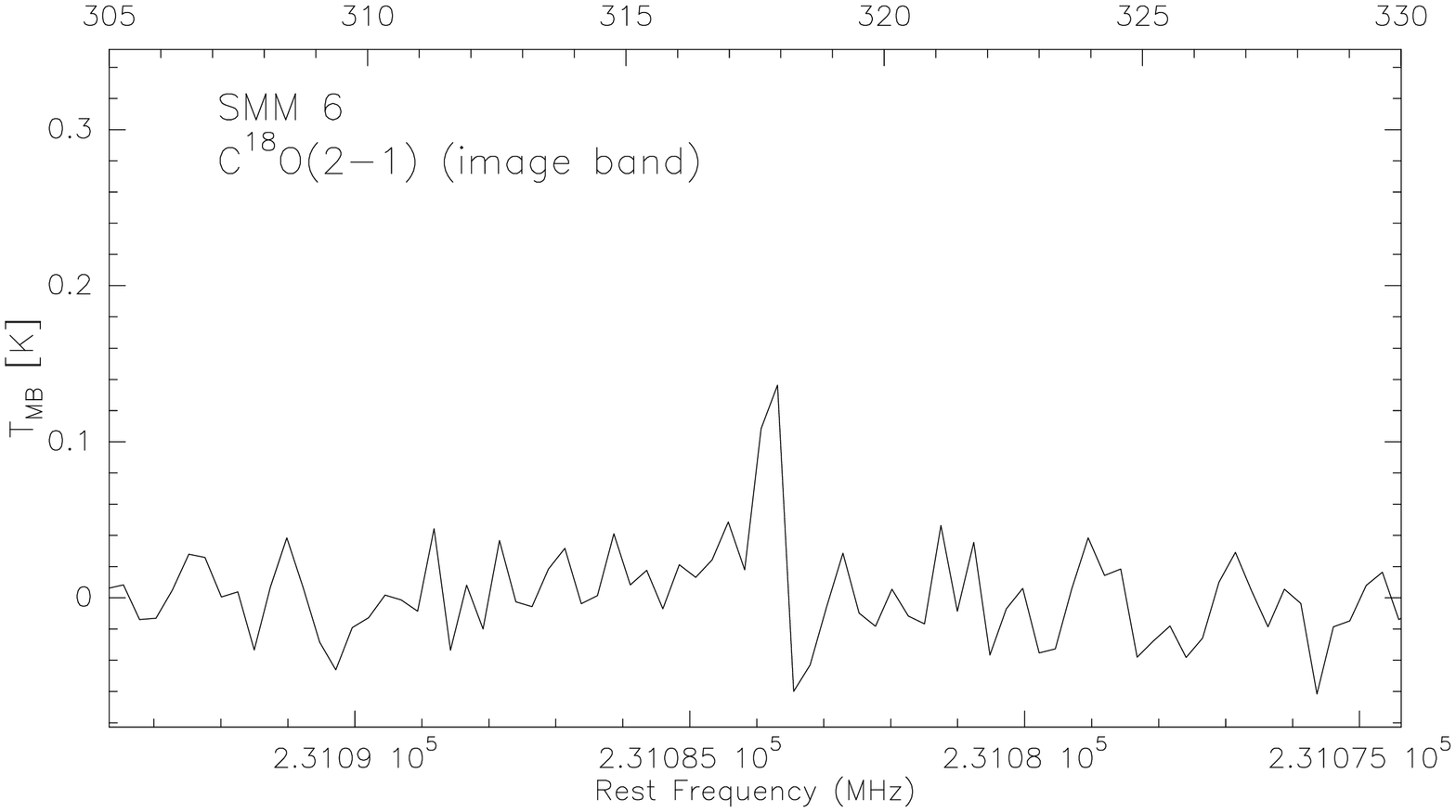}
\includegraphics[width=2.5cm,angle=-90]{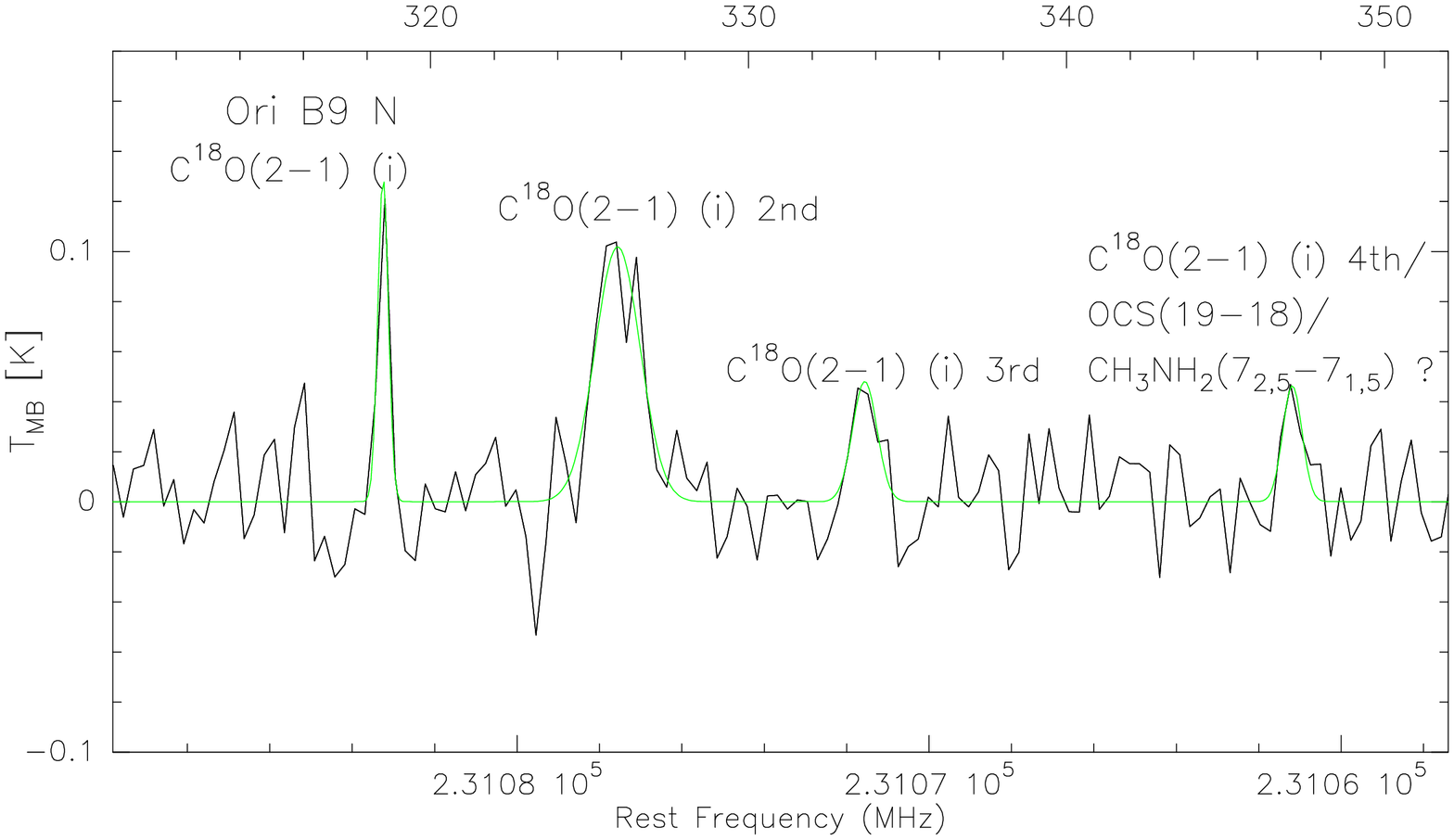}
\includegraphics[width=2.5cm,angle=-90]{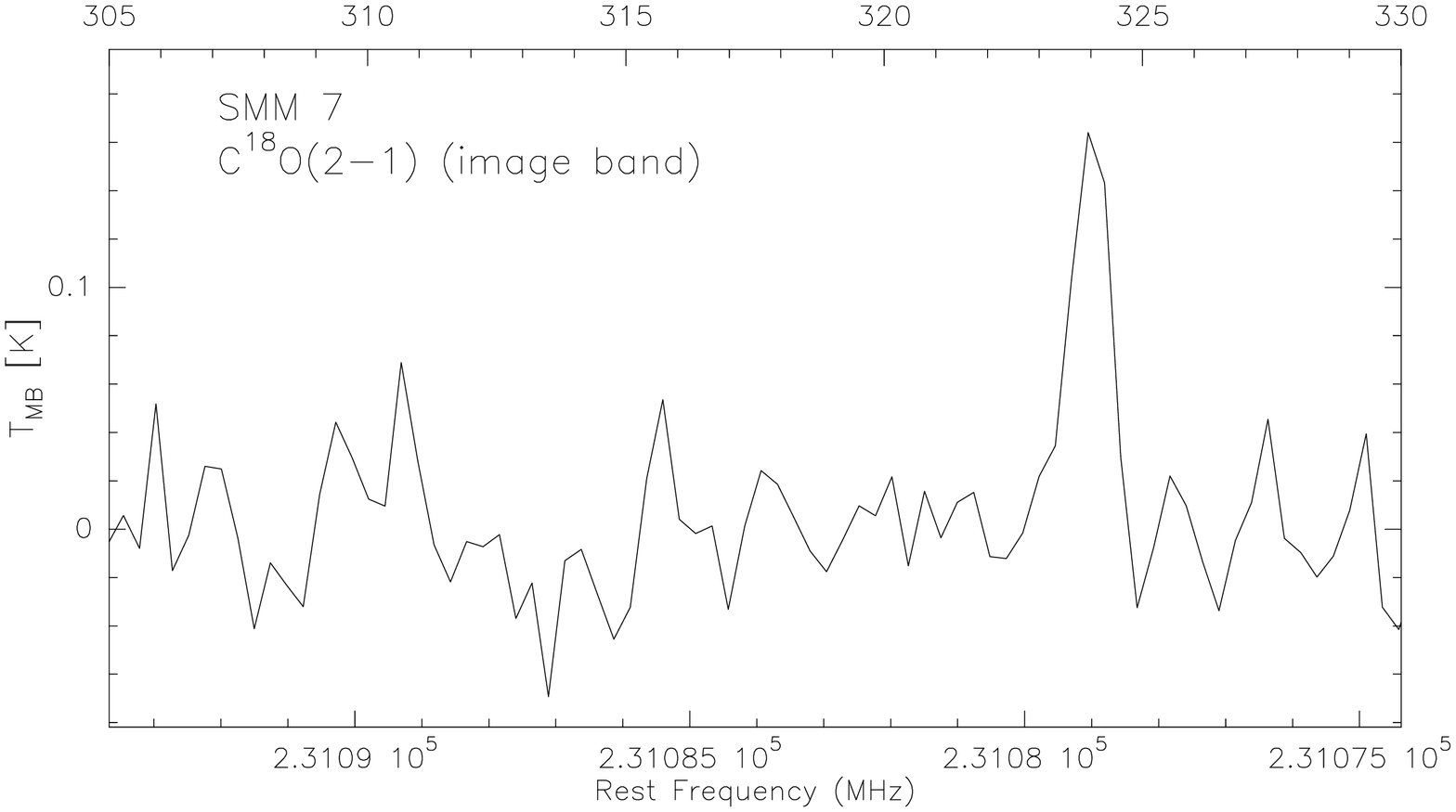}
\caption{C$^{18}$O$(2-1)$ lines at 219\,560.3568 MHz (JPL) arising from the 
image sideband. The x-axis range is different in the spectrum 
towards Ori B9 N, and the lines are overlaid with Gaussian fits for a better 
illustration. The blended OCS and CH$_3$NH$_2$ lines at $\sim231.06$ GHz 
marked on the spectrum towards Ori B9 N can be excluded on the basis of high 
upper-state energies and formation chemistry of the species (see text). The 
line is likely to be caused by an additional C$^{18}$O$(2-1)$ velocity 
component.}
\label{figure:C18O}
\end{center}
\end{figure*}

The $J_{K_a,K_c}=4_{0,4}-3_{0,3}$ transition of \textit{ortho}-D$_2$CO at 
$\sim231.4$ GHz was detected towards IRAS05399 and SMM 1 
in the LSB (see Fig.~\ref{figure:D2CO}). Also, the $J=2-1$ transition of 
C$^{18}$O at $\sim219.56$ GHz (LSB) was detected in the image sideband towards 
all sources except SMM 5 [marked with ``(i)''; Fig.~\ref{figure:C18O}]. 
An additional velocity component is detected towards IRAS05405 and SMM 4. 
Moreover, in the spectrum towards Ori B9 N several distinct velocity 
components are detected. The possible fourth velocity component of C$^{18}$O 
could, in principle, be blended with the line at $\sim231.06$ GHz, which can 
be assigned to either OCS$(J=19-18)$ (231\,060.9830 MHz, 
$E_{\rm u}/k_{\rm B}=110.9$ K) or CH$_3$NH$_2$-E$(J_{K_a,K_c}=7_{2,5}-7_{1,5})$ 
(231\,060.6041 MHz, $E_{\rm u}/k_{\rm B}=75.6$ K) because these two transitions 
are also blended. However, these can almost certainly be excluded because of 
the high transition energies involved, and because both species are expected 
to be formed by grain-surface chemistry. Detection of these species in a 
cold core is therefore very unlikely. Note that the lines 
``leaking'' from the rejected image band are heavily attenuated by the 
sideband filter. Therefore, we cannot establish the correct intensity scale 
for the C$^{18}$O$(2-1)$ lines.

In Table~\ref{table:otherparameters} we list the detected extra transitions, 
their rest frequencies, and upper-state energies. The rest frequencies and 
upper-state ener\-gies were assigned using JPL, CDMS, and 
Splatalogue\footnote{{\tt http://www.splatalogue.net/}} (\cite{remijan2007}) 
spectroscopic databases. The D$_2$CO detections are discussed further 
in Sect.~5.6.

\begin{table}
\caption{Other detected species/transitions.}
{\small
\begin{minipage}{1\columnwidth}
\centering
\renewcommand{\footnoterule}{}
\label{table:otherparameters}
\begin{tabular}{c c c}
\hline\hline 
Species/ & $\nu$ & $E_{\rm u}/k_{\rm B}$\\  
transition\tablefootmark{a}  
        & [MHz] & [K]\\
\hline
C$^{18}$O$(J=2-1)$\tablefootmark{b} & 219\,560.3600 (JPL) & 15.8\\
\textit{ortho}-D$_2$CO$(J_{K_a,K_c}=4_{0,4}-3_{0,3})$ & 231\,410.234 (CDMS) & 27.9\\
\hline 
\end{tabular} 
\tablefoot{
\tablefoottext{a}{For asymmetric top molecules such as D$_2$CO, $K_a$ and 
$K_c$ refer to the projection of the angular momentum along the $a$ and $c$ 
principal axes. The detected \textit{o}-D$_2$CO transition exbits $a$-type 
selection rules ($\Delta K_a=0,\pm2,\ldots$ and 
$\Delta K_c=\pm1,\pm3,\ldots$) (\cite{gordy1984}).} \tablefoottext{b}{Seen in 
the image band.}}
\end{minipage} }
\end{table}

\subsection{Spectral-line parameters}

The spectral line parameters are given in Table~\ref{table:lineparameters}. 
In this table we give the radial velocity (${\rm v}_{\rm LSR}$), FWHM 
linewidth ($\Delta {\rm v}$), peak intensity ($T_{\rm MB}$), integrated 
intensity ($\int T_{\rm MB}{\rm dv}$), peak optical thickness of the line
($\tau_0$), and excitation temperature ($T_{\rm ex}$). For non-detections, 
the $3\sigma$ upper limit on the line intensity is given. The va\-lues of 
${\rm v}_{\rm LSR}$ and $\Delta {\rm v}$ for C$^{17}$O, DCO$^+$, N$_2$H$^+$, 
and N$_2$D$^+$ were derived through fitting the hf structure. 
For the other lines these parameters were derived by fitting a single 
Gaussian to the line profile. Also, the values of $T_{\rm MB}$ and 
$\int T_{\rm MB}{\rm dv}$ were determined from Gaussian fits. 
The integrated C$^{17}$O$(2-1)$ line 
intensities of IRAS05399, SMM 1, 3, and 7, and Ori B9 N were calculated over 
the velocity range given in Col.~(6) of Table~\ref{table:lineparameters} 
because the hf structure of the line is partially resolved in these 
cases. The uncertainties reported in ${\rm v}_{\rm LSR}$ and  
$\Delta {\rm v}$ represent the formal $1\sigma$ errors determined by the 
fitting routine, whereas those in $T_{\rm MB}$ and $\int T_{\rm MB}{\rm dv}$ also 
include the 10\% calibration uncertainty. We used 
{\tt RADEX}\footnote{{\tt http://www.strw.leidenuniv.nl/$\sim$moldata/radex.html}} (\cite{vandertak2007}) to determine the va\-lues of $\tau_0$ and 
$T_{\rm ex}$ for the lines of C$^{17}$O, H$^{13}$CO$^+$, DCO$^+$, and 
N$_2$H$^+$. {\tt RADEX} modelling is described in more detail in Sect.~4.4 
where we discuss the determination of molecular column densities. 
For the rest of the lines, $\tau_0$ was determined through Weeds modelling 
(Sect.~4.4), and the associated error was estimated from the 10\% 
calibration uncertainty. The lines appear to be optically thin in most cases, 
except that the N$_2$H$^+$ lines are mostly optically thick ($\tau_0 >1$). 
The $T_{\rm ex}$ values for C$^{17}$O$(2-1)$ are close to $T_{\rm kin}$, 
indicating that the lines are nearly thermalised. 
The $T_{\rm ex}[\rm H^{13}CO^+(4-3)]$ va\-lues are found to be similar to 
$T_{\rm ex}[\rm DCO^+(4-3)]$. We note that the $T_{\rm ex}$ values obtained for 
the $J=3-2$ transition of N$_2$H$^+$ (3.8--5.0 K) are mostly comparable to 
the va\-lues $T_{\rm ex}=4.6-5.5$ K determined or assumed in Paper II. 
For N$_2$D$^+(3-2)$ we adopt as $T_{\rm ex}$ the value derived for 
N$_2$H$^+(3-2)$.

\begin{table*}
\caption{Spectral-line parameters.}
{\scriptsize
\begin{minipage}{2\columnwidth}
\centering
\renewcommand{\footnoterule}{}
\label{table:lineparameters}
\begin{tabular}{c c c c c c c c}
\hline\hline 
Source & Transition & ${\rm v}_{\rm LSR}$ & $\Delta {\rm v}$ & $T_{\rm MB}$ & $\int T_{\rm MB} {\rm dv}$\tablefootmark{a} & $\tau_0$\tablefootmark{b, c} & $T_{\rm ex}$\tablefootmark{c}\\
     & & [km~s$^{-1}$] & [km~s$^{-1}$] & [K] & [K~km~s$^{-1}$] & & [K]\\
\hline
IRAS 05399-0121 & C$^{17}$O$(2-1)$ & $8.68\pm0.02$ & $0.74\pm0.06$ & $0.79\pm0.18$ & $1.27\pm0.14$ [6.40, 9.64] & $0.10\pm0.05$ & $12.8\pm1.5$\\
                    & H$^{13}$CO$^+(4-3)$ & $8.62\pm0.08$ & $0.82\pm0.16$ & $0.09\pm0.02$ & $0.08\pm0.02$ & $0.11\pm0.05$ & $5.5\pm0.5$\\
		    & DCO$^+(4-3)$ & $8.58\pm0.03$ & $0.69\pm0.06$ & $0.33\pm0.04$ & $0.25\pm0.03$ & $0.51\pm0.10$ & $4.9\pm0.3$\\
                   & N$_2$H$^+(3-2)$ & $8.73\pm0.02$ & $0.71\pm0.07$ & $0.88\pm0.11$ & $1.00\pm0.11$ [92.6\%] & $2.80\pm0.20$ & $5.0\pm0.2$\\
                    & N$_2$D$^+(3-2)$ & $8.46\pm0.04$ & $0.53\pm0.02$ & $0.28\pm0.03$ & $0.14\pm0.02$ [91.6\%] & $0.68\pm0.10$ & $5.0\pm0.2$\tablefootmark{d}\\
               & \textit{o}-D$_2$CO$(4_{0,4}-3_{0,3})$ & $8.70\pm0.22$ & $2.20\pm0.46$ & $0.08\pm0.03$ & $0.18\pm0.04$ & $0.01\pm0.001$ & 18.6 \\
SMM 1 & C$^{17}$O$(2-1)$ & $9.20\pm0.01$ & $0.93\pm0.03$ & $0.96\pm0.17$ & $1.85\pm0.20$ [7.00, 11.33]& $0.15\pm0.05$ & $11.3\pm1.0$\\
      & H$^{13}$CO$^+(4-3)$ & \ldots & \ldots & $<0.13$ & \ldots & \ldots & \ldots\\
      & DCO$^+(4-3)$ & $9.31\pm0.01$ & $0.47\pm0.05$ & $0.55\pm0.07$ & $0.42\pm0.05$ & $1.89\pm0.20$ & $4.5\pm0.3$\\
                  & N$_2$H$^+(3-2)$ & $9.31\pm0.02$ & $0.66\pm0.09$ & $0.65\pm0.07$ & $0.70\pm0.08$ [92.6\%] & $2.43\pm0.10$ & $4.6\pm0.3$\\
                  & N$_2$D$^+(3-2)$ & $9.20\pm0.03$ & $0.75\pm0.07$ & $0.38\pm0.04$ & $0.40\pm0.04$ [92.6\%] & $1.71\pm0.20$ & $4.6\pm0.3$\tablefootmark{d}\\
                  & \textit{o}-D$_2$CO$(4_{0,4}-3_{0,3})$ & $9.21\pm0.06$ & $0.67\pm0.16$ & $0.15\pm0.02$ & $0.10\pm0.02$ & $0.02\pm0.002$ & 18.6 \\
SMM 3 & C$^{17}$O$(2-1)$ & $8.68\pm0.06$ & $0.58\pm0.09$ & $0.35\pm0.05$ & $0.51\pm0.09$ [6.56, 9.52] & $0.05\pm0.03$ & $11.0\pm1.0$\\
               & DCO$^+(4-3)$ & $8.54\pm0.03$ & $0.42\pm0.16$ & $0.20\pm0.02$ & $0.09\pm0.01$ & $0.24\pm0.03$ & $5.1\pm0.3$\\
               & N$_2$H$^+(3-2)$ & $8.57\pm0.03$ & $0.85\pm0.09$ & $0.62\pm0.09$ & $0.67\pm0.08$ [92.6\%] & $1.23\pm0.10$ & $4.9\pm0.3$\\   
                & N$_2$D$^+(3-2)$ & $8.39\pm0.04$ & $0.53\pm0.11$ & $0.21\pm0.03$ & $0.17\pm0.03$ [91.6\%] & $0.59\pm0.10$ & $4.9\pm0.3$\tablefootmark{d}\\
IRAS 05405-0117 & C$^{17}$O$(2-1)$ & $9.20\pm0.04$ & $0.54\pm0.04$ & $0.30\pm0.03$ & $0.17\pm0.03$ [78.7\%] & $0.05\pm0.03$ & $10.5\pm0.5$\\
2nd v-comp. & C$^{17}$O$(2-1)$  & $1.30\pm0.33$ & $0.58\pm1.08$ & $0.17\pm0.02$ & $0.14\pm0.02$ [78.7\%] & \ldots & \ldots \\
3rd v-comp. & C$^{17}$O$(2-1)$ & $2.98\pm0.33$ & $0.54\pm1.08$ & $0.20\pm0.02$ & $0.13\pm0.02$ [78.7\%] & \ldots & \ldots \\           
            & DCO$^+(4-3)$ & \ldots & \ldots & $<0.08$ & \ldots & \ldots & \ldots\\      
            & N$_2$H$^+(3-2)$ & $9.30\pm0.03$ & $0.69\pm0.06$ & $0.66\pm0.09$ & $0.56\pm0.07$ [90.9\%] & $4.26\pm0.10$ & $4.5\pm0.3$\\
                 & N$_2$D$^+(3-2)$ & $8.93\pm0.15$ & $0.53\pm0.34$ & $0.06\pm0.01$ & $0.04\pm0.03$ [91.5\%] & $0.16\pm0.02$ & $4.5\pm0.3$\tablefootmark{d}\\
3rd v-comp. & N$_2$D$^+(3-2)$ & $2.98\pm0.20$ & $0.53\pm0.56$ & $0.05\pm0.01$ & $0.05\pm0.02$ [92.4\%] & \ldots & \ldots \\
SMM 4 (2nd v-comp.) & C$^{17}$O$(2-1)$ & $1.67\pm0.05$ & $0.92\pm0.19$ & $0.32\pm0.06$ & $0.61\pm0.08$ & $0.06\pm0.03$ & $9.7\pm1.2$\\
2nd v-comp. & H$^{13}$CO$^+(4-3)$ & $1.53\pm0.08$ & $0.43\pm0.15$ & $0.15\pm0.02$ & $0.07\pm0.02$ & $1.02\pm0.60$ & $4.0\pm0.2$\\
2nd v-comp. & DCO$^+(4-3)$ & $1.65\pm0.03$ & $0.49\pm0.09$ & $0.21\pm0.04$ & $0.19\pm0.03$ & $0.76\pm0.50$ & $4.0\pm0.3$\\
2nd v-comp. & N$_2$H$^+(3-2)$ & $1.62\pm0.78$ & $0.61\pm21.10$ & $0.32\pm0.07$ & $0.27\pm0.07$ [92.6\%] & $1.59\pm0.50$ & $4.0\pm0.3$\\   
2nd v-comp. & N$_2$D$^+(3-2)$ & $1.49\pm0.04$ & $0.54\pm0.10$ & $0.24\pm0.03$ & $0.17\pm0.03$ [90.9\%] & $1.50\pm0.20$ & $4.0\pm0.3$\tablefootmark{d}\\
SMM 5 & C$^{17}$O$(2-1)$ & $9.37\pm0.03$ & $0.54\pm0.01$ & $0.22\pm0.02$ & $0.14\pm0.02$ [78.7\%] & $0.04\pm0.03$ & $10.1\pm0.5$\\
   & DCO$^+(4-3)$ & \ldots & \ldots & $<0.07$ & \ldots & \ldots & \ldots\\      
& N$_2$H$^+(3-2)$ & $9.43\pm0.07$ & $0.44\pm0.17$ & $0.16\pm0.03$ & $0.14\pm0.02$ [92.6\%] & $0.67\pm0.04$ & $3.8\pm0.1$\\
& N$_2$D$^+(3-2)$ & \ldots & \ldots & $<0.11$ & \ldots & \ldots & \ldots\\
SMM 6 & C$^{17}$O$(2-1)$ & $9.51\pm0.04$ & $0.54\pm0.04$ & $0.28\pm0.03$ & $0.15\pm0.03$ [69.4\%] & $0.05\pm0.03$ & $9.7\pm0.5$\\
      & DCO$^+(4-3)$ & $9.47\pm0.02$ & $0.42\pm0.04$ & $0.37\pm0.05$ & $0.19\pm0.02$ & $3.47\pm0.20$ & $4.0\pm0.3$\\
      & N$_2$H$^+(3-2)$ & $9.52\pm0.03$ & $0.65\pm0.09$ & $0.51\pm0.06$ & $0.44\pm0.06$ [92.6\%] & $5.75\pm0.35$ & $4.2\pm0.2$\\
      & N$_2$D$^+(3-2)$ & $9.25\pm0.03$ & $0.56\pm0.10$ & $0.33\pm0.03$ & $0.26\pm0.03$ [90.9\%] & $2.92\pm0.30$ & $4.2\pm0.2$\tablefootmark{d}\\
Ori B9 N & C$^{17}$O$(2-1)$ & $9.29\pm0.33$ & $0.54\pm1.08$ & $0.30\pm0.05$ & $0.20\pm0.03$ [96.0\%] & $0.04\pm0.03$ & $11.6\pm1.0$\\
2nd v-comp. & C$^{17}$O$(2-1)$ & $1.85\pm0.33$ & $1.39\pm1.08$ & $0.29\pm0.05$ & $0.70\pm0.08$ & $0.04\pm0.03$ & $11.7\pm2.0$\\     
            & DCO$^+(4-3)$ & $9.42\pm0.19$ & $0.47\pm1.38$ & $0.04\pm0.01$ & $0.02\pm0.02$ & $0.05\pm0.02$ & $4.9\pm0.3$\\
2nd v-comp. & DCO$^+(4-3)$ & $2.11\pm0.11$ & $0.78\pm0.31$ & $0.09\pm0.01$ & $0.09\pm0.02$ & $0.14\pm0.08$ & $4.6\pm0.5$\\
           & N$_2$H$^+(3-2)$ & $8.80\pm0.16$ & $0.44\pm0.24$ & $0.05\pm0.02$ & $0.05\pm0.02$ [92.6\%] & $0.12\pm0.03$ & $4.1\pm0.2$\\
2nd v-comp. & N$_2$H$^+(3-2)$ & $1.81\pm0.11$ & $0.85\pm0.11$ & $0.09\pm0.02$ & $0.13\pm0.03$ [92.6\%] & $0.23\pm0.07$ & $4.1\pm0.2$\\
            & N$_2$D$^+(3-2)$ & \ldots & \ldots & $<0.07$ & \ldots & \ldots & \ldots\\
SMM 7  & C$^{17}$O$(2-1)$ & $3.75\pm0.02$ & $1.04\pm0.05$ & $0.66\pm0.09$ & $0.82\pm0.09$ [1.38, 4.95] & $0.15\pm0.05$ & $8.9\pm1.0$\\
``9 km~s$^{-1}$''-comp. & C$^{17}$O$(2-1)$ & $9.04\pm0.07$ & $0.55\pm1.74$ & $0.10\pm0.06$ & $0.07\pm0.03$ [69.4\%] &  \ldots & \ldots \\ 
        & DCO$^+(4-3)$ & $3.70\pm0.04$ & $0.42\pm0.10$ & $0.22\pm0.02$ & $0.10\pm0.02$ & $0.94\pm0.20$ & $3.9\pm0.3$\\
        & N$_2$H$^+(3-2)$ & $4.01\pm0.14$ & $0.67\pm0.27$ & $0.14\pm0.02$ & $0.17\pm0.04$ [92.6\%] & $0.49\pm0.07$ & $3.9\pm0.2$\\
     & N$_2$D$^+(3-2)$ & \ldots & \ldots & $<0.07$ & \ldots & \ldots & \ldots\\
\hline 
\end{tabular} 
\tablefoot{
\tablefoottext{a}{Integrated intensity is derived from a Gaussian fit or, in 
the case of some C$^{17}$O lines, by integrating over the velocity range 
indicated in brackets. The percentage in brackets indicates the contribution 
of hf component's intensity lying within the Gaussian fit.}\tablefoottext{b}{For C$^{17}$O, H$^{13}$CO$^+$, DCO$^+$, N$_2$H$^+$, and N$_2$D$^+$ $\tau_0$ is the optical thickness in the centre of a hypothetical unsplit line (see text).}\tablefoottext{c}{For C$^{17}$O, H$^{13}$CO$^+$, DCO$^+$, and N$_2$H$^+$ the 
values of $\tau_0$ and $T_{\rm ex}$ were estimated using {\tt RADEX} with the 
kinetic temperature ($T_{\rm kin}$) and the H$_2$ density 
($\langle n({\rm H_2}) \rangle$). The value of $\tau_0$ for N$_2$D$^+$ and 
\textit{o}-D$_2$CO was estimated using CLASS/Weeds; see Sect.~4.4 for details.}\tablefoottext{d}{$T_{\rm ex}$ is assumed to be the same as for N$_2$H$^+(3-2)$.}}
\end{minipage} }
\end{table*}

\section{Analysis and results}

\subsection{Core properties derived from 350 $\mu$m emission}

We used the 350-$\mu$m dust continuum data to determine the mass, peak 
beam-averaged column density of H$_2$, and volume-averaged H$_2$ number 
density of the cores and their condensations. The formulas for the mass and 
H$_2$ column density can be found in Paper I [Eqs.~(2) and (3) therein]. 
Note that in Eq.~(3) of Paper I, the peak surface brightness is 
$I_{\lambda}^{\rm dust}=S_{\lambda}^{\rm peak}/\Omega_{\rm beam}$, 
where $\Omega_{\rm beam}$ is the solid angle of the telescope beam.
As the dust temperature of the sources, we used the gas kinetic temperatures 
listed in Col.~(4) of Table~\ref{table:sources}, and assumed that 
$T_{\rm dust}=T_{\rm kin}$. For the subcondensations, such as SMM 6b, 6c, 
etc., it was assumed that $T_{\rm dust}$ is the same as in the ``main'' core. 
This assumption may not be valid, however, because in most cases the 
subcondensations lie outside the $40\arcsec$ beam of the NH$_3$ measurements 
used to derive $T_{\rm kin}$ in the main cores. Moreover, the temperature of 
an individual small condensation may be lower than the tempe\-rature 
of the parent core because of more effective shielding from the 
external radiation field. If the temperature is lower than assumed the mass 
and column density will be underestimated. As the dust opa\-city per unit dust 
mass at 350 $\mu$m, $\kappa_{\rm 350\,\mu m}$, we used the value 
1.0 m$^2$~kg$^{-1}$ which was extrapolated from the Ossenkopf \& Henning 
(1994, hereafter OH94) model describing graphite-silicate dust grains that 
have coagulated and accreted thick ice mantles over a period of $10^5$ yr at 
a gas density of 
$n_{\rm H}=n({\rm H})+2n({\rm H_2})\simeq 2n({\rm H_2})=10^5$ cm$^{-3}$. The 
same dust model was adopted in Paper I ($\kappa_{\rm 870\,\mu m}\simeq0.17$ 
m$^2$~kg$^{-1}$), and is expected to be a rea\-sonable model for cold, 
dense mole\-cular cloud cores\footnote{For comparison, for dust grains 
covered by thin ice mantles at a density $10^5$ cm$^{-3}$, 
$\kappa_{\rm 350\,\mu m}\simeq0.78$ m$^2$~kg$^{-1}$. 
The $\kappa_{\rm 350\,\mu m}$ value we have adopted is the same as in the 
OH94 model of grains with thin ice mantles at a density $10^6$ cm$^{-3}$. 
At $10^6$ cm$^{-3}$ with thick ice mantles $\kappa_{\rm 350\,\mu m}$ rises to 
about 1.1 m$^2$~kg$^{-1}$.}. We note that the sub-mm opacities recently 
calculated by Ormel et al. (2011) are comparable to the OH94 values (after 
$10^5$ yr of coagulation). However, the dust opacities 
are likely to be uncertain by a factor of $\gtrsim2$ (e.g., OH94; 
\cite{motte2001}; \cite{ormel2011}). For the ave\-rage dust-to-gas mass ratio, 
$R_{\rm d}\equiv \langle M_{\rm dust}/M_{\rm gas}\rangle$, we adopted the canonical 
value 1/100. Finally, we assumed a He/H abundance ratio of 0.1, which leads 
to the mean molecular weight per H$_2$ molecule of $\mu_{\rm H_2}=2.8$.

The integrated flux densities used to calculate the masses refer to core 
areas, $A$. Thus, in order to properly calculate the volume-average H$_2$ 
number density, $\langle n({\rm H_2}) \rangle$, we use the 
effective radius, $R_{\rm eff}=\sqrt{A/\pi}$, in the formula

\begin{equation}
\label{eq:density}
\langle n({\rm H_2}) \rangle=\frac{\langle \rho \rangle}{\mu_{{\rm H_2}}m_{\rm H}}\,,
\end{equation}
where $\langle \rho \rangle=M/\left(4\pi/3 \times R_{\rm eff}^3\right)$ is 
the mass density, and $m_{\rm H}$ is the mass of a hydrogen atom.

The results of the above calculations are presented in 
Table~\ref{table:properties}. The uncertainties in the derived parameters
were propagated from the uncertainties in $T_{\rm kin}$. We note that the 
$1\sigma$ rms noise on our SABOCA map, $\sim0.06$ Jy~beam$^{-1}$, corresponds 
to a $3\sigma$ mass detection limit of $\sim0.1$ M$_{\sun}$ assuming 
$T_{\rm dust}=10$ K. In terms of column density, the $3\sigma$ detection limit is 
about $2.8\times10^{21}$ cm$^{-2}$.

\begin{table}
\caption{The effective radius, mass, H$_2$ column density, and 
volume-averaged H$_2$ density derived from the 350 $\mu$m emission.}
{\tiny
\begin{minipage}{1\columnwidth}
\centering
\renewcommand{\footnoterule}{}
\label{table:properties}
\begin{tabular}{c c c c c}
\hline\hline 
Source & $R_{\rm eff} $& $M$ & $N({\rm H_2})$ & $\langle n({\rm H_2}) \rangle$ \\  
       & [pc] & [M$_{\sun}$] & [$10^{22}$ cm$^{-2}$] & [$10^5$ cm$^{-3}$]\\ 
\hline
SMM 3 & 0.03 & $2.1\pm0.8$ & $10.4\pm2.7$ & $4.0\pm1.5$\\
SMM 3b & \ldots & \ldots & $0.7\pm0.2$\tablefootmark{a} & \ldots \\
SMM 3c & \ldots & \ldots & $0.7\pm0.2$\tablefootmark{a} & \ldots \\
IRAS 05405-0117 & 0.02 & $0.2\pm0.1$ & $1.3\pm0.3$ & $2.0\pm1.0$\\
SMM 4 & 0.01 & $0.04\pm0.02$ & $0.4\pm0.1$ & $0.9\pm0.5$\\
SMM 4b & 0.02 & $0.1\pm0.04$\tablefootmark{a} & $0.7\pm0.1$\tablefootmark{a} & $1.0\pm0.4$\\
SMM 5 & \ldots & \ldots & $0.6\pm0.1$ & \ldots \\
SMM 6 & 0.02 & $0.2\pm0.1$ & $1.0\pm0.1$ & $1.6\pm0.8$\\
SMM 6b & 0.02 & $0.1\pm0.05$\tablefootmark{a} & $0.9\pm0.1$\tablefootmark{a} & $1.3\pm0.6$\\
SMM 6c & 0.01\tablefootmark{b} & $0.1\pm0.04$\tablefootmark{a,b} & $0.7\pm0.1$\tablefootmark{a} & $1.9\pm0.7$\tablefootmark{b}\\
SMM 6d & \ldots & \ldots & $0.7\pm0.1$\tablefootmark{a} & \ldots\\
Ori B9 N & \ldots & \ldots & $0.3\pm0.1$ & \ldots\\
SMM 7 & 0.02 & $0.6\pm0.3$ & $1.9\pm1.0$ & $2.6\pm1.3$\\
SMM 7b & \ldots & \ldots & $1.3\pm0.7$\tablefootmark{a} & \ldots\\
\hline 
\end{tabular} 
\tablefoot{
\tablefoottext{a}{Calculated by assuming $T_{\rm dust}$ is the same as for the ``main'' core.}\tablefoottext{b}{These values include contributions from both SMM 6b and 6c.}}
\end{minipage}  }
\end{table}

\subsection{Dust properties determined from 350 and 870 $\mu$m data}

We can use our observations at two submm wavelengths, 350 and 870 $\mu$m, to 
estimate the dust colour temperature, $T_{\rm dust}$, and dust 
emissivity spectral index, $\beta$ 
[$\kappa_{\lambda}=\kappa_0(\lambda_0/\lambda)^{\beta}$]. Note that according to 
the Wien displacement law, the wavelength of the peak of the blackbody 
radiation curve is about 290 $\mu$m at 10 K. Therefore, the Rayleigh-Jeans 
(R-J) approximation, $h\nu \ll k_{\rm B}T$, is not valid for our 350 and 870 
$\mu$m data. 

In the following analysis, it is assumed that $T_{\rm dust}$ and 
$\beta$ are constant across the source. We first smoothed the SABOCA 
map to the resolution of our LABOCA data, and then calculated the flux 
densities at both wavelengths in a fixed $40\arcsec$ diameter aperture 
[Col.~(7) of Table~\ref{table:cores}]. The resulting flux density ratios, 
$S_{350}^{40\arcsec}/S_{870}^{40\arcsec}$, are given in Col.~(2) of 
Table~\ref{table:colour}. A dust colour temperature can be determined 
by fixing the value of $\beta$ [see, e.g., Eq.~(3) of Shetty et al. (2009)]. 
The value of $\beta$ in the OH94 thick-ice dust model we adopted earlier is 
about 1.9 over the wavelength range $\lambda \in [250,\, 1300\, \mu{\rm m}]$ 
(see also \cite{shirley2005}). By adop\-ting the value $\beta=1.9$, we derive 
the $T_{\rm dust}$ values in the range $\sim7.9-10.8$ K [see Col.~(3) of 
Table~\ref{table:colour}].

To calculate $\beta$ from the ratio of two flux 
densities at different wavelengths, an estimate for the dust temperature is 
needed. We assumed that $T_{\rm dust}=T_{\rm kin}$, and 
applied Eq.~(5) of Shetty et al. (2009). The resulting values, 
$\beta\simeq0.5-1.8$, are listed in Col.~(4) of 
Table~\ref{table:colour}. We note that $T_{\rm kin}$ measurements were 
obtained with a $40\arcsec$ resolution, whereas the above flux density 
ratio was determined at about $20\arcsec$ resolution. %However, angular 
%resolution should not significantly affect the flux density \textit{ratio}. 
However, flux densities were measured in a $40\arcsec$ aperture, which 
matches the resolution of our NH$_3$ data. 

\begin{table}
\caption{Dust colour temperature and emissivity index derived from 
350-to-870 $\mu$m flux density ratio.}
\begin{minipage}{1\columnwidth}
\centering
\renewcommand{\footnoterule}{}
\label{table:colour}
\begin{tabular}{c c c c}
\hline\hline 
Source & $S_{350}^{40\arcsec}/S_{870}^{40\arcsec}$ & $T_{\rm dust}$\tablefootmark{a} [K] & $\beta$\tablefootmark{b}\\  
\hline
SMM 3 & $2.4\pm1.1$ & $10.8^{+5.7}_{-2.6}$ & $1.8\pm0.6$\\
IRAS 05405-0117 & $1.5\pm0.6$ & $8.7^{+2.6}_{-1.7}$ & $1.1\pm0.6$ \\
SMM 4 & $1.4\pm0.8$ & $8.4^{+4.5}_{-2.0}$ & $0.5\pm0.8$ \\
SMM 5 & $1.7\pm0.9$ & $9.0^{+4.3}_{-2.2}$ & $1.2\pm0.8$ \\
SMM 6 & $1.2\pm0.6$ & $7.9^{+3.0}_{-1.8}$ & $0.8\pm0.9$ \\
Ori B9 N & $1.7\pm0.9$ & $9.0^{+4.3}_{-2.2}$ & $0.8\pm0.8$ \\
SMM 7 & $1.6\pm0.9$ & $8.9^{+4.8}_{-2.1}$ & $1.7\pm0.8$\\
\hline 
\end{tabular} 
\tablefoot{
\tablefoottext{a}{Derived by assuming $\beta=1.9$ (see text).}\tablefoottext{b}{Calculated by assuming $T_{\rm dust}=T_{\rm kin}$ as derived from NH$_3$ observations.}}
\end{minipage} 
\end{table}

\subsection{SEDs}

\begin{figure*}
\begin{center}
\includegraphics[scale=0.33]{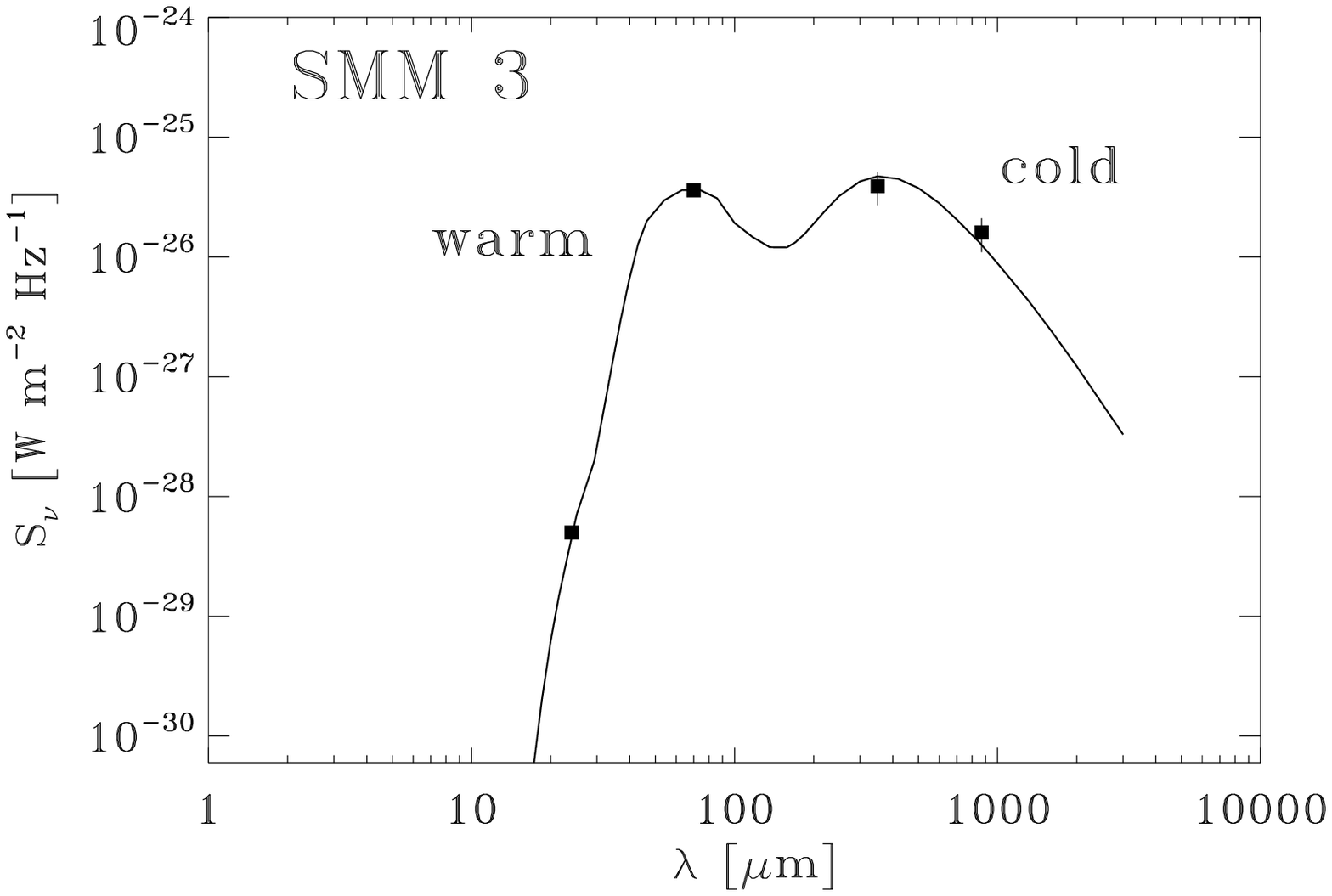}
\includegraphics[scale=0.33]{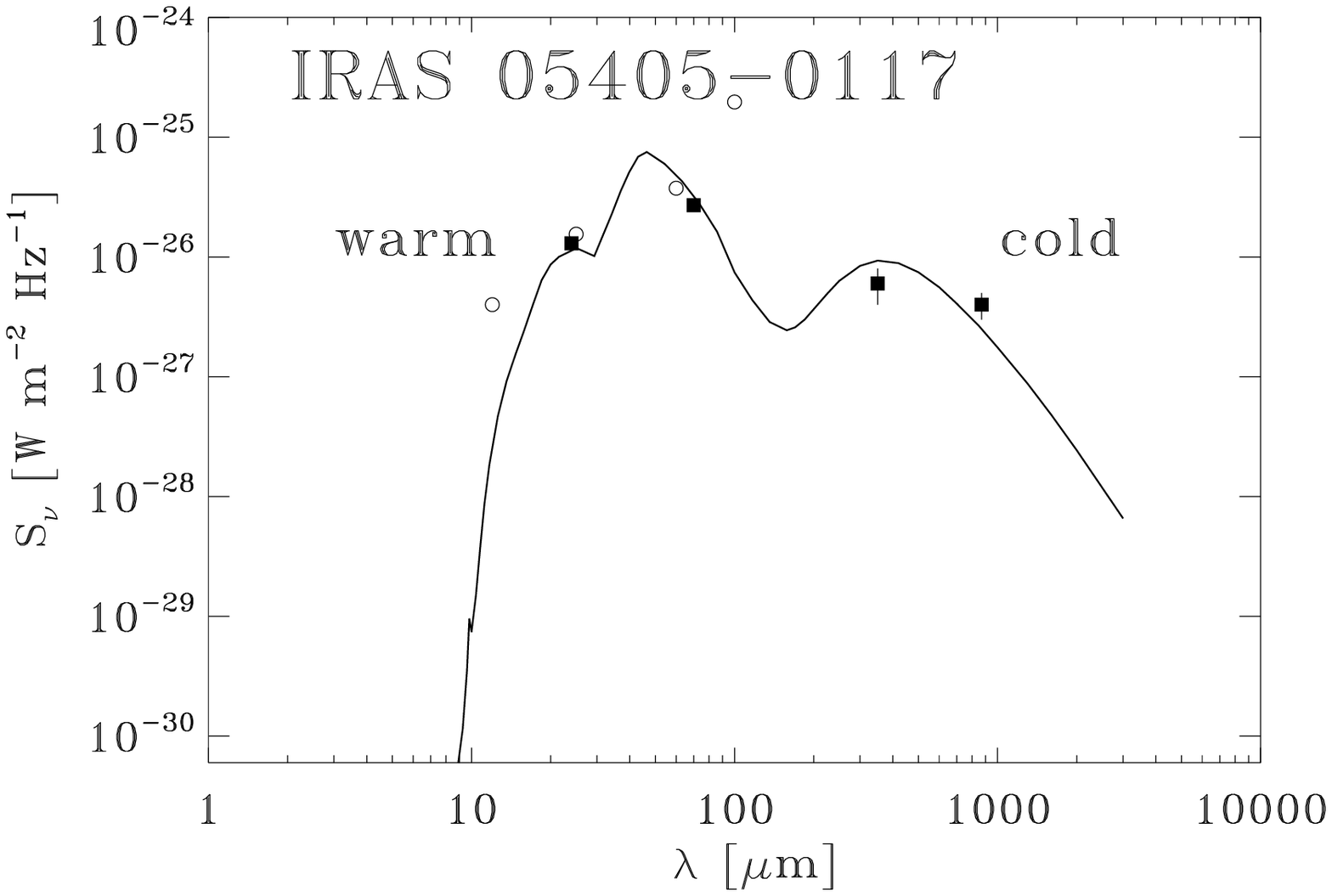}
\includegraphics[scale=0.33]{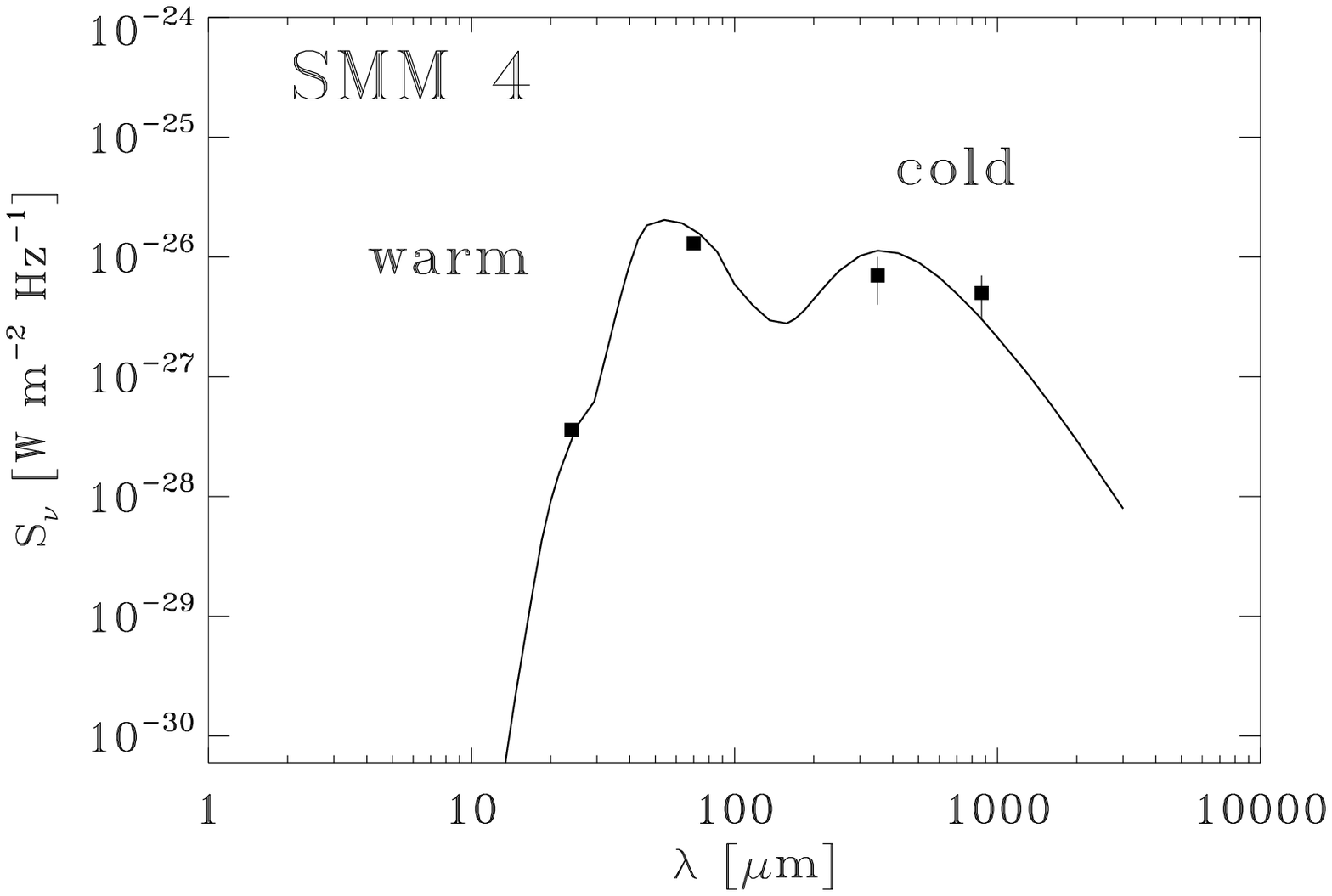}
\caption{Spectral energy distributions of three protostellar cores in 
Orion B9 built from \textit{Spitzer} 24 and 70 $\mu$m, SABOCA 350 $\mu$m, 
and LABOCA 870-$\mu$m flux densities. Open circles in the middle panel 
represent IRAS data points at 12, 25, 60, and 100 $\mu$m (not used in the 
fit). The solid lines correspond to the two-temperature model fits to the 
data. Error bars ($1\sigma$) are indicated for all data points, but are mostly 
smaller than the symbol size. Note the appearance of a 10-$\mu$m silicate 
absorption feature in the SED of IRAS05405, and the absorption ``knee'' at 
$\sim30$ $\mu$m (most notably towards IRAS05405) where a considerable change 
of $\kappa_{\nu}$ occurs at the dust model used (OH94 and references therein).}
\label{figure:sed}
\end{center}
\end{figure*}

With the aid of our new 350 $\mu$m data, we were able to refine some 
of the protostellar SEDs presented in Paper I. As in Paper I, the SEDs 
constructed from the 24, 70, 350, and 870-$\mu$m flux densities were fitted 
by the sum of two modified blackbody curves of cold and warm temperatures 
(the original version of the fitting routine was written by J.~Steinacker).
Again, the dust model used to fit the SEDs is the OH94 model of thick 
ice mantles as adopted earlier in Sect.~4.1, incorporating a dust-to-gas mass 
ratio of 1/100. The SEDs of SMM 3, IRAS05405, and SMM 4 are shown in 
Fig.~\ref{figure:sed}. Note that in Paper I we utilised the IRAS flux 
densities to build the SED of IRAS05405. In the present paper we have 
ignored the IRAS data from the fit because the $\sim$arcminute resolution 
of the IRAS observations also includes emission from the nearby core SMM 4, 
confusing the emission from IRAS05405. This probably explains why the 
100-$\mu$m flux density of IRAS05405 appears so high ($\sim19.7$ Jy). 

The SED fitting results are shown in Table~\ref{table:sed}. 
Column~(2) of Table~\ref{table:sed} gives the mass of the cold 
component, which represents the mass of the cold envelope, 
$M_{\rm cold}\equiv M_{\rm env}$. In Cols.~(3) -- (5), we list the 
luminosities of the cold and warm component, and the bolometric luminosity, 
$L_{\rm bol}=L_{\rm cold}+L_{\rm warm}$. The temperature of the cold component, 
$T_{\rm cold}$, is given in Col.~(6). Columns~(7) and (8) give the 
$L_{\rm cold}/L_{\rm bol}$ and $L_{\rm submm}/L_{\rm bol}$ ratios, where the 
submm luminosity, $L_{\rm submm}$, is defined to be the luminosity longward of 
350 $\mu$m. In the last co\-lumn of Table~\ref{table:sed} we give the 
normalised envelope mass, $M_{\rm env}/L_{\rm bol}^{0.6}$ (\cite{bontemps1996}). 
The latter parameter, which is related to the outflow activity, decreases with 
time and can be used to further constrain the core evolu\-tionary stage.
We do not report the temperature and mass of the warm component, 
because \textit{i)} the OH94 dust model of grains covered by thick ice mantles 
we have adopted may not be appropriate for the warm dust component; and 
\textit{ii)} dust emission may not be optically thin at 24 and 70 $\mu$m, so 
the masses and temperatures of the warm component are not well constrained. 
In contrast, the luminosity of the warm component is not affected by these 
opacity effects.

\begin{table*}
\caption{Fitting results of the SEDs.}
\begin{minipage}{2\columnwidth}
\centering
\renewcommand{\footnoterule}{}
\label{table:sed}
\begin{tabular}{c c c c c c c c c}
\hline\hline 
Source & $M_{\rm cold}$\tablefootmark{a} & $L_{\rm cold}$ & $L_{\rm warm}$ & $L_{\rm bol}$\tablefootmark{b} & $T_{\rm cold}$ & $L_{\rm cold}/L_{\rm bol}$ & $L_{\rm submm}/L_{\rm bol}$ & $M_{\rm env}/L_{\rm bol}^{0.6}$\\ 
 & [M$_{\sun}$] & [L$_{\sun}$] & [L$_{\sun}$] & [L$_{\sun}$] & [K] &  &  & [M$_{\sun}$/L$_{\sun}^{0.6}$]\\
\hline
SMM 3 & $8.2\pm2.6$ & $0.3\pm0.1$ & $0.9\pm0.1$ & $1.2\pm0.1$ & 8.0 & $0.3\pm0.1$ & 0.1 & $7.4\pm3.2$\\
IRAS 05405-0117 & $1.6\pm0.5$ & $0.06\pm0.02$ & $2.3\pm0.2$ & $2.3\pm0.2$ & 8.0 & $0.03\pm0.01$ & 0.01 & $1.0\pm0.3$\\
SMM 4 & $2.0\pm0.8$ & $0.07\pm0.03$ & $0.5\pm0.02$ & $0.6\pm0.04$ & 8.0 & $0.1\pm0.05$ & 0.05 & $2.7\pm1.1$\\
\hline 
\end{tabular} 
\tablefoot{
\tablefoottext{a}{$M_{\rm cold}\equiv M_{\rm env}$.}\tablefoottext{b}{$L_{\rm bol}=L_{\rm warm}+L_{\rm cold}$.}}
\end{minipage}
\end{table*}

\subsection{Molecular column densities and fractional abundances}

The beam-averaged column densities of C$^{17}$O, H$^{13}$CO$^+$, DCO$^+$, and 
N$_2$H$^+$ were derived using a one-dimensional spherically symmetric 
non-LTE radiative transfer code called {\tt RADEX} (see Sect.~3.3). 
{\tt RADEX} uses the method of mean escape probabi\-lity for an isothermal and 
homogeneous medium. The molecular data files (collisional rates) used in the 
{\tt RADEX} excitation analysis were taken from the LAMDA database 
(\cite{schoier2005}). The C$^{17}$O, H$^{13}$CO$^+$, DCO$^+$, and N$_2$H$^+$ 
transitions are treated as a hypothetical unsplit transition.
The input parameters in the off-line mode of {\tt RADEX} are the gas kinetic 
temperature, H$_2$ number density, and the width (FWHM) and intensity of the 
spectral line. We used the values of $T_{\rm kin}$ and 
$\langle n({\rm H_2}) \rangle$ listed in Table~\ref{table:sources}. 
However, we multiplied the densities by 1.2 (He/H$_2=0.2$) to take the 
collisions with He into account (see Sect.~4.1 of the {\tt RADEX} 
manual\footnote{{\tt http://www.sron.rug.nl/$\sim$vdtak/radex/radex$_{-}$manual.pdf}}). As the input line intensity we used the main-beam brightness 
temperature, $T_{\rm MB}$. When the source is resolved, $T_{\rm MB}$ is equal to 
the R-J equivalent radiation temperature, $T_{\rm R}$. The 
simulations aim to reproduce the observed line intensity and yield the values 
of $\tau_0$ and $T_{\rm ex}$, and the total column density of the molecule 
($N_{\rm tot}$). We varied $T_{\rm kin}$ and $\langle n({\rm H_2}) \rangle$ 
according to their errors to estimate the uncertainties associated with 
$\tau_0$, $T_{\rm ex}$, and $N_{\rm tot}$. The uncertainties in 
$T_{\rm MB}$ and $\Delta {\rm v}$ were not taken into account, but test 
calculations showed that they lead to errors in column density that are
comparable to those derived from the errors in temperature and density. We 
also note that the use of higher H$_2$ densities for the cores derived from 
the SABOCA map [Col.~(4), Table~\ref{table:properties}] would lead to lower 
column densities of the molecules because then the excitation would be closer 
to thermalisation.

\begin{figure}[!h]
\begin{center}
\resizebox{0.7\columnwidth}{!}{\includegraphics[angle=-90]{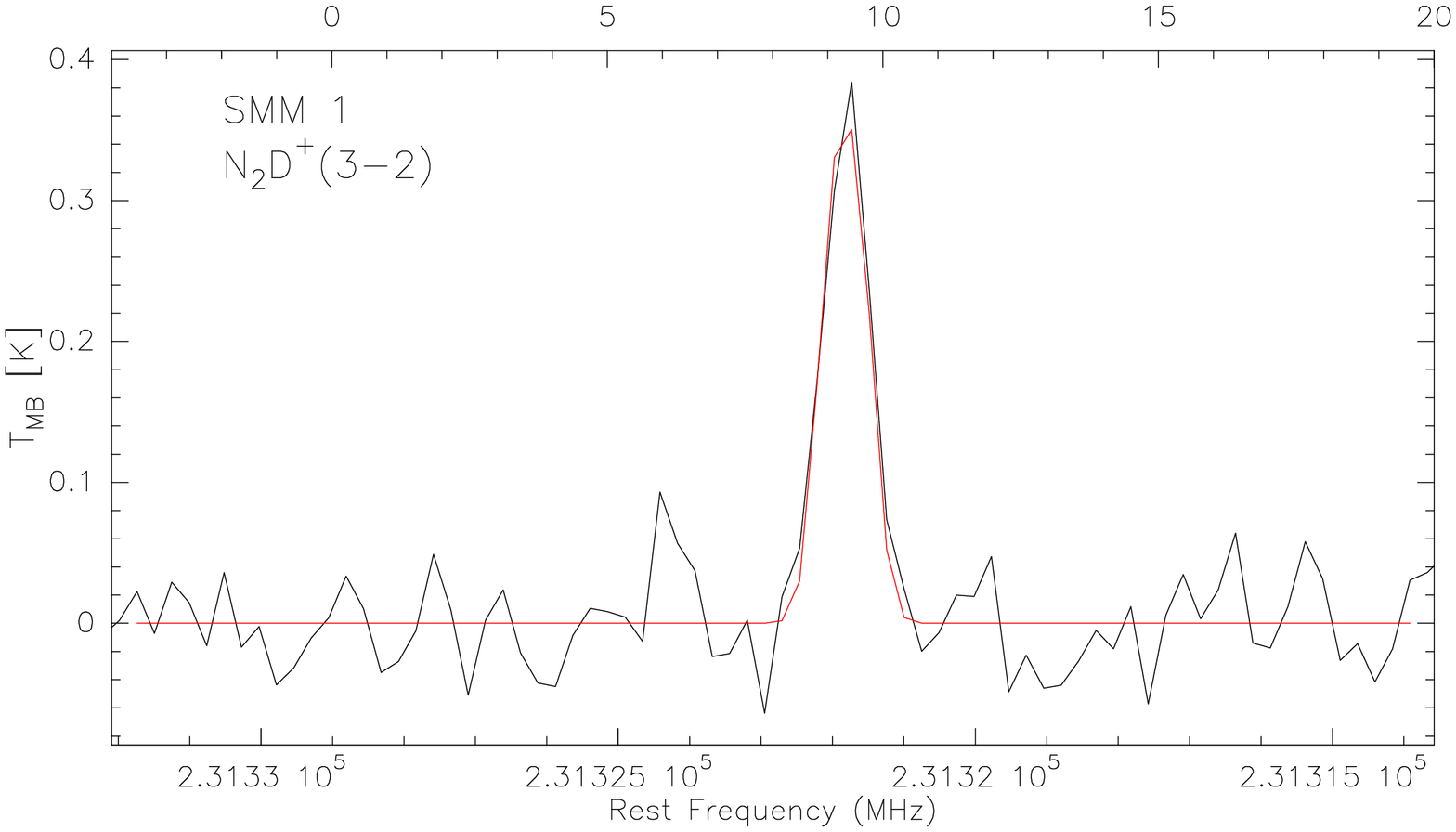}}
\end{center}
\caption{Example of the Weeds LTE modelling outlined in Sect.~4.4. The 
synthetic model spectrum is overlaid as a red line.}
\label{figure:model}
\end{figure}

For N$_2$D$^+$ and \textit{o}-D$_2$CO there are no molecular data files 
available in the LAMDA database. The line optical thicknesses and total 
beam-averaged column densities of these molecules were determined through LTE 
modelling with CLASS/Weeds. The input parameters for a Weeds model are 
$N_{\rm tot}$, $T_{\rm ex}$, source size ($\theta_{\rm s}$), linewidth (FWHM), 
and offset from the reference-channel velocity. The linewidth is directly 
determined from the observed line profile, so there are basically three free 
parameters left ($N_{\rm tot}$, $T_{\rm ex}$, $\theta_{\rm s}$). Some of the model 
parameters may be degenerate, and cannot be determined independently 
(\cite{schilke2006}; \cite{maret2011}). The source size is degenerate with 
excitation temperature in the case of completely optically thick lines 
($\tau \gg 1$), and with column density if the lines are completely optically 
thin ($\tau \ll 1$). We assumed that the source fills the telescope beam, 
i.e., that the beam filling factor is unity\footnote{The $27\arcsec$ 
beam of our N$_2$D$^+$ observations is comparable to the extent of the 
strongest dust emission region within the cores. On the other hand, 
N$_2$D$^+$ emission has been found to trace dust emission very well in low-mass 
dense cores (e.g., \cite{crapsi2005}). Therefore, the assumption of unity 
beam filling factor is reasonable. For \textit{o}-D$_2$CO, the beam filling 
factor may be $<1$. For example, Bergman et al. (2011), using the same 
resolution as we ($27\arcsec$), derived the filling factor of $\simeq0.5$ for 
the \textit{o}-D$_2$CO$(4_{0,\,4}-3_{0,\,3})$ emission region in $\rho$ 
Oph A. Moreover, the filling factor is lower if the cores contain unresolved 
small-scale structure.}, 
and thus the line brightness temperature is $T_{\rm B} \simeq T_{\rm MB}$ 
[see Eqs.~(1) and (2) in Maret et al. (2011)]. For N$_2$D$^+$, we used as 
$T_{\rm ex}$ the values obtained for N$_2$H$^+$ from {\tt RADEX} simulations. 
For the asymmetric top rotor \textit{o}-D$_2$CO we adopted 
the value $T_{\rm ex}=2/3\times E_{\rm u}/k_{\rm B}$, which gives a lower limit 
to the column density (\cite{hatchell1998}). The input $N_{\rm tot}$ was then 
varied until a reasonable fit to the line was obtained 
(see Fig.~\ref{figure:model}). The associated error estimate is based 
on the 10\% calibration uncertainty. In the derivation of column densities, 
the line-strength contribution of hf components within the detected lines 
was taken into account, i.e., the column densities were corrected for the 
fraction of the total line strength given in brackets in Col.~(6) of 
Table~\ref{table:lineparameters}.

We also determined the HCO$^+$ column density from the column density of 
H$^{13}$CO$^+$. For this calculation, it was assumed that the carbon-isotope 
ratio is $[^{12}{\rm C}]/[^{13}{\rm C}]=60$ (\cite{wilson1994}; 
\cite{savage2002}). This value has been used in several previous studies, 
e.g., by Bergin et al. (1999) in their study of dense cores in Orion 
(including IRAS05399)\footnote{Spectral lines of 
$^{12}$C-isotopologue of HCO$^+$ are likely to be optically thick. Therefore, 
the HCO$^+$ deuteration can be better investigated through the 
DCO$^+$/H$^{13}$CO$^+$ column density ratio. However, a caveat should be noted 
here. The HCO$^+$ molecules are produced directly from CO (Sect.~4.6). 
On the other hand, at low temperature, CO is 
susceptible to the exothermic isotopic charge exchange reaction 
${\rm ^{13}C^+}+{\rm ^{12}CO}\rightarrow {\rm ^{12}C^+}+{\rm ^{13}CO}+\Delta E$, where $\Delta E/k_{\rm B}=35$ K (\cite{watson1976}). This is expected to 
cause considerable $^{13}$C-fractionation in cold and dense gas, which 
complicates the deuteration analysis.}. 

\begin{figure}[!h]
\begin{center}
%\resizebox{0.6\columnwidth}{!}{\includegraphics[]{CDs.eps}}
\resizebox{0.7\columnwidth}{!}{\includegraphics[]{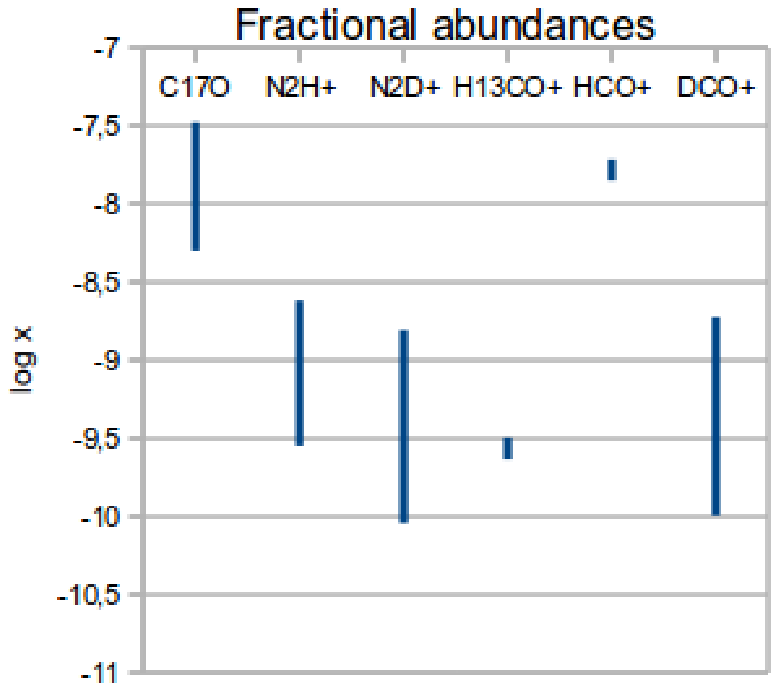}}
\end{center}
\caption{A bar representation of fractional abundances on a 
logarithmic scale. The H$^{13}$CO$^+$ and HCO$^+$ abundances derived towards 
IRAS05399 are shown by a slightly stretch stock for a better illustration.}
\label{figure:abundances}
\end{figure}

We calculated the fractional abundances of the molecules by dividing the 
molecular column density by the H$_2$ column density: 
$x({\rm mol})=N({\rm mol})/N({\rm H_2})$. For this purpose, the va\-lues of 
$N({\rm H_2})$ were derived from the LABOCA dust continuum map smoothed to 
the corresponding resolution of the line observations. The resolution of 
the H$^{13}$CO$^+$ observations (18\arcsec) is slightly better, but comparable, 
to that of the original LABOCA data, and thus no smoothing was done in this 
case. The derived column densities and abundances are listed in 
Table~\ref{table:column}. We stress that the reported uncertainties 
are formal and optimistic, and probably underestimate the true 
uncertainties. A stock chart showing the fractional 
abundances (excluding the additional velocity components) is presented in 
Fig.~\ref{figure:abundances}. The abundance errors were derived by 
propagating the errors in $N({\rm mol})$ and $N({\rm H_2})$. 

\begin{table*}
\caption{Molecular column densities and fractional abundances with respect to 
H$_2$.}
%{\tiny
\begin{minipage}{2\columnwidth}
\centering
\renewcommand{\footnoterule}{}
\label{table:column}
\begin{tabular}{c c c c c c c c}
\hline\hline 
Source & $N({\rm C^{17}O})$ & $N({\rm N_2H^+})$ & $N({\rm N_2D^+})$\tablefootmark{a} & $N({\rm H^{13}CO^+})$ & $N({\rm HCO^+})$ & $N({\rm DCO^+})$ & $N(o-{\rm D_2CO})$\tablefootmark{a}\\
        & [$10^{14}$ cm$^{-2}$] & [$10^{13}$ cm$^{-2}$] & [$10^{12}$ cm$^{-2}$] & [$10^{12}$ cm$^{-2}$] & [$10^{14}$ cm$^{-2}$] & [$10^{12}$ cm$^{-2}$] & [$10^{12}$ cm$^{-2}$]\\
\hline
IRAS 05399-0121 & $3.4\pm0.5$ & $1.5\pm0.3$ & $3.1\pm0.3$ & $6.4\pm1.0$ & $3.8\pm0.6$ & $7.6\pm0.9$ & $2.0\pm0.2$\\
SMM 1 & $5.6\pm0.5$ & $1.2\pm0.3$ & $11.9\pm1.2$ & \ldots & \ldots & $13.1\pm1.0$ & $1.0\pm0.1$\\
SMM 3 & $1.3\pm0.3$ & $0.8\pm0.2$ & $2.7\pm0.3$ & \ldots & \ldots & $2.3\pm0.3$ & \ldots\\
IRAS 05405-0117 & $1.3\pm0.4$ & $2.3\pm0.3$ & $0.8\pm0.1$ & \ldots & \ldots & \ldots & \ldots\\
SMM 4 & \ldots & \ldots & \ldots & \ldots & \ldots & \ldots & \ldots\\
2nd v-comp.\tablefootmark{b} & $2.0\pm0.5$ & $1.0\pm0.3$ & $9.5\pm1.0$ & $18.2\pm7.8$ & $10.9\pm4.7$ & $10.4\pm5.0$ & \ldots\\
SMM 5 & $1.0\pm0.4$ & $0.4\pm0.1$ & \ldots & \ldots & \ldots & \ldots & \ldots\\
SMM 6 & $1.4\pm0.4$ & $3.0\pm0.3$ & $17.6\pm1.8$ & \ldots & \ldots & $25.9\pm2.0$ & \ldots\\  
Ori B9 N & $0.9\pm0.3$ & $0.1$\tablefootmark{c} & \ldots & \ldots & \ldots & $1.8\pm0.3$ & \ldots\\
2nd v-comp.\tablefootmark{b} & $2.3\pm0.5$ & $0.4\pm0.1$ & \ldots & \ldots & \ldots & $6.4\pm2.0$& \ldots \\
SMM 7 & $5.2\pm0.8$ & $0.5\pm0.1$ & \ldots & \ldots & \ldots & $10.0\pm2.0$ & \ldots\\
\hline
& $x({\rm C^{17}O})$ & $x({\rm N_2H^+})$ & $x({\rm N_2D^+})$ & $x({\rm H^{13}CO^+})$ & $x({\rm HCO^+})$ & $x({\rm DCO^+})$ & $x(o-{\rm D_2CO})$\\
& [$10^{-8}$] & [$10^{-10}$] & [$10^{-10}$] & [$10^{-10}$] & [$10^{-8}$] & [$10^{-10}$] & [$10^{-11}$]\\
\hline 
IRAS 05399-0121 & $1.7\pm0.4$ & $6.5\pm1.9$ & $1.5\pm0.3$ & $2.3\pm0.6$ & $1.4\pm0.4$ & $3.6\pm0.9$ & $9.8\pm2.2$\\
SMM 1 & $2.9\pm0.5$ & $5.4\pm1.6$ & $6.1\pm1.1$ & \ldots & \ldots & $5.8\pm0.9$ & $5.1\pm0.9$\\
SMM 3 & $0.5\pm0.1$ & $3.5\pm1.0$ & $1.1\pm0.2$ & \ldots & \ldots & $1.0\pm0.2$& \ldots\\
IRAS 05405-0117 & $1.5\pm0.5$ & $23.9\pm4.0$ & $0.9\pm0.2$ & \ldots & \ldots & \ldots & \ldots\\
SMM 4 & \ldots & \ldots & \ldots & \ldots & \ldots & \ldots & \ldots\\
2nd v-comp.\tablefootmark{b} & $1.5\pm0.5$ & $6.2\pm2.5$ & $6.8\pm2.0$ & $9.6\pm4.8$ & $5.8\pm2.9$ & $6.3\pm3.5$ & \ldots\\
SMM 5 & $1.5\pm0.6$ & $5.5\pm1.5$ & \ldots & \ldots & \ldots & \ldots & \ldots\\
SMM 6 & $1.3\pm0.4$ & $22.3\pm2.7$ & $15.4\pm1.9$ & \ldots & \ldots & $18.7\pm1.9$ & \ldots\\
Ori B9 N & $2.6\pm1.0$ & $3.3\pm0.6$ & \ldots & \ldots & \ldots & $6.3\pm1.5$ & \ldots\\
2nd v-comp.\tablefootmark{b} & $6.7\pm2.6$ & $13.4\pm5.4$ & \ldots & \ldots & \ldots & $22.9\pm10.2$ & \ldots\\
SMM 7 & $3.3\pm1.0$ & $2.8\pm0.9$ & \ldots & \ldots & \ldots & $5.5\pm1.8$ & \ldots\\
\hline 
\end{tabular} 
\tablefoot{\tablefoottext{a}{The N$_2$D$^+$ and \textit{o}-D$_2$CO column 
densities were derived from Weeds models. For \textit{o}-D$_2$CO we assumed 
that $T_{\rm ex}=2/3\times E_{\rm u}/k_{\rm B}$.}\tablefoottext{b}{For SMM 4-LVC 
($\sim1.6$ km~s$^{-1}$) and Ori B9 N-LVC ($\sim1.9$ km~s$^{-1}$), 
$T_{\rm kin}=10.4\pm1.4$ K and $13.6\pm2.5$ K, respectively 
(Paper II).}\tablefoottext{c}{The associated error derived by varying 
$T_{\rm kin}$ and $\langle n({\rm H_2}) \rangle$ in the {\tt RADEX} 
calculation is negligible.}}
\end{minipage} 
\end{table*}

\subsection{CO depletion and deuterium fractionation}

To estimate the amount of CO depletion in the cores, we 
calculated the CO depletion factor, $f_{\rm D}$. If $x({\rm CO})_{\rm can}$ is 
the ``canonical'' (undepleted) abundance, and $x({\rm CO})_{\rm obs}$ is the 
observed CO abundance, $f_{\rm D}$ is given by 

\begin{equation}
\label{eq:depletion}
f_{\rm D}=\frac{x({\rm CO})_{\rm can}}{x({\rm CO})_{\rm obs}}\, .
\end{equation}
We adopted the standard value $x({\rm CO})_{\rm can}=9.5\times10^{-5}$ 
(\cite{frerking1982}). To calculate the ``canonical'' C$^{17}$O abundance we 
assumed the oxygen-isotopic ratio of $[^{16}{\rm O}]/[^{18}{\rm O}]=500$ 
(e.g., \cite{wilson1994}; \cite{williams1998}). When this is combined with the 
$[^{18}{\rm O}]/[^{17}{\rm O}]$ ratio, for which we use 
the standard value 3.52 (\cite{frerking1982}), the value of 
$x({\rm C^{17}O})_{\rm can}$ can be calculated as 

\begin{equation}
x({\rm C^{17}O})_{\rm can} = \frac{x({\rm CO})_{\rm can}}{[^{18}{\rm O}]/[^{17}{\rm O}]\times [^{16}{\rm O}]/[^{18}{\rm O}]}= \frac{x({\rm CO})_{\rm can}}{1760}\, .
\end{equation}
The depletion factor $f_{\rm D}$ is then calculated from
$f_{\rm D}=x({\rm C^{17}O})_{\rm can}/x({\rm C^{17}O})_{\rm obs}$. The results are 
listed in Col.~(2) of Table~\ref{table:depletion}, where 
the $\pm$-errors quoted were calculated by pro\-pagating the uncertainty in 
$x({\rm C^{17}O})_{\rm obs}$. 

The degree of deuterium fractionation in HCO$^+$ and N$_2$H$^+$ was calculated 
by dividing the column density of the deuterated isotopologue by its normal 
hydrogen-bearing form as 
$R_{\rm D}({\rm HCO^+})\equiv N({\rm DCO^+})/N({\rm HCO^+})$ and 
$R_{\rm D}({\rm N_2H^+})\equiv N({\rm N_2D^+})/N({\rm N_2H^+})$. The error in 
$R_{\rm D}$ was derived from the errors in the corresponding column densities 
[see Cols.~(3) and (4) of Table~\ref{table:depletion}]. We note that 
the N$_2$H$^+$ and N$_2$D$^+$ column densities were calculated using the 
non-LTE and LTE models, respectively. Although not taken into account here, 
this may introduce an additional error of a factor of a few in $R_{\rm D}$, 
which should be bear in mind. Note, however, that in all cases the 
normal and deuterated forms of the molecule show line emission at similar 
radial velocities and with similar linewidths. Therefore, the two transitions 
are probably tracing the same gas which makes the derived deuteration levels 
reasonable.

\subsection{Fractional ionisation}

Fractional abundances of all the observed ionic species and their 
different isotopologues could be determined only towards IRAS05399 and 
SMM 4-LVC. Therefore, the fractional ionisation and cosmic-ray 
ionisation rate of H$_2$ were estimated only for these two sources. 
A rough estimate of the lower limit to the ionisation degree can be obtained by 
simply summing up the abundances of the ionic species (e.g., 
\cite{casellietal2002a}; Paper I):
\begin{equation}
\label{eq:ion1}
x({\rm e})>x({\rm HCO^+})+x({\rm H^{13}CO^+})+x({\rm DCO^+})+x({\rm N_2H^+})+x({\rm N_2D^+}) \, .
\end{equation}
This is based on the gas quasi-neutrality: the electron abundance equals the 
difference between the total abundances of the cations and anions.
The resulting values are $x({\rm e})>1.5\times10^{-8}$ for IRAS05399, 
and $x({\rm e})>6.1\times10^{-8}$ for SMM 4-LVC. In the outer 
envelope where CO is not heavily depleted, HCO$^+$ is expected to be the main 
molecular ion. On the other hand, we do not have observational constraints on 
the abundances of H$^+$, H$_3^+$ (and its deuterated isotopologues), H$_3$O$^+$, 
and metal ions (such as C$^+$), all of which could play an important role in 
the ionisation level.

When the fractional ionisation in the source is determined, the 
abundance ratio $R_{\rm H}\equiv [{\rm HCO^+}]/[{\rm CO}]$ can be used to infer 
the cosmic-ray ionisation rate of H$_2$, $\zeta_{\rm H_2}$. By deriving a 
steady-state equation for the H$_3^+$ abundance, and applying it in the 
corresponding equation for HCO$^+$, it can be shown that

\small
\begin{equation}
\label{eq:RH}
R_{\rm H}=\frac{[\zeta_{\rm H_2}/n({\rm H_2})]k_{\rm H_3^+}}{\left[\beta_{\rm H_3^+} x({\rm e}) + k_{\rm H_3^+} x({\rm CO})+k_{\rm gr,\,H_3^+} x({\rm g})\right]\left [\beta_{\rm HCO^+} x({\rm e})+k_{\rm gr,\,HCO^+}x({\rm g}) \right]} \, ,
\end{equation}
\normalsize
where $k_{\rm H_3^+}$ is the rate coefficient for the reaction 
${\rm H_3^+}+{\rm CO}\Arrow^{k_{\rm H_3^+}}{\rm HCO^+}+{\rm H_2}$, 
$\beta_{\rm H_3^+}$ and $\beta_{\rm HCO^+}$ are the dissociative 
recombination rate coefficients of H$_3^+$ and HCO$^+$, and $k_{\rm gr,\,H_3^+}$ 
and $k_{\rm gr,\,HCO^+}$ are the rate coefficients for the recombination of 
H$_3^+$ and HCO$^+$ onto dust grains. The values of $k_{\rm H_3^+}$ and 
$\beta_{\rm HCO^+}$ were taken from the UMIST 
database\footnote{{\tt http://www.udfa.net/}} (\cite{woodall2007}), 
whereas $\beta_{\rm H_3^+}$, $k_{\rm gr,\,H_3^+}$, and $k_{\rm gr,\,HCO^+}$ were 
interpolated from Pagani et al. (2009a; Tables A.1 and B.1 therein). For the 
grain abundance, $x({\rm g})$, we used the value $2.64\times10^{-12}$, which is 
based on the grain radius $a=0.1$ $\mu$m, density $\rho_{\rm grain}=3$ 
g~cm$^{-3}$, and the dust-to-gas mass ratio $R_{\rm d}=1/100$ [see, e.g., 
Eq.~(15) in Pagani et al. (2009a)]. To calculate $\zeta_{\rm H_2}$, we adopted 
as $x({\rm e})$ the summed abundance of ionic species. As $n({\rm H_2})$ of 
SMM 4-LVC, we used the density of SMM 4. The obtained values are 
$\zeta_{\rm H_2}\sim 2.6\times10^{-17}$ s$^{-1}$ towards IRAS05399, and 
$\sim 4.8\times10^{-16}$ s$^{-1}$ towards SMM 4-LVC. These 
values should be taken as lower limits in the sense that we have used 
the lower limits to $x({\rm e})$.

\begin{table}
\caption{The degree of CO depletion and deuterium fractionation.}
\begin{minipage}{1\columnwidth}
\centering
\renewcommand{\footnoterule}{}
\label{table:depletion}
\begin{tabular}{c c c c}
\hline\hline 
Source & $f_{\rm D}$ & $R_{\rm D}({\rm HCO^+})$ & $R_{\rm D}({\rm N_2H^+})$\\
\hline
IRAS 05399-0121 & $3.2\pm0.7$ & $0.020\pm0.004$ & $0.207\pm0.046$\\
SMM 1 & $1.9\pm0.3$ & \ldots & $0.992\pm0.267$\\
SMM 3 & $10.8\pm2.2$ & \ldots & $0.338\pm0.092$\\
IRAS 05405-0117 & $3.6\pm1.2$ & \ldots & $0.035\pm0.006$\\
SMM 4 & \ldots & \ldots & \ldots \\
2nd v-comp. & $3.6\pm1.2$ & $0.010\pm0.006$ & $0.950\pm0.302$\\
SMM 5 & $3.6\pm1.4$ & \ldots & \ldots\\
SMM 6 & $4.2\pm1.3$ & \ldots & $0.587\pm0.084$\\
Ori B9 N & $2.1\pm0.8$ & \ldots & \ldots \\
2nd v-comp. & $0.8\pm0.3$ & \ldots & \ldots \\
SMM 7 & $1.6\pm0.5$ & \ldots & \ldots\\
\hline 
\end{tabular} 
\end{minipage}
\end{table}

\section{Discussion}

\subsection{Dust properties}

By fixing the value of $\beta$ to 1.9, we derived the dust tempe\-ratures
in the range $\sim7.9-10.8$ K. These are about 0.5--5.5 K lower than the 
gas temperatures in the same objects as derived from NH$_3$. 
Due to the large uncertainties associated with $T_{\rm dust}$, the 
near equa\-lity $T_{\rm kin}\simeq T_{\rm dust}$ seems possible, as expected at 
high densities where collisional coup\-ling between the gas and dust becomes 
efficient. Theoretical models have shown that in the dense interiors of 
starless cores [$n({\rm H_2})\gtrsim3\times10^4$ cm$^{-3}$], the gas and dust 
temperatures are si\-milar, although the gas can be slightly warmer due to 
cosmic-ray heating (e.g., \cite{galli2002}).

Some of the dust temperatures we derived are very low. In particular, for SMM 6 
we obtained $T_{\rm dust}\simeq 7.9_{-1.8}^{+3.0}$ K. Theoretical models 
(e.g., \cite{evans2001}) and previous observational studies (e.g., 
\cite{schnee2007a}; \cite{crapsi2007}; \cite{harju2008}) have indicated that 
very low gas and dust temperatures ($\sim6-7$ K) can be reached in the dense 
interiors of dense cores.

The dust emissivity spectral indices we derived, $\beta\sim0.5-1.8$, 
are physically reasonable, and suggest that the assumption 
$T_{\rm dust}=T_{\rm kin}$ is valid (cf.~\cite{schnee2005}). 
Furthermore, the $\beta$ values derived towards SMM 3, 5, and 7 are
close to 1.9. We note that the dense gas and dust associated with the 
additional velo\-city components along the line of sight also affect the 
observed dust continuum pro\-perties (e.g., SMM 4). Therefore, the derived 
$\beta$ values for these sources should be taken with caution.

We note, however, that a decrease of $\beta$ from the 'fiducial' 
value 2 to a shallower emissivity spectral index could be related to dust 
grain coagulation in the inner parts of dense cores (e.g., \cite{miyake1993}; 
OH94). Observational stu\-dies have found evidence that $\beta$ decreases as 
a result of grain growth at high densities (e.g., \cite{goldsmith1997}; 
\cite{visser1998}). More recently, Kwon et al. (2009) found that 
$\beta \lesssim1$ for their sample of three Class 0 sources, 
resembling the values we found towards the Class 0 candidates 
IRAS05405 and SMM 4. These results suggest that dust grains in the envelopes 
of Class 0 protostars can grow in size leading to a shallower spectral index 
of dust emissivity. Based on \textit{Herschel} observations 
of cold interstellar clouds detected with the \textit{Planck} satellite,  
Juvela et al. (2011) found that $\beta$ decreases down to $\sim1$ near 
internal heating sources. Radiative transfer modelling, however, suggests 
that such a decrement is due to line-of-sight temperature variations rather 
than changes in the grain properties. On the other hand, some studies of 
low-mass dense cores suggest that $\beta$ could be larger than 2 
(e.g., \cite{shirley2005}; \cite{schnee2010a}; \cite{shirley2011}). 
Also, recent studies of \textit{Planck}-detected cold Galactic clumps indicate 
that $\beta>2$ (\cite{planck2011}). It is 
possible that $\beta$ is anticorrelated with $T_{\rm dust}$, so that in cold, 
dense cores the emissivity spectral index is steeper than 2. 

\subsection{Refined SEDs}

Inclusion of the new 350 $\mu$m data to the source SEDs confirmed 
our earlier protostellar classifications. In Paper I, we classified the 
sources SMM 3, SMM 4, and IRAS05405 as Class 0 objects. For such objects, 
the bolometric temperature is $T_{\rm bol}<70$ K, $L_{\rm submm}/L_{\rm bol}$ ratio 
is $>0.005$, and they are characterised by the ratio 
$M_{\rm env}/L_{\rm bol}^{0.6}\gtrsim0.4$ M$_{\sun}$/L$_{\sun}^{0.6}$ 
(\cite{andre1993}; \cite{bontemps1996}; \cite{andre2000}). All these 
conditions are fulfilled after the refined SED analysis.  

The SED models presented in Paper I suggested 350-$\mu$m flux densities of 
21.7 Jy for SMM 3, 13.5 Jy for IRAS05405, and 11.1 Jy for SMM 4 -- much higher 
than determined in the present work. 
%How the SED results presented in this paper differ from those presented 
%in Paper I ? Inclusion of the 350-$\mu$m flux densities led 
Compared to the SED results in Paper I, the envelope mass of SMM 3 and 
4 is 1.0 M$_{\sun}$ higher and 1.8 M$_{\sun}$ lower, respectively; the mass of 
IRAS05405 remains the same. The refined luminosity is lower by a factor of 
$\sim3$ in all three cases. The temperature 
of the envelope, $T_{\rm cold}$, resulting from the new SEDs is only 8 K for all 
sources. This is much lower than the va\-lues 11.6--16.1 K derived in Paper I. 
%The $T_{\rm warm}$ values are quite similar to the previous estimates except 
%for IRAS05405 where it becomes about 35 K lower due to omitting the IRAS flux 
%densities. 
The $L_{\rm cold}/L_{\rm bol}$ ratios derived here are also much smaller 
than estimated previously. Unlike deduced in Paper I, the cold component does 
not appear to dominate the total source lumino\-sity. 
The $L_{\rm submm}/L_{\rm bol}$ and $M_{\rm env}/L_{\rm bol}^{0.6}$ ratios derived 
here are mostly comparable to those obtained in Paper I. 

%There are a few points related to the generated SEDs that should be raised.
The angular resolution of the \textit{Spitzer} images used 
is about $6\arcsec$ at 24 $\mu$m, and $18\arcsec$ at 70 $\mu$m. Photometry was 
done by point-source fitting, and the aperture size for 
photometry was $13\arcsec$ at 24 $\mu$m, and $35\arcsec$ at 70 $\mu$m 
(Paper I). The submm data points used in the SEDs were obtained at about 
$20\arcsec$ resolution within a $40\arcsec$ aperture. Therefore, the 70, 350, 
and 870-$\mu$m flux densities refer to a similar spatial scale ($\sim0.09$ pc). 
Direct comparison of these data with the 24 $\mu$m data is reasonable because 
we are dealing with 24-$\mu$m point sources. Our model SEDs therefore produce 
an approximation to the core parameters on $\sim0.09$ pc spatial scale.
Note that the sources SMM 4 and 4b are treated as one source and thus the 
corresponding SED should be taken with caution (Sect.~5.7.2). 

%Secondly, the OH94 dust model of grains covered by thick ice mantles we have 
%adopted may not be appropriate for the warm dust component. An embedded 
%protostar is expected to heat the inner envelope causing the ice-mantle 
%composition to change. To test the effect of the adopted 
%dust model, we also fit the SEDs with the thin-ice model. 
%This causes $M_{\rm cold}$ to increase by a factor of $\sim1.2$, 
%\textbf{whereas} $T_{\rm cold}$ does not change. 
%The luminosities are the same within the errors. Thus, 
%the warm-component parameters do not change very dramatically if the grains 
%are assumed to be covered by thin ice mantles.

%Thirdly, the dust emission may not be optically thin at 24 and 70 $\mu$m. 
%Therefore, the masses and temperatures of the warm component should be taken 
%as lower limits only. Moreover, besides opacity effects, the SED is more 
%sensitive to detailed source geo\-metry at short IR wavelengths than at submm 
%wavelengths (e.g., \cite{indebetouw2006}).

\subsection{Molecular column densities and abundances}

The present N$_2$H$^+$ column densities are about 0.6--5 times the 
corresponding LTE column densities presented in Paper II; within the errors, 
the two values are comparable in the case of IRAS05399, SMM 1, 3, and 7.  
Bergin et al. (1999) found the co\-lumn densities 
$N({\rm N_2H^+})\sim 4.0\pm0.3\times10^{12}$ cm$^{-2}$, 
$N({\rm H^{13}CO^+})\sim 6.9\pm1.3\times10^{11}$ cm$^{-2}$, and
$N({\rm DCO^+})\sim 1.4\pm0.2\times10^{12}$ cm$^{-2}$ towards IRAS05399 
using a statistical equilibrium model. Our values, which are derived 
towards a position of about $12\arcsec$ northeast from the target position of 
Bergin et al. (1999), are about $3.8\pm0.8$, $9.3\pm2.3$, and 
$5.4\pm1.0$ times higher, respectively. Moreover, the C$^{18}$O co\-lumn 
density of $\sim 5.8\pm0.2\times10^{15}$ cm$^{-2}$ derived by Bergin 
et al. (1999) towards IRAS05399 implies the value 
$N({\rm C^{17}O})\sim 1.6\pm0.1\times10^{15}$ cm$^{-2}$. The latter value is 
$4.7\pm0.3$ times higher than the value we have obtained. Besides a slightly 
different target position, these discrepancies are likely caused by the fact 
that Bergin et al. (1999) assumed optically thin emission, and the values 
$n({\rm H_2})\sim10^5$ cm$^{-3}$ and $T_{\rm kin}=15$ K in their non-LTE 
excitation analysis.

The C$^{17}$O column densities and abundances we have derived are comparable 
to those found by Bacmann et al. (2002) for their sample of seven 
prestellar cores, and by Schnee et al. (2007b) towards the 
starless core TMC-1C. The N$_2$H$^+$ column densities are also comparable 
to those determined towards several other low-mass starless cores and 
Class 0 protostellar objects (\cite{casellietal2002b}; \cite{crapsi2005}; 
\cite{roberts2007}; \cite{daniel2007}; \cite{emprechtinger2009}; 
\cite{friesen2010a},b). On the other hand, the N$_2$D$^+$ column densities we 
have obtained are generally larger than those derived for other low-mass cores 
in the studies mentioned above. Also, the H$^{13}$CO$^+$ and DCO$^+$ 
column densities and abundances we derive are higher than those obtained by 
Anderson et al. (1999) in the R CrA region, and by Frau et al. (2010) for a 
sample of starless cores in the Pipe Nebula.

The HCO$^+$ abundance of $\sim1.4\times10^{-8}$ we have derived towards 
IRAS05399 is relatively high. HCO$^+$ could, in principle, increase 
in abundance at later evolutionary stages, when the CO gas-phase abundance is 
enhanced, and the reaction 
${\rm H_3^+}+{\rm CO}\rightarrow {\rm HCO^+}+{\rm H_2}$ becomes efficient. 
However, the derived CO depletion factor in IRAS05399 
is relatively large, $f_{\rm D}\sim 3.2$, and thus the high HCO$^+$ 
abundance may be the result of an alternative production pathway, 
${\rm C}^+ + {\rm H_2O}\rightarrow {\rm HCO^+}+{\rm H}$, viable in the shocks 
(\cite{rawlings2000}; \cite{viti2002}). We note that IRAS05399 drives the HH92 
jet, which lies close to the plane of the sky, and the presence of shock is 
therefore plausible (\cite{gredel1992}; \cite{bally2002}).

\subsection{Depletion and deuteration}

The CO depletion factors we have derived are in the range 
$f_{\rm D}\sim 1.6\pm0.5 -10.8\pm2.2$ ($f_{\rm D}=3.6\pm1.2$ 
for SMM4-LVC and $0.8\pm0.3$ for Ori B9 N-LVC). Interestingly, 
the strongest depletion is observed towards the protostellar core SMM 3, and 
the lowest $f_{\rm D}$ value is found in the starless core SMM 7. This could 
be caused by the fact that we do not observe exactly towards the submm peak 
of SMM 7, whereas we probe the dense envelope of SMM 3 
(see Figs.~\ref{figure:laboca} and 
\ref{figure:saboca}). The second lowest $f_{\rm D}$ value, $1.9\pm0.3$, 
is seen towards the starless core SMM 1. Bergin et al. (1999) mapped the 
IRAS05399/SMM 1-system in C$^{18}$O, CS, H$^{13}$CO$^+$, and DCO$^+$, 
and found that the emission peaks coincide with the position of SMM 1, 
in agreement with our finding of low CO depletion. In general, the gas-phase 
CO abundance is expected to decrease during the prestellar phase of core 
evolution as a result of freeze out onto grain surfaces. 
During the protostellar phase, on the other hand, CO is expected to be 
released back into the gas phase via ice-mantle sublimation in the warmer 
envelope surrounding the protostar.

Caselli et al. (2008) estimated the value $f_{\rm D}({\rm CO})=3.6$ towards a 
position at the edge of Ori B9 N using the data from Caselli \& Myers 
(1995) and Harju et al. (2006). This is higher than the value 
$2.1\pm0.8$ we have derived towards our target position near Ori B9 N (also 
at the core edge). For comparison, Bacmann et al. (2002) studied the level 
of CO depletion in prestellar cores, and found the values in the range 
4.5--15.5. These are comparable to the values we have obtained. 
Moreover, Bacmann et al. (2002) found a positive correlation between 
$f_{\rm D}$ and $n({\rm H_2})$ of the form 
$f_{\rm D}\propto n({\rm H_2})^{0.4-0.8}$. No correlation was found 
between $f_{\rm D}$ and $n({\rm H_2})$ in the present study.
Emprechtinger et al. (2009) derived the values $f_{\rm D}\sim0.3-4.4$ 
towards a sample of Class 0 protostars. 
%The high value of $f_{\rm D}$ measured 
%in SMM 3 is similar to that recently observed in the very low luminosity 
%object (VeLLO) IRAM 04191+1522 by Takakuwa et al. (2011), 
%$f_{\rm D}=20.0\pm2.8$.

We note that in the depletion analysis we used the value $9.5\times10^{-5}$ 
for the undepleted CO abundance. However, this value is known to vary by a 
factor of $\sim2-3$ between different star-forming regions. For example, Lacy 
et al. (1994) determined the CO abundance of $\sim2.7\times10^{-4}$ towards 
NGC 2024 in Orion B. Adopting the latter value would result in $\sim2.8$ times 
larger depletion factors.

The N$_2$H$^+$ deuteration  degree in the Orion B9 cores is found to be in the 
range $R_{\rm D}({\rm N_2H^+})\simeq0.035-0.992$. The extreme value of 0.992 
measured towards SMM 1 is, to our know\-ledge, the highest level of deuteration 
in N$_2$H$^+$ observed so far. This suggests that the core is chemically 
highly evolved but is in contradiction with the low $f_{\rm D}$ value which 
points towards a younger evolutionary stage. SMM 1 could have been affected 
by the outflow driven by IRAS05399, releasing CO into the gas phase and 
effectively resetting the chemical clock. In this case, the very high 
$R_{\rm D}({\rm N_2H^+})$ value would be remnant of the earlier CO-depleted 
phase, and not yet affected by the gas-phase CO which destroys 
N$_2$H$^+$ and N$_2$D$^+$. An opposite si\-tuation was found by Crapsi et al. 
(2004) towards the chemically evolved dense core L1521F which harbours 
a very low luminosity object or VeLLO (\cite{bourke2006}), where 
$f_{\rm D}\sim15$ but $R_{\rm D}({\rm N_2H^+})$ is only $\sim0.1$. 
For comparison, Crapsi et al. (2005) derived the va\-lues of 
$R_{\rm D}({\rm N_2H^+})$ in the range $\lesssim0.02-0.44$ for their sample of 
starless cores, and Daniel et al. (2007) derived comparable values of 
$R_{\rm D}({\rm N_2H^+})\sim0.07-0.53$ towards another sample of starless 
cores. Also, Roberts \& Millar (2007) found $R_{\rm D}({\rm N_2H^+})$ values 
of $<0.01-0.31$ for low-mass cores. Emprechtinger et al. (2009) 
found the va\-lues $R_{\rm D}({\rm N_2H^+})<0.029-0.271$ towards Class 0 objects.
Pagani et al. (2009a) and Fontani et al. (2011) derived the value 
$R_{\rm D}({\rm N_2H^+})\sim0.7$ at the centre of the starless core L183 and 
towards the high-mass prestellar core candidate G034-G2 MM2, respectively, 
which were the highest fractionations reported before the present work. 

The degree of deuterium fractionation in HCO$^+$ could be derived only towards 
IRAS05399 and SMM 4-LVC with the values 
$R_{\rm D}({\rm HCO^+})\simeq0.020$ and 0.010, respectively. These are 
significantly lower than the corresponding $R_{\rm D}({\rm N_2H^+})$ values, 
in agreement with the results by Roberts \& Millar (2007) and 
Emprechtinger et al. (2009). Such a trend is believed to be caused 
by the role of CO, which is the parent species of HCO$^+$ and DCO$^+$, 
but the main destroyer of H$_2$D$^+$, N$_2$H$^+$, and N$_2$D$^+$ molecules. 
Therefore, molecular deuteration proceeds most efficiently in regions where CO 
is depleted. In the warmer envelope layers where CO is not depleted, 
HCO$^+$ can have a relatively high abundance, resulting in a lower 
line-of-sight average value of $R_{\rm D}({\rm HCO^+})$ compared to 
that of N$_2$H$^+$ (\cite{emprechtinger2009}). Note that we have employed 
the DCO$^+(4-3)$ transition, which is expected to trace the inner part of 
the high-density envelope, where heating by the embedded protostar may 
affect the deuterium chemistry [by lowering the $R_{\rm D}({\rm HCO^+})$ 
ratio]. For comparison, J{\o}rgensen et al. (2004) determined similar  
$R_{\rm D}({\rm HCO^+})$ values of $\lesssim0.001-0.05$ for a sample of 18 
low-mass pre- and protostellar cores.
 
Previous studies of low-mass starless and protostellar cores
have found correlations between the degree of CO depletion and deuterium 
fractionation (\cite{bacmann2003}; \cite{jorgensen2004}; 
\cite{crapsi2005}; \cite{emprechtinger2009}). No correlation was found between 
$f_{\rm D}$ and $R_{\rm D}({\rm N_2H^+})$ in the present study. However, 
the values of $f_{\rm D}$ and $R_{\rm D}({\rm N_2H^+})$ for the protostellar 
cores SMM 3 and IRAS05399 are in rough agreement with 
the finding of Emprechtinger et al. (2009), i.e., that deuteration in the 
envelopes of Class 0 protostars increases as a function of $f_{\rm D}$.

\subsection{Fractional ionisation}

In Paper I, we determined the lower limits to $x({\rm e})$ of a few times 
$10^{-8}$, and upper limits of about six times $10^{-7}$ towards two target 
positions near the clump associated with IRAS05405. 
The lower limits to $x({\rm e})$ derived in the present work are comparable 
to those from Paper I.

The standard relation between the electron abundance and the H$_2$ number 
density is $x({\rm e})\sim1.3\times10^{-5}n({\rm H_2})^{-1/2}$ 
(\cite{mckee1989}). This is based on the pure cosmic-ray ionisation with the 
rate $1.3\times10^{-17}$ s$^{-1}$ and includes no depletion of heavy elements. 
The standard relation yields the values $x({\rm e})\simeq5.5\times10^{-8}$ 
for IRAS05399, and $\simeq6.7\times10^{-8}$ for SMM 4-LVC.
These are roughly comparable to the estimated lower limits to $x({\rm e})$. 
The values of $\zeta_{\rm H_2}$ were found in Paper I to be 
$\sim1-2\times10^{-16}$ s$^{-1}$. These resemble the value derived here 
towards SMM 4-LVC, but are an order of magnitude 
higher than obtained for IRAS05399. Instead, the value 
$\zeta_{\rm H_2}\sim 2.6\times10^{-17}$ s$^{-1}$ derived towards IRAS05399 
is within a factor of two of the standard value. For comparision, 
observations of the Horsehead Nebula in Orion B by Goicoechea et al. (2009) 
could only be reproduced with $\zeta_{\rm H_2}=7.7\pm4.6\times10^{-17}$ s$^{-1}$.

Caselli et al. (1998) determined $x({\rm e})$ in a sample of 24 low-mass 
cores consisting of both starless and protostellar objects. 
Their analysis was based on observations of CO, HCO$^+$, and DCO$^+$, and the 
resulting values were in the range $10^{-8}$ to $10^{-6}$, bracketing the 
values for the Orion B9 cores. They argued that the 
variation in $x({\rm e})$ among the sources is due to variations in metal 
abundance and $\zeta_{\rm H_2}$; the latter was found to span a 
range of two orders of magnitude between $10^{-18}-10^{-16}$ s$^{-1}$. 
Some of this variation could be due to different cosmic-ray flux in the source 
regions. Williams et al. (1998) used observations of 
C$^{18}$O, H$^{13}$CO$^+$, and DCO$^+$ to determine the values 
$10^{-7.5}\lesssim x({\rm e}) \lesssim 10^{-6.5}$ in a similar sample of 
low-mass cores as Caselli et al. (1998), but using a slightly different 
analysis. They deduced a mean value of $\zeta_{\rm H_2}=5\times10^{-17}$ s$^{-1}$.
Applying the same analysis as Williams et al. (1998), 
Bergin et al. (1999) found the ionisation levels of 
$10^{-7.3}\lesssim x({\rm e})\lesssim 10^{-6.9}$ towards more massive cores 
(in Orion) than to those studied by Williams et al.. Bergin et al. (1999) 
determined the fractional ionisation towards IRAS05399 to lie in the range 
$x({\rm e})\sim9.3\times10^{-8}-1.8\times10^{-7}$. 
The lower limit we have derived, $x({\rm e})\sim 1.5\times10^{-8}$, is 
about six times lower than the corresponding value derived by Bergin 
et al. (1999).

Recently, Padovani \& Galli (2011) suggested that the magnetic field threading 
the core affects the penetration of cosmic rays, and can decrease the 
ionisation rate by a factor of $\sim3-4$. The values of $\zeta_{\rm H_2}$ 
determined through observations (so far) would then underestimate the 
inter-core values by the above factor, making the $\zeta_{\rm H_2}$ values more 
comparable with those measured in diffuse clouds. 

\subsection{Deuterated formaldehyde}

We have detected the \textit{ortho} form of doubly deuterated formaldehyde, 
D$_2$CO, towards the prestellar core SMM 1 and the protostellar core  
IRAS05399. We derived the column densities and abundances of this molecule 
to be $1-2\times10^{12}$ cm$^{-2}$ and $\sim 5-10\times10^{-11}$, 
respectively. We note that the high-temperature statistical ortho/para ratio 
of D$_2$CO is 2, and appears to be similar even in cold star-forming cores 
(see \cite{roberts2007}). 

Using the IRAM 30-m telescope, Ceccarelli et al. (1998) detected D$_2$CO 
towards IRAS 16293-2422. It was the first reported detection of D$_2$CO
towards a low-mass star-forming core. A lower limit to the D$_2$CO column 
density they obtained, $\sim10^{14}$ cm$^{-2}$, is much higher than in our 
sources. Ceccarelli et al. (1998) speculated that such a high D$_2$CO column 
density requires eva\-poration of D$_2$CO from the grain mantles, where it is 
expected to be formed during the prestellar phase. Later, Ceccarelli et al. 
(2001) demonstrated that heating by the central protostar in IRAS 16293-2422 
is responsible for the mantle evaporation, and injection of D$_2$CO into 
the gas phase. Loinard et al. (2002) detected D$_2$CO towards 19 
low-mass protostellar cores, and found the D$_2$CO/H$_2$CO abundance ratios 
of $\sim0.02-0.4$. Bacmann et al. (2003) and Roberts \& Millar (2007) found 
D$_2$CO column densities of $\sim0.5-2.7\times10^{12}$ cm$^{-2}$ towards a 
sample of low-mass prestellar and protostellar cores. These column densities 
are similar to those we have derived. However, the D$_2$CO abundances 
derived by Bacmann et al. (2003) are somewhat lower than those we have derived.
Recently, Bergman et al. (2011) found the D$_2$CO column densities of 
$\sim2\times10^{12}$ cm$^{-2}$ towards a few positions in $\rho$ Oph A, which 
are very similar to our values; towards the D-peak of $\rho$ Oph A, however, 
they derived a high value of $\sim3.2\times10^{13}$ cm$^{-2}$.   

D$_2$CO is expected to evaporate from the CO-rich grain mantles when the
dust temperature exceeds about 25 K (\cite{ceccarelli2001}). This would 
agree with the low CO depletion factor of 1.9 derived towards SMM 1, 
because the CO sublimation temperature is about 20 K 
(e.g., \cite{aikawa2008}). It is uncertain, however, why we see higher degree 
of CO depletion towards the nearby protostellar core IRAS05399. 
Moreover, the gas kinetic temperature is only 11.9 K in SMM 1 and 13.5 K in 
IRAS05399. It is possible, that the presence of 
gas-phase D$_2$CO in these two sources is due to a non-thermal desorption 
mechanism (\cite{roberts2007}). Bergman et al. (2011) suggested that 
cosmic-ray heating and the formation energy of the newly formed species could 
play an important role in the release of D$_2$CO from the grain mantles in 
starless cores. In addition to these two mechanisms, shocks caused by the 
protostellar outflow driven by IRAS05399 could be responsible in releasing 
D$_2$CO from the icy grain mantles in the IRAS05399/SMM 1-region.
Towards IRAS05399 the emission is perhaps dominated by the cool 
envelope whereas in SMM 1 we might see a component which is interacting with 
the outflow (causing the high gas-phase CO abundance).

\subsection{Fragmentation in the Orion B9 cores}

\subsubsection{SMM 6 -- a fragmented prestellar core}

\begin{figure}[!h]
\resizebox{1.0\hsize}{!}{\includegraphics[angle=0]{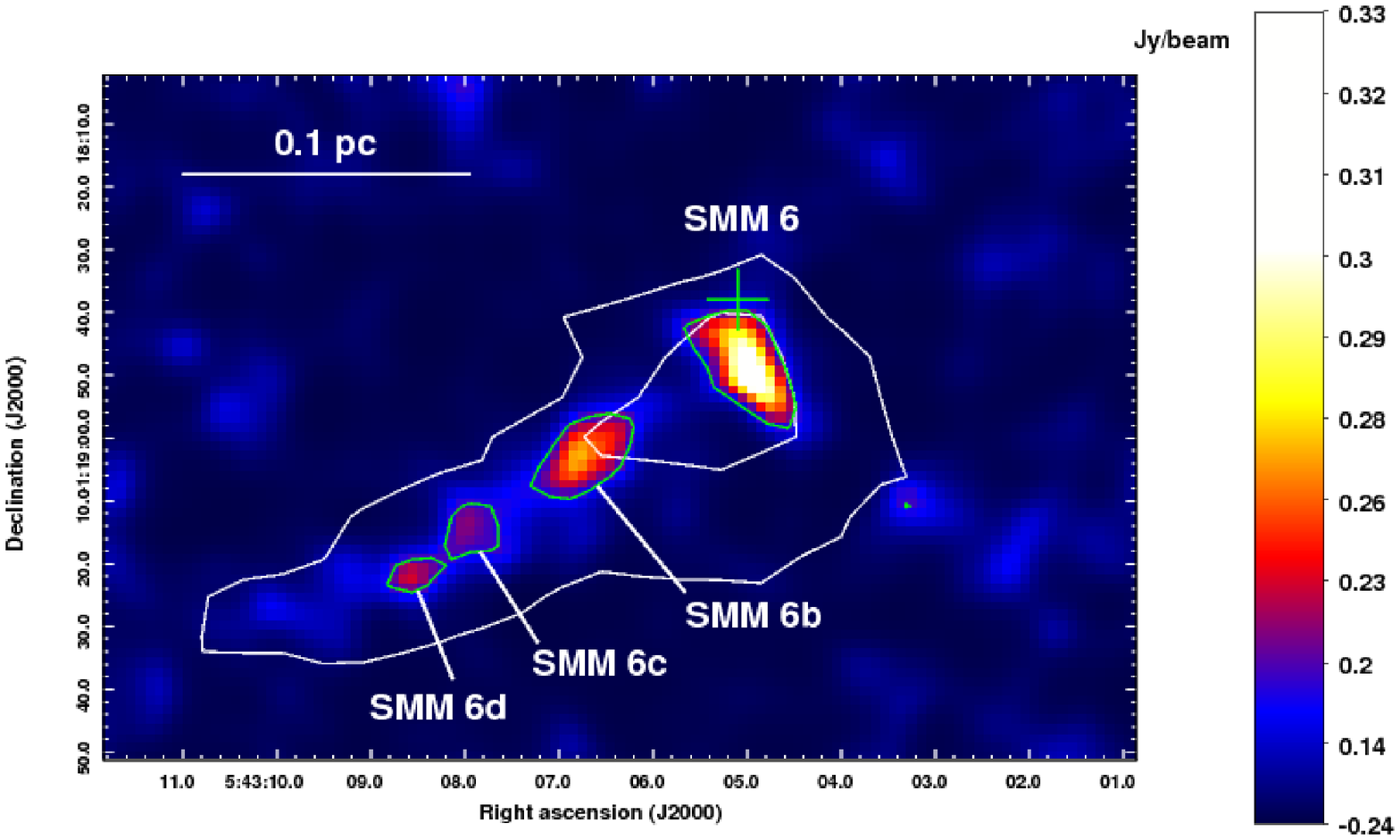}}
\caption{Detailed 350-$\mu$m SABOCA image of the fragmented prestellar 
core SMM 6 with power-law scaling. The contours are as in 
Fig.~\ref{figure:saboca}. The green plus sign indicates the position 
of our molecular-line observations. The colour-bar scale corresponds to 
Jy~beam$^{-1}$.}
\label{figure:SMM6}
\end{figure}

Figure~\ref{figure:SMM6} shows the SABOCA 350-$\mu$m image of the prestellar 
core SMM 6 overlaid with the LABOCA 870-$\mu$m contours. The core is 
filamentary in shape, and it is resolved into three to four 
subcondensations at 350 $\mu$m. The projected linear 
extent of core's long axis is about 1\farcm9 or 0.25 pc, and the core's
mass-per-length is $M_{\rm line}\simeq33$ M$_{\sun}$~pc$^{-1}$. 
The length-to-width ratio increases from about 1.9 at the NW end to about 7.4 
at the SE end. The projected separation between the condensations is 
$\sim29\arcsec$ or $\sim0.06$ pc, where the sources 6c and 6d are treated as 
a single fragment. The measured separations should be taken as lower 
limits because of the possible projection effects.  

To analyse the fragmentation of SMM 6 in more detail, we calculate its
thermal Jeans length from $\lambda_{\rm J}=c_{\rm s}^2/(G \Sigma_0)$, where 
$c_{\rm s}=\sqrt{k_{\rm B}T_{\rm kin}/\mu m_{\rm H}}$ is the 
isothermal sound speed (at $T_{\rm kin}=11$ K), $G$ is the gravitational 
constant, and $\Sigma_0=\mu m_{\rm H}N({\rm H_2})$ is the surface density; 
$\mu$ is the mean molecular weight per free particle (2.33 for He/H$=0.1$), 
and $N({\rm H_2})$ refers to the central column density for which we use the 
value $\sim10^{22}$ cm$^{-2}$ [see Col.~(3) of Table~\ref{table:properties}]. 
The resulting Jeans length is $\sim0.05$ pc, very si\-milar to the 
observed separation between the condensations. In Paper II, we found from 
NH$_3$ measurements that turbulent motions within SMM 6 are subsonic, so 
their contribution to the effective sound speed would not increase the 
calculated Jeans length much. Assuming spherical geometry, the local Jeans 
mass of SMM 6 is 
$M_{\rm J}=4\pi/3\times \langle \rho \rangle (\lambda_{\rm J}/2)^3\sim2.2$ 
M$_{\sun}$. The cor\-responding Jeans number is $n_{\rm J}=M/M_{\rm J}\sim4$, 
which is similar to the number of observed subfragments. If the core 
substructure is caused by gravitational fragmentation, $n_{\rm J}$ is 
expected to give an approximate number of subfragments within the core (e.g., 
\cite{rathborne2007}).

The measured radii of the individual condensations, $\sim0.01-0.02$ pc, and 
their masses, $\sim0.1-0.2$ M$_{\sun}$, are smaller than the Jeans lengths of 
$\lambda_{\rm J}=\sqrt{\pi c_{\rm s}^2/G \rho}\sim0.05-0.06$ pc\footnote{For the 
condensations we utilise the $\lambda_{\rm J}$-formula which assumes spherical 
symmetry. For the filamentary parent core, we calculated the central Jeans 
length $\lambda_{\rm J}\propto1/\Sigma_0$.}, and Jeans 
masses of $\sim0.8-0.9$ M$_{\sun}$ [assuming $T_{\rm kin}=11$ K and densities 
from Col.~(4) of Table~\ref{table:properties}]. Because the subcondensations 
are presumably colder than 11 K, their masses and densities are likely to be 
higher. For instance, using the dust temperature $T_{\rm dust}=7.9$ K derived 
earlier for SMM 6 would result in about 4.4 times higher masses and densities.
Lower temperature would also cause the Jeans length and mass to 
be smaller. At 7.9 K, these would be $\sim0.02$ pc and $\sim0.2-0.3$ 
M$_{\sun}$, respectively. Therefore, the observed condensation properties can 
well be comparable to the corresponding Jeans values. 

The above analysis suggests that thermal Jeans instability is 
the dominant process responsible for the core fragmentation. Moreover, 
the process of Jeans-fragmentation appears to have reached 
its final state at scale of the observed subcondensations. Thus, the 
condensations are potential sites to give birth to 
individual stars or small stellar systems. The condensations can  
grow in mass through (competitive) accretion from the pa\-rent core. 
In general, core fragmentation is believed to be the principal me\-chanism 
for the formation of binary and multiple stellar systems (e.g., 
\cite{tohline2002}; \cite{goodwin2007}).

Besides the Jeans analysis, it is interesting to examine whether 
SMM 6, owing to its filamentary shape, is unstable to axisymmetric 
perturbations. For an unmagnetised isothermal fi\-lament the instability is 
reached if its $M_{\rm line}$ value exceeds the critical equilibrium value of 
$M_{\rm line}^{\rm crit}=2c_{\rm s}^2/G$ (e.g., \cite{ostriker1964}; 
\cite{inutsuka1992}). For SMM 6, 
$M_{\rm line}^{\rm crit}\approx18$ M$_{\sun}$~pc$^{-1}$, and 
$M_{\rm line}/M_{\rm line}^{\rm crit}\approx1.8$. 
Thus, SMM 6 appears to be a thermally supercritical filament susceptible 
to fragmen\-tation, in agreement with the detected substructure. 
We note that in the case of a magnetised molecular-cloud filament, 
$M_{\rm line}^{\rm crit}$ differs by only a factor of order unity from that 
of an unmagnetised case (\cite{fiege2000}).

The three-dimensional velocity dispersion in SMM 6 is 
$\sigma_{\rm 3D}=0.384$ km~s$^{-1}$ (from the NH$_3$ data in Paper II). 
This can be used to estimate the perturbation timescale, or the signal 
crossing time, $\tau_{\rm cross}\equiv D/\sigma_{\rm 3D}$, where the diameter 
is $D=0.25$ pc for SMM 6. The value of $\tau_{\rm cross}$ is about 
$6.3\times10^5$ yr or $\sim2.8$ times the free-fall timescale, 
$\tau_{\rm ff}$. This timescale is comparable to the lifetime of the prestellar 
phase of core evolution of a few times $\tau_{\rm ff}$ (see Papers I and II and 
references therein).

The observed substructure within SMM 6 shows that core fragmentation can take 
place already during the early prestellar phase of evolution, i.e., before the 
formation of embedded protostar(s). Recently, Chen \& Arce (2010) 
suggested that the three subcondensations in the prestellar core R CrA SMM 1A 
were formed through the fragmentation of the elongated parent core in the 
isothermal phase. It seems likely that SMM 6 has undergone a similar 
fragmentation process. The universality of fragmentation of the starless cores 
is unclear, however. For example, Schnee et al. (2010b) found that none of 
their 11 starless cores in Perseus are fragmented into smaller subunits. 
This is important knowledge when comparing the core mass function (CMF) to the 
stellar initial mass function (IMF), because the presence of substructure is 
believed to be one reason for the mass shift between the CMF and the IMF.

\subsubsection{SMM 3, 4, and 7}

\begin{figure}[!h]
\resizebox{1.0\hsize}{!}{\includegraphics[angle=0]{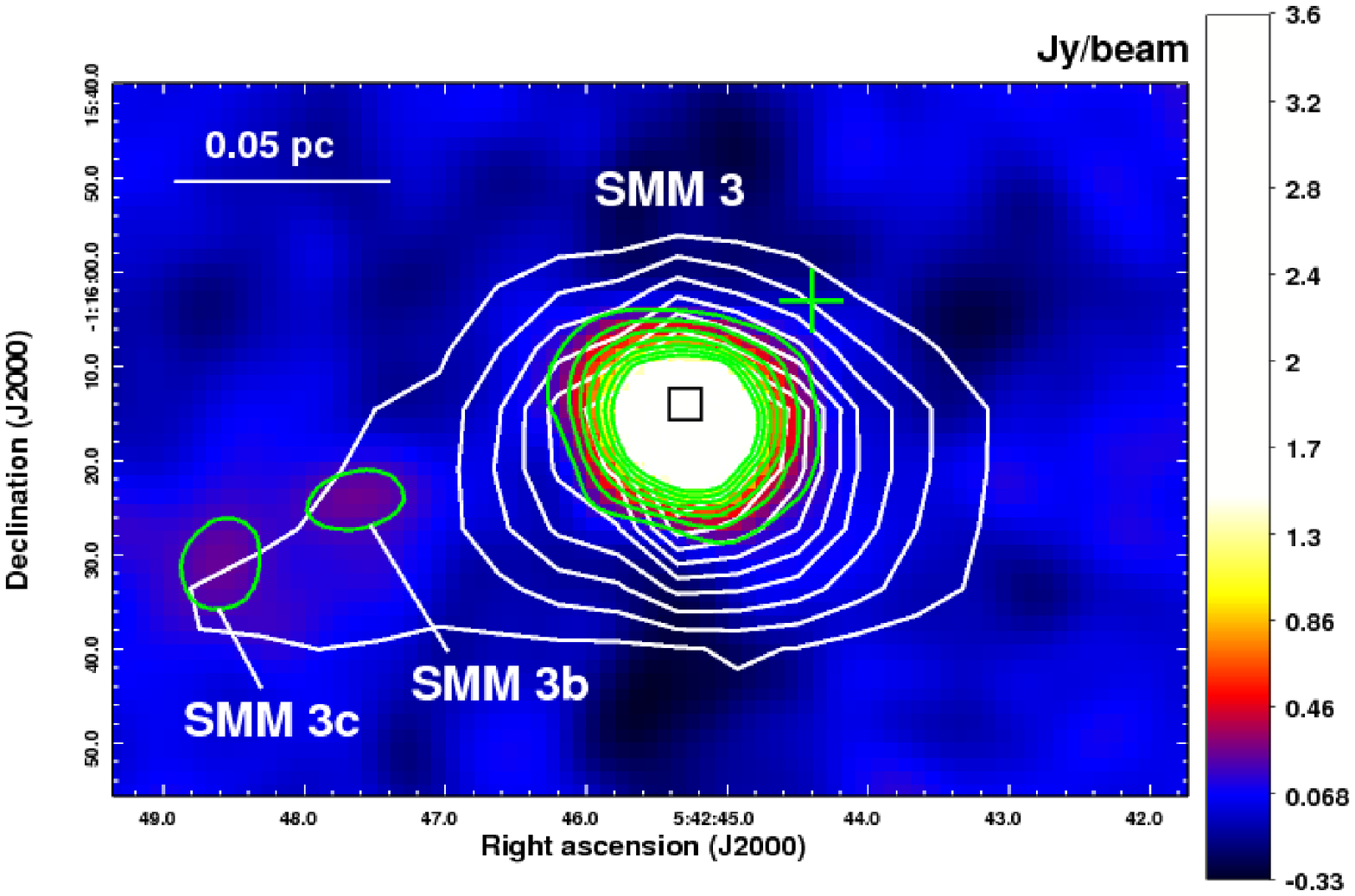}}
\resizebox{1.0\hsize}{!}{\includegraphics[angle=0]{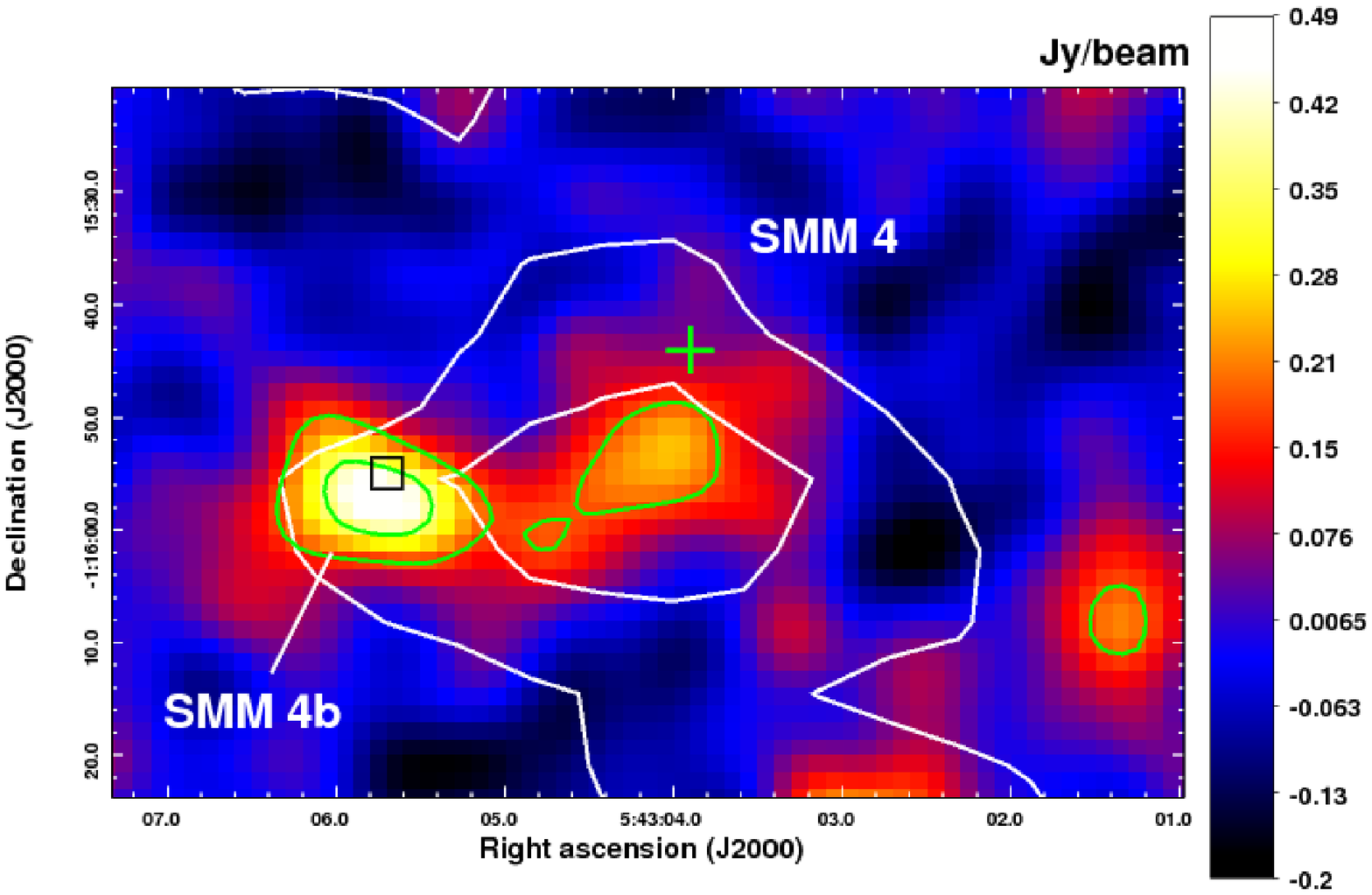}}
\resizebox{1.0\hsize}{!}{\includegraphics[angle=0]{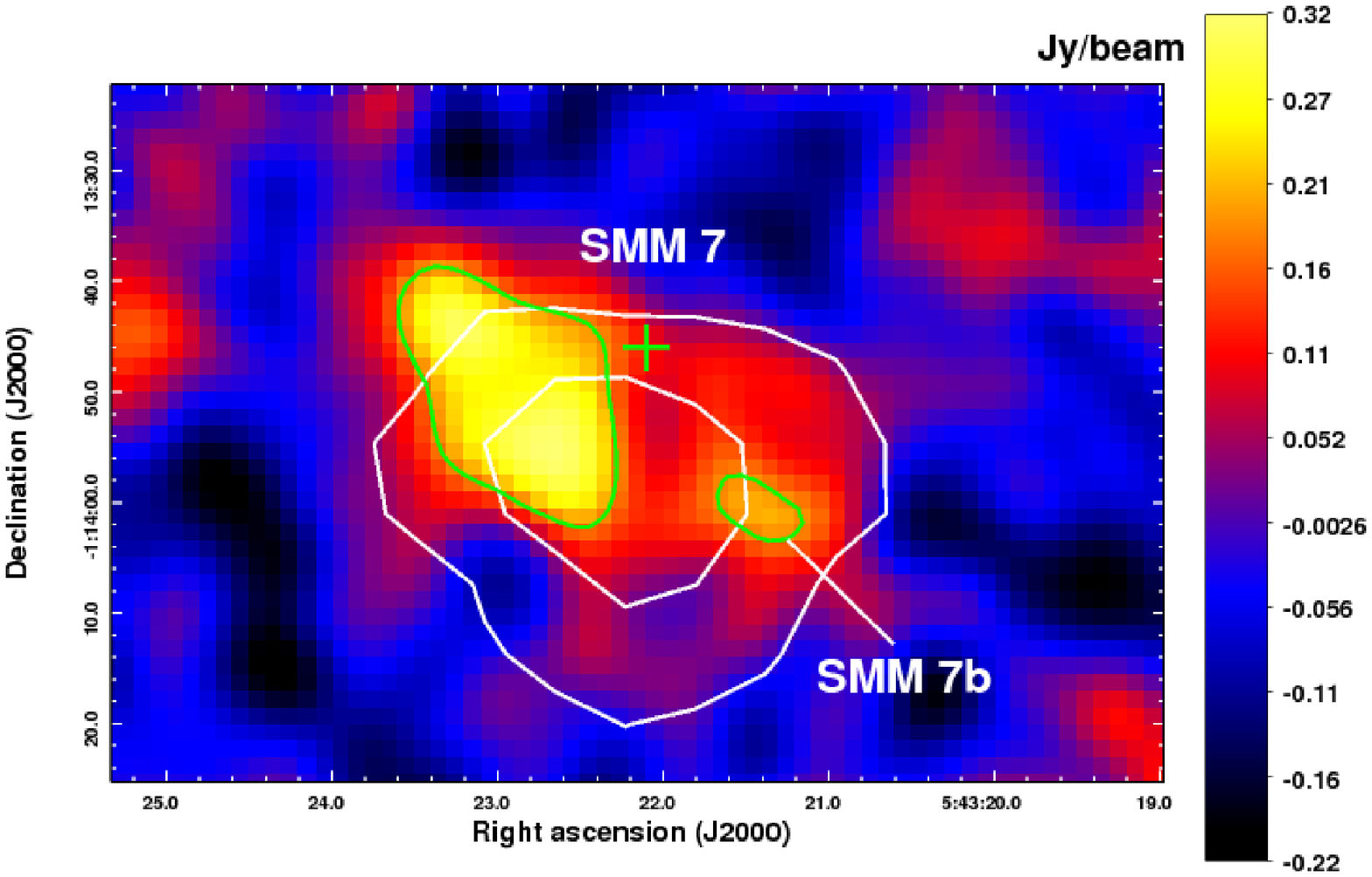}}
\caption{Zoom-in views of Fig.~\ref{figure:saboca} with linear scaling 
towards SMM 3, 4, and 7. The box symbols mark the positions of the 
\textit{Spitzer} 24-$\mu$m point sources. The green plus signs 
show the positions of our molecular-line observations. In each panel, the 
colour-bar scale corresponds to Jy~beam$^{-1}$.}
\label{figure:sources}
\end{figure}

The SABOCA 350-$\mu$m images of SMM 3, 4, and 7 overlaid with 870-$\mu$m 
contours are shown in more detail in Fig.~\ref{figure:sources}. 
In SMM 3, there is a subcondensation, we called SMM 3b, at about 
$36\arcsec$ from the protostar position. Next to that, at the borderline of 
the $3.3\sigma$ LABOCA 870-$\mu$m contour, there is another 
subcondensation, SMM 3c. The projected se\-paration between SMM 3 and 3b 
is about 0.08 pc. This is very close to the thermal Jeans length for the whole 
SMM 3 core of 0.07 pc. In SMM 7, the subcondensation SMM 7b is at 
$\sim26\arcsec$ or 0.06 pc from the 'main' submm peak. Again, this is 
quite close to the local Jeans length of 0.09 pc. It thus seems possible, 
that the substructure within SMM 3 and 7 is caused by thermal Jeans 
fragmentation. We note that the SMM 3/3b-system is qualitatively similar to 
L1448 IRS2/2E, where there is a dust condensation (IRS 2E) next to a Class 0 
protostar; L1448 IRS2E is the most promising candidate of the ``first 
hydrostatic core'' detected so far (\cite{chen2010}).

SMM 4 is resolved into two fragments at 350 $\mu$m. The eastern one, SMM 4b, 
is associated with a \textit{Spitzer} source at 24 and 70 $\mu$m. 
In Papers I and II, we proposed that N$_2$H$^+$ could be depleted in the 
dense envelope of SMM 4. This was based on two observational results: 
\textit{i)} the N$_2$H$^+(1-0)$ map of Caselli \& Myers (1994) shows no 
emission peaks near SMM 4; and \textit{ii)} the N$_2$H$^+(3-2)$ line was 
not detected towards SMM 4 at the systemic velocity $\sim9$ km~s$^{-1}$ 
(see also Fig.~\ref{figure:spectra}). The NH$_3$ lines, on the other hand, 
can be seen at $\sim9$ km~s$^{-1}$ with an additional velocity component at 
about 1.6 km~s$^{-1}$ (Paper II). The present molecular-line observations, 
however, revealed a somewhat surprising result: all the observed lines are 
at an LSR velocity of about 1.5--1.7 km~s$^{-1}$, not at $\sim9$ km~s$^{-1}$. 
The absence of molecular-line emission at $\sim9$ km~s$^{-1}$ is probably not 
due to chemistry, because it is not reasonable to think that all the observed 
species, in particular DCO$^+$ and N$_2$D$^+$, would have been depleted 
(e.g., \cite{lee2003}). Instead, it seems that SMM 4 is a member of the 
“low-velocity” part of Orion B discussed in Paper II. Such a chance 
line-of-sight alignment is surprising, because the nearby sources SMM 5, 
IRAS05405, and Ori B9 N show line emission at $\sim9$ km~s$^{-1}$. 
The reason why we detected the 9-km~s$^{-1}$ NH$_3$ lines towards SMM 4
could be due to the large beam size (40\arcsec) of the observations; NH$_3$ 
emission could have been captured from the nearby cores by the beam. Also, the 
morpho\-logy of the N$_2$H$^+$ integrated intensity map of Caselli \& Myers 
(1994) can be understood if the velocity range used to construct the map 
only covers emission near 9 km~s$^{-1}$; SMM 4 emits at lower velocity and 
does not show up on the map. The dust pro\-perties of SMM 4, such as its SED, 
should be taken with caution as the core consists of subcondensations. It 
also seems possible that the emission from the clump near SMM 4 is 
'contaminated' by an unrelated object(s) along the line of sight.

%\begin{figure*}
%\begin{center}
%\includegraphics[width=0.33\textwidth]{SMM3_new.eps}
%\includegraphics[width=0.33\textwidth]{SMM4_new.eps}
%\includegraphics[width=0.33\textwidth]{SMM7_new.eps}
%\caption{Zoom-in views of Fig.~\ref{figure:saboca} with linear scaling 
%towards SMM 3, 4, and 7. The box symbols mark the positions of the 
%\textit{Spitzer} 24-$\mu$m point sources. \textbf{The green plus signs 
%show the positions of our molecular-line observations.} In each panel, the 
%colour-bar scale corresponds to Jy~beam$^{-1}$.}
%\label{figure:sources}
%\end{center}
%\end{figure*}

\section{Summary and conclusions}

We have carried out a (sub)mm study of dense cores in Orion B9. 
We used APEX to map the region at 350 $\mu$m, and to observe the
transitions of C$^{17}$O, H$^{13}$CO$^+$, DCO$^+$, N$_2$H$^+$, and N$_2$D$^+$. 
These data were compared with our previous LABOCA 870-$\mu$m data.
The principal aim of this study was to investigate the dust emission of the 
cores near the peak of their SEDs, and the degrees of CO depletion, deuterium 
fractionation, and ionisation in the sources. Our primary results are 
summarised as follows:

\begin{enumerate}
\item All the 870-$\mu$m cores within the boundaries of our SABOCA map were 
detected at 350 $\mu$m. The strongest 350-$\mu$m source in the region is SMM 3, 
a candidate Class-0 protostellar core.
\item Four of the 870-$\mu$m cores, namely SMM 3, 4, 6, and 7, were resolved 
into at least two condensations at 350 $\mu$m. In particular, the elongated 
prestellar core SMM 6 was resolved into three to four very low-mass 
subcondensations, showing that core fragmentation can take place 
during the prestellar phase of evolution. In all cases, the origin 
of substructure can be explained by thermal Jeans-type fragmentation. 
\item The dust temperatures derived from the 350-to-870-$\mu$m flux 
density ratio are very low, only 
$T_{\rm dust}\approx 7.9_{-1.8}^{+3.0}-10.8_{-2.6}^{+5.7}$ K. The 
corresponding gas temperatures are typically a few kelvins higher.
We also derived the submm dust emissivity spectral indices using the 
assumption $T_{\rm dust}=T_{\rm gas}$, and they are in the range 
$\beta\approx 0.5\pm0.8-1.8\pm0.6$. The uncertainties in $\beta$ are 
large, and within the errors the values are comparable to the fiducial value 
$\beta=2$.
\item We refined some of the protostellar SEDs presented in Paper I by adding 
the observed 350-$\mu$m flux densities. No ra\-dical changes were found, and 
the sources IRAS 05405-0117 and SMM 3 and 4 are classified as Class 0 objects, 
in agreement with our previous results.
\item The CO depletion factors were found to lie in the range 
$f_{\rm D}\sim 1.6\pm0.5-10.8\pm2.2$. We found no systematic difference in 
$f_{\rm D}$ between the starless and protostellar cores. In accordance with 
previous observations and theoretical predictions the most severe CO depletion 
is seen towards the core with highest average density, SMM 3. The degree of 
deuteration in N$_2$H$^+$ was found to be in the range 
$N({\rm N_2D^+})/N({\rm N_2H^+})\simeq 0.035\pm0.006-0.992\pm0.267$. 
The $N({\rm DCO^+})/N({\rm HCO^+})$ ratio was found to be about 1--2\%, 
comparable to those seen in other low-mass star-forming regions. 
\item The fractional ionisation could only be derived towards IRAS 05399-0121 
and SMM 4-LVC with the lower limits of $x({\rm e})>1.5\times10^{-8}$ and 
$>6.1\times10^{-8}$, respectively. These values are comparable to the 
fractional ionisations we derived earlier towards a few target positions near 
IRAS 05405-0117 and SMM 4. The cosmic-ray ionisation rate of H$_2$ implied by 
the derived lower $x({\rm e})$ limit is 
$\zeta_{\rm H_2}\sim 2.6\times10^{-17}$ s$^{-1}$ towards IRAS 05399-0121, 
and $\sim 4.8\times10^{-16}$ s$^{-1}$ towards 
SMM 4-LVC. The former value, which does not suffer from line-of-sight 
contamination, is within a factor of two of the 'standard' value 
$1.3\times10^{-17}$ s$^{-1}$.
\item The highest degree of deuteration in N$_2$H$^+$, $\sim 0.99$, 
was derived towards the prestellar core SMM 1. To our know\-ledge, this is 
the most extreme level of N$_2$H$^+$ deuteration reported so far. We also 
detected D$_2$CO emission towards SMM 1 with an abundance of several times 
$10^{-11}$. Because D$_2$CO is expected to be formed through the grain-surface 
chemistry, its presence in the gas phase in SMM 1 could be due to shocks 
driven by the jet from IRAS 05399-0121, resulting in the release of D$_2$CO 
from the grain mantles. This conforms to the low CO depletion factor of 
1.9 derived towards SMM 1. The very high N$_2$H$^+$ deuteration could 
be remnant of the earlier CO-depleted phase, and not yet affected by the 
destroying effect of gas-phase CO.
\item It seems likely that the elongated clump associated with IRAS 05405-0117 
and SMM 4 consists of physically independent objects that form a single clump 
only in projection along the line of sight.
\end{enumerate}

In Papers I and II we discussed the origin of dense cores in Orion B9. 
The region could have been affected by the feedback from the nearby Ori OB1b 
subgroup of the Orion OB1 association [resembling the process termed ``cloud 
shuffling'' by Elmegreen (1979)]. This feedback process may have played 
a role in sweeping up the gas into dense shells and filaments out of which 
the dense cores were later fragmented. This picture is supported by the 
kinematics of Orion B9 (low-velocity line-of-sight components). The present 
observational results suggest that the further fragmentation of cores into 
smaller condensations is caused by the thermal Jeans instability. 
The observations presented in this paper also reveal the intricate chemistry 
of the cores, such as heterogeneous depletion and deuteration among the cores. 
This suggests that the individual Orion-B9 cores, though possibly formed 
collectively in the parent filaments, are evolving chemically at different 
rates.

\begin{acknowledgements}

We acknowledge the staff at the APEX telescope for performing the 
service-mode observations presented in this paper. We would also like to 
thank the people who maintain the CDMS and JPL molecular spectroscopy 
databases, and the \textit{Splatalogue} Database for Astronomical Spectroscopy. 
The authors would like to thank the anonymous referee for useful comments. 
We acknowledge support from the Academy of Finland through grants 
127015 and 132291. This work makes use of observations made with the Spitzer 
Space Telescope, which is operated by the Jet Propulsion Laboratory (JPL), 
California Institute of Technology under a contract with NASA. Moreover, this 
research has made use of NASA's Astrophysics Data System and the NASA/IPAC 
Infrared Science Archive, which is operated by the JPL, California Institute 
of Technology, under contract with the NASA.

\end{acknowledgements}

\end{document}